\documentclass[%
superscriptaddress,
reprint,
 amsmath,amssymb,
 aps,
floatfix]{revtex4-2}

\usepackage{bbold}
\usepackage{graphicx}
\usepackage{dcolumn}
\usepackage{bm}
\usepackage{setspace}
\usepackage{mathrsfs}
\usepackage{hyperref}

\usepackage{graphicx}
\usepackage{dcolumn}
\usepackage{bm}
\usepackage[normalem]{ulem}

\usepackage{amsmath}
\usepackage[caption=false]{subfig}

\usepackage{braket}

\usepackage{pgfplots}
\pgfplotsset{width=8.6cm,compat=1.5}

\usepackage{textcomp}

\usepackage{xcolor}

\usepackage{hyperref}
\hypersetup{
    colorlinks=true,
    linkcolor=blue,
    citecolor=blue,      
    urlcolor=blue,
}

\usepackage{csquotes}

\usepackage{bbold}

\usepackage{mathrsfs}

\graphicspath{{Pictures/}}

\begin{document}

\preprint{APS/123-QED}

\title{Feasibility of a trapped atom interferometer with accelerating optical traps}

\author{Gayathrini Premawardhana}
\email{gtp6626@umd.edu}
\affiliation{Joint Center for Quantum Information and Computer Science, University of Maryland-NIST, College Park, Maryland 20742, USA}
\author{Jonathan Kunjummen}
\affiliation{Joint Center for Quantum Information and Computer Science, University of Maryland-NIST, College Park, Maryland 20742, USA}
\author{Sarthak Subhankar}
\affiliation{Joint Quantum Institute, University of Maryland-NIST, College Park, Maryland 20742, USA}
\author{Jacob M. Taylor}
\email{jmtaylor@umd.edu}
\affiliation{Joint Center for Quantum Information and Computer Science, University of Maryland-NIST, College Park, Maryland 20742, USA}
\affiliation{Joint Quantum Institute, University of Maryland-NIST, College Park, Maryland 20742, USA}





\date{\today}


\begin{abstract}

In order to increase the measured phase of an atom interferometer and improve its sensitivity, researchers attempt to increase the enclosed space-time area using two methods: creating larger separations between the interferometer arms and having longer evolution times. However, increasing the evolution time reduces the bandwidth that can be sampled, whereas decreasing the evolution time worsens the sensitivity. In this paper, we attempt to address this by proposing a setup for high-bandwidth applications, with improved overall sensitivity. This is realized by accelerating and holding the atoms using optical dipole traps. We find that accelerations of up to $10^{3}$-$10^{5}$ m/s$^2$ can be achieved using acousto-optic deflectors (AODs) to move the traps. By comparing the sensitivity of our approach to acceleration as a baseline to traditional atom interferometry, we find a substantial improvement to the state of the art. In the limit of appropriate beam and optics stabilization, sensitivities approaching 10$^{-14}$ (m/s$^2$)/$\sqrt{\rm Hz}$ may be achievable at 1 Hz, while detection at 1 kHz with a sensitivity an order of magnitude better than traditional free-fall atom interferometers is possible with today's systems.

\end{abstract}

\maketitle


\section{Introduction}

Atom interferometers can be used for diverse applications, ranging from the exploration of fundamental aspects of physics, such as verifying theoretical predictions, to employment as a measurement device, for instance, an accelerometer or in rotation sensing~\cite{JakeAcceler,RotSensWaveguide,RotSens1,RotSens2}. More concretely, they can be used to extract the values of field parameters, such as the acceleration of free fall $g$~\cite{AtomInterExpHolger}, detect dark matter~\cite{DarkMatterInter} and gravitational waves~\cite{Miga}, and test theories of gravity~\cite{PRXQuantum.2.030330}. Signals can be obtained from volcanoes and earthquakes~\cite{VolcanoInter,MobAtomInter}, and variations in the Earth's gravitational field, resulting from melting ice, can be measured in order to understand climate change~\cite{ClimateInterESA}. The structures of planetary bodies can be investigated and spacecraft navigation improved~\cite{InterNASA,NASAgradiometer,NASAinterInSpace,BECinSpace}. 

Since the measured phase of an atom interferometer is proportional -- and the sensitivity of the interferometer inversely proportional -- to the total space-time area enclosed by the interferometry arms, longer free evolution periods and further distance between the arms is beneficial. However, for high-bandwidth applications, such as in inertial platforms, the total time evolution has to be limited due to the need to observe rapid changes. Thus, there is a necessity to rapidly accelerate atoms in order to create a large distance in a short fixed time. Techniques making use of laser pulses to achieve such accelerations have been developed by the community~\cite{102hk,BlochOscillInterFast,141hk}. 

Rather than using pulses, in this paper we propose achieving such high accelerations by employing highly accelerating optical traps in the middle of the interferometer sequence. This is an extension to existing proposals and/or experiments where interferometry includes translating atoms in a trapping potential ~\cite{EnnoGravRedShift,TractorAtomInter,TractorAtomInterPrinciples,RotSensWaveguide,PhysRevA.108.043305}. As a particular subset of optical traps, tweezer traps, first proposed  by Arthur Ashkin~\cite{ArthurAshkin1,ArthurAshkin2}, have been used in various experiments and their capabilities are well understood. For instance, Kaufman et al. employed manipulable tweezers, a characteristic that would be practically important to us, in their investigation of the interference between atoms placed in two optical tweezers~\cite{TweezerInterference}. Using tweezers to extract and transport various states of one or more atoms have been explored, both theoretically, such as by Diener et al. who provide a proposal for a single atom~\cite{Diener}, and experimentally~\cite{Gustavson,Roberts,BEUGNON}. Experiments have been conducted with an assortment of multi-tweezer setups~\cite{Endres,Lorenz,BEUGNON,TweezerArraySchlosser1,TweezerArraySchlosser2}, and cooling atoms in tweezers~\cite{Lorenz,Kaufman} and loading atoms into tweezers have been given detailed attention~\cite{Lester}. 

\begin{figure*}
    \centering
      \subfloat[]{       
      \includegraphics[width=8.6 cm]{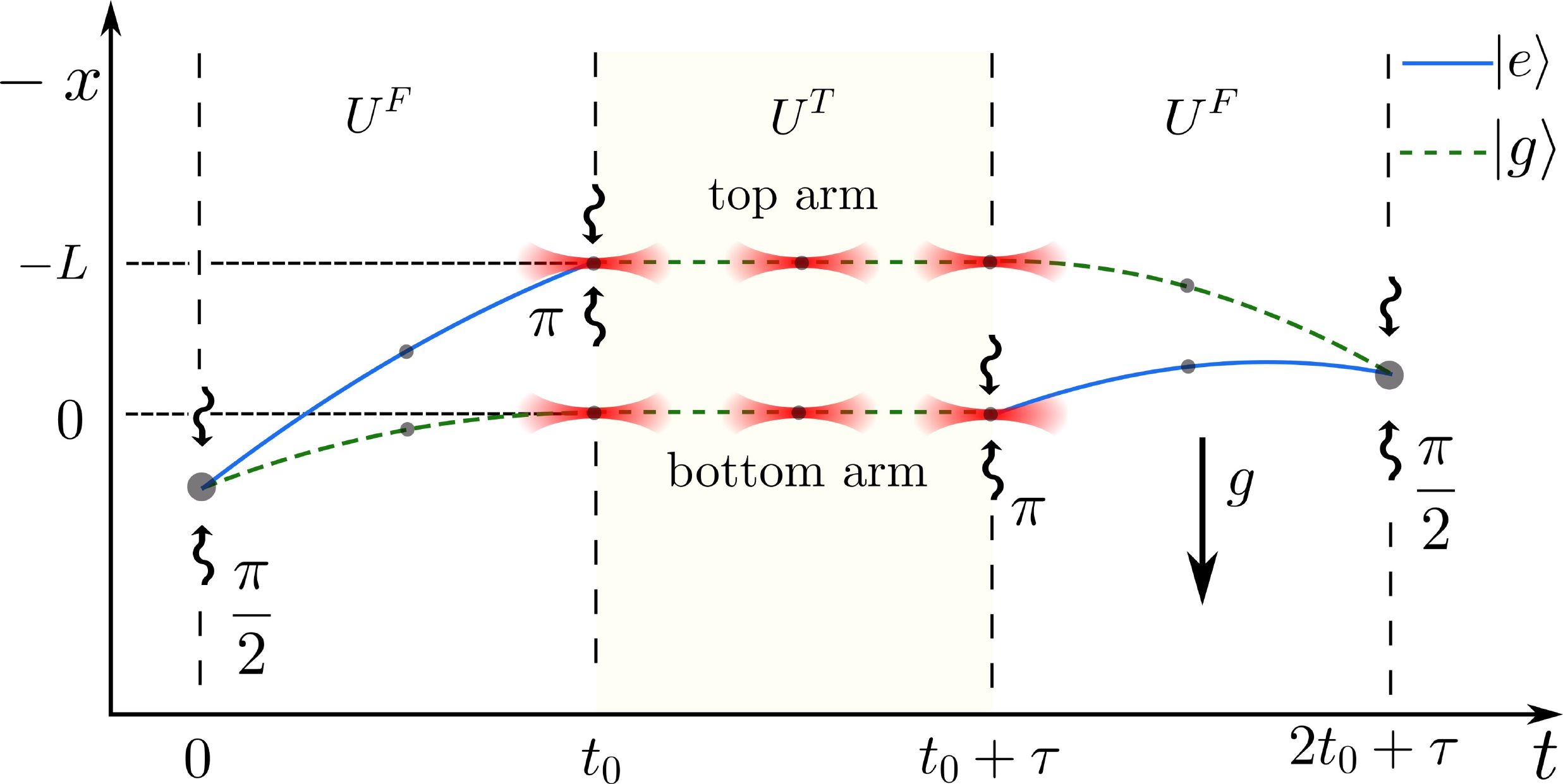}
      \label{fig:InterSequence_a}
     }
     \hfill
     \subfloat[]{     
     \includegraphics[width=8.6 cm]{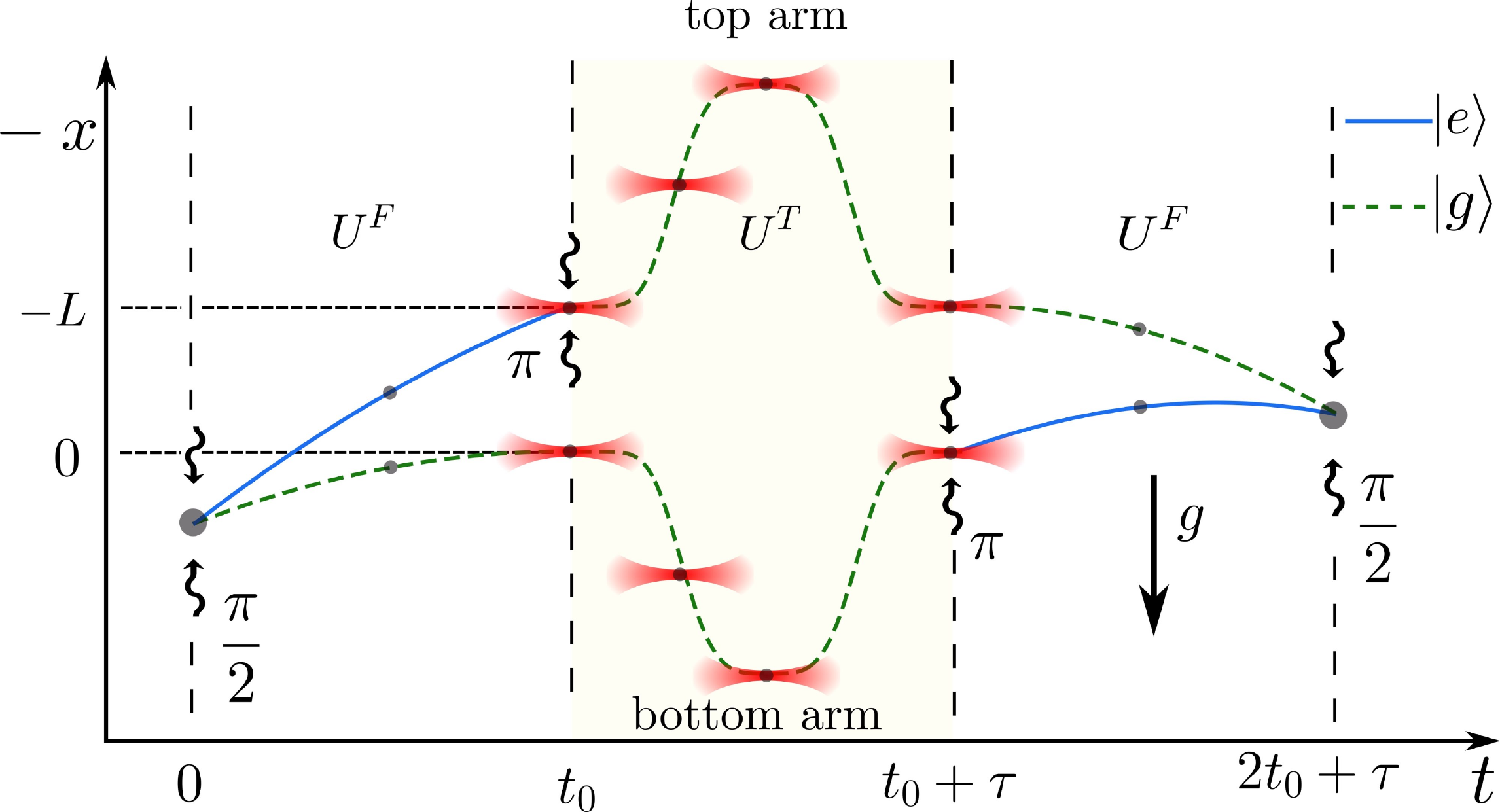}
      \label{fig:InterSequence_b}
    }
    \caption{The interferometer sequence including stationary and accelerating traps~\cite{AtomInterExpHolger}. $U^{F}$ and $U^{T}$ are the free fall and atom trap unitaries. A cloud of atoms is launched at $t<0$. The $\pi/2$ pulse results in two arms, with the top arm having a higher momentum and a different internal state. At $t=t_{0}$, the bottom arm reaches zero momentum. The $\pi$ pulse acts on the top arm to place it in zero momentum and the same internal state as the bottom arm. The atoms spend a total time of $\tau$ in the traps, after which they free fall for time $t_{0}$ and are recombined by another set of $\pi$ and $\pi/2$ pulses. Note that the direction of gravity is along the direction of positive $x$. Hence, the gravitational potential with reference to this diagram will be given by -$mgx$. The diagrams are not to scale. \textbf{(a)} The traps are held stationary for time $\tau$. \textbf{(b)} The traps are given a constant acceleration for some time in the middle of the sequence.}
    \label{fig:InterSequence}
\end{figure*}

We find that advances in optical trapping, largely driven by the creation of programmable neutral atom arrays~\cite{CsTweezerWavelength}, translate to rapid atomic acceleration capabilities. In particular, we propose an interferometer sequence where, mid-sequence, the atoms are trapped in two optical dipole traps, one located at each of the two interferometer arms. The traps are then given a constant acceleration and translated a particular distance before being accelerated back to their original locations. We suggest the traps be accelerated using acousto-optic deflectors (AODs), for which we find that accelerations of up to $10^{3}$-$10^{5}$ m/s$^2$ can be achieved. However, the combination of an AOD with lenses dictates that only a relatively small field of view (FOV) can be covered, because diffraction-limited performance is achievable only within a lens's FOV. 
In this case, we find that at $1$ Hz, for a trap waist of 38 µm ($10^{6}$ atoms), a sensitivity of nearly $10^{-10}$ (m/s$^2)/\sqrt{\rm Hz}$ can be achieved, approximately an order of magnitude better than a purely free-fall interferometer using the same number of atoms. For unrestricted maximum distance, the sensitivity at $1$ Hz can be improved up to $10^{-14}$ (m/s$^2)/\sqrt{\rm Hz}$ for a waist of 38 µm. 

In terms of noise, we expect that due to uncorrelated traps to be the most significant. Recently, researchers have been achieving increased interrogation times for atom interferometers by holding the atoms in a trap, such as an optical lattice, as an alternative to increasing their free fall time~\cite{AtomInterExpHolger}. Our approach, in which each trap is controlled separately in contrast to using a single optical lattice, is unlikely to achieve those long hold times when the trapping beams are not highly correlated due to intensity noise-induced dephasing of the atomic interferometer signal. In particular, the variance in the measured phase due to noise from uncorrelated traps is of the order $10-100$ s$^{-1}$, potentially limiting trapping times to as short as 0.01 s. Nevertheless, if the traps are made to be highly correlated, much longer coherence times are possible, enabling compact devices with long hold times.

We can now try to understand where our proposal stands when applied to certain investigations of fundamental physics. We first look at weak equivalence principle tests, such as that by R. Geiger et al \cite{WEPentangle} and the Q-WEP team \cite{WEPspace} which both involve two atom interferometers, each employing a different rubidium isotope. Specifically, R. Geiger et al \cite{WEPentangle} entangle the two interferometers together. For the case where each interferometer's arms reaches largest separation after 50 ms, the achieved vertical separation (comparing the same interferometric path of the different interferometers) between the different isotopes is 5 µm. Our accelerating traps may be used to increase the distance between the different isotopes in a much shorter period of time, since even with an AOD configuration, a tweezer can be moved up to 200 µm. If the atoms were additionally held stationary in the midst of the sequence, it may be possible for our proposed accelerating trap interferometer to achieve better sensitivities than the stated $5 \times 10^{-7}$ m/s$^{2}$ in the reference. However, looking at another application, such as in the search for gravitational redshift violations as in the paper by Di Pumpo et al \cite{EnnoGravRedShift}, it is unlikely our proposal reaches the stated sensitivities of the described free-fall and guided interferometers. This is because for finding gravitational redshift violations, there is no requirement for a relatively large space-time area to be achieved in a very short period of time, as is the focus of the scheme in this paper; hence, the space-time areas stated in the reference are orders of magnitude larger than those achievable by our proposal because the free-fall and guided interferometers are allowed to have prolonged interferometer times.

In this paper, we characterize the feasibility of this accelerating-trap-interferometer setup. The layout is as follows. Section~\ref{ExpSetup} details the experimental procedure and mathematical description of the interferometer sequence for stationary traps. This is adapted to moving traps in Section~\ref{MovingTrapConcept}, where we also explore the limits of trap acceleration in terms of atom loss, capabilities of existing technology, and the achievable sensitivities. In Section~\ref{NoisSection}, we investigate the effects of various sources of noise, such as that due to laser intensity and trap center fluctuations.

\section{Experimental setup for stationary traps}
\label{ExpSetup}

\begin{table}[htb]
\caption{\label{tab:Parameters}
List of notation.
}
\begin{ruledtabular}
\begin{tabular}{cc}
\textrm{Symbol}&
\textrm{Description}\\
\colrule
$t_{0}$ & atom free fall duration\\
$t_{\rm ru}$ & trap potential ramp up duration\\
$t_{\rm rd}$ & trap potential ramp down duration\\
$t_{t}$ & ``steady" trap potential duration\\
$t_{m}$ & duration of traps moving\\
$\tau$ & total duration trap is on, e.g., $\tau = t_{\rm ru} + t_{m} + t_{\rm rd} $\\
$(b/t)$ & subscript indicating either bottom or top trap\\
$w_{0,(b/t)}$ & trap waist\\
$\omega_{(b/t)}$ & frequency of potential\\
$\omega_{0(,b/t)}$ & unperturbed frequency of potential\\
$k_{(b/t)}$ & trap spring constant\\
$k_{0(,b/t)}$ & unperturbed trap spring constant\\
$\delta k_{(b/t)}$ & noise perturbation of spring constant\\
$\nu_{(b/t)}$ & frequency of trap light\\
$\kappa$ & inverse coherence time of light in cavity\\
\end{tabular}
\end{ruledtabular}
\end{table}

In this section, we present a compact understanding of a typical interferometer sequence, using a mathematical description that leverages the language of quantum circuits rather than path integrals. Due to the generality of definitions, the formulae in this section apply to both stationary and moving traps. Nevertheless, for conceptual simplicity, we will illustrate here the case of a stationary trap and elaborate on the complexities of moving traps in later sections. For convenience, a list of notation used throughout this paper is given in Table~\ref{tab:Parameters}.

The basic experimental sequence is shown in Fig.~\ref{fig:InterSequence_a} and follows that proposed by Xu et al~\cite{AtomInterExpHolger}.The key differences between the two setups are that we include optical dipole traps instead of an optical lattice in the midst of the sequence, and we employ two pairs of $\pi/2$ and $\pi$ pulses instead of pairs of two $\pi/2$ pulses as done by Xu et al. As such, our interferometer sequence is as follows. A launched cloud of atoms is split into a superposition of two paths (arms) by a $\pi/2$ laser pulse, which provides an $\hbar q$ momentum kick, along with a transition to a higher internal state, to the atoms in one path. After time $t_{0}$, a $\pi$ pulse acting on only the top arm places these atoms in a zero momentum and lower internal state, such that the atoms in both arms now have zero momentum and the same internal state (for details about how the pulses act, see Appendix~\ref{InterFinState}). This path-dependent $\pi$ pulse can be achieved using two methods. The first is by adopting a technique experimentally implemented by Xu et al~\cite{AtomInterExpHolger}, which is to act with a global $\pi/2$ pulse and discard the atoms of the unneeded state in both arms. The second is to switch on the trap of the bottom arm before that of the top and use the trap laser to AC Stark shift the bottom atoms such that they are unaffected by a global $\pi$ pulse. After this effective $\pi$ pulse the two arms are separated by length $L = \hbar q t_{0}/m$. Next, two optical dipole traps, their centers also separated by $L$, are turned on, trapping the atoms in the corresponding arms. The atoms remain in the traps for time $\tau$, after which the traps are turned off. The atoms will once again free fall for time $t_{0}$, during which the atomic paths are recombined by a path-dependent $\pi$ pulse and a final $\pi/2$ pulse. The coherence, a quantity that measures the difference in phase, is finally estimated by counting the number of atoms at each of the two output ports at the end.

\begin{figure}[htb]
         \includegraphics[width = 8.6 cm]{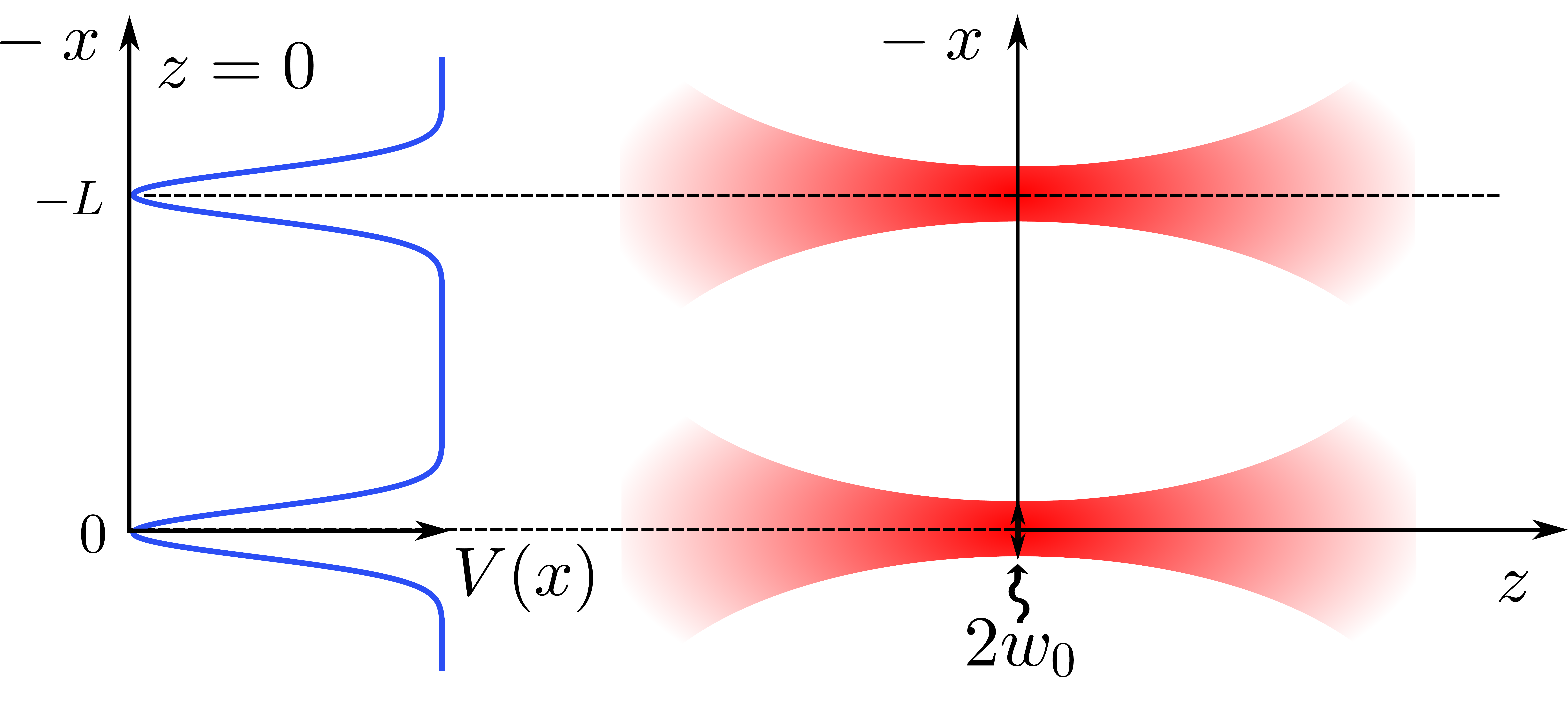}
       \caption{The trap potential, with $w_{0}$ as the beam waist~\cite{TweezerInterference}. The diagram is not to scale. Note that the direction of the focal plane of the trap (i.e. the direction of the beam waist cross-section) is perpendicular to the direction of gravity.} 
       \label{fig:TweezerPotential}
\end{figure}

For a two-level atom with states $\ket{g}$ and $\ket{e}$ (these can be Raman-separated hyperfine states, or other combinations), the probability of measuring state $\ket{g}$ of the atoms at the end of the interferometer sequence is given by $P_{g} = \frac{1}{2}+\frac{1}{2}\Re[C]$~\cite{JonInterPublished}, where 
\begin{equation}
\ C = \bra{\psi_{t=0}}\hat{U}_{\textrm{bot}}^{\dagger}\hat{U}_{\textrm{top}}\ket{\psi_{t=0}}
\label{eq:Coherence}
\end{equation}
is the coherence, $\Re$ symbolizes the real part, and $\ket{\psi_{t=0}}$ is the initial spatial component. Here, $\hat{U}_{\textrm{bot}}$ and $\hat{U}_{\textrm{top}}$ contain information about the operations experienced by the atoms in the two corresponding paths and the order in which they occur (see Appendix~\ref{InterFinState} and Ref.~\cite{JonInterPublished,UnitaryFormalism1,UnitaryFormalism2} for a longer discussion of this formalism). Namely,
\begin{equation}
    \ \hat{U}_{\textrm{top}} = \hat{U}^{F}_{(t_{0}+\tau,2t_{0}+\tau)} \hat{U}^{T}_{(t_{0},t_{0}+\tau)} e^{iq\hat{x}} \hat{U}^{F}_{(0,t_{0})} e^{-iq\hat{x}}
\label{eq:UTop}
\end{equation}
\begin{equation}
    \ \hat{U}_{\textrm{bot}} = e^{iq\hat{x}} \hat{U}^{F}_{(t_{0}+\tau,2t_{0}+\tau)} e^{-iq\hat{x}} \hat{U}^{T}_{(t_{0},t_{0}+\tau)} \hat{U}^{F}_{(0,t_{0})}
\label{eq:UBot}
\end{equation}
where $\hat{U}^{F}$ is the spatial free fall unitary and $\hat{U}^{T}$ is the spatial unitary of the optical dipole trap. $\hat{U}^{T}$ includes the process of turning the trap on and off, also termed the ``ramping up/down" of the trap, in addition to the holding of the atom in the trap.

In our case, the coherence can be evaluated to show that
\begin{equation}
    \begin{split}
        C 
         & = \exp\left[-\frac{i}{\hbar}mgLt_{0}\right]\bra{\tilde{\psi}}\hat{U}^{T\dagger}_{(t_{0},t_{0}+\tau)} ~e^{-\frac{i}{\hbar}\hat{p}L}~\hat{U}^{T}_{(t_{0},t_{0}+\tau)} \\ & ~~~~~~~~~~~~~~~~~~~~~~~~~~~~~~~~~~~~~~~~~~~~~~~~~~~~~~~\times e^{\frac{i}{\hbar}\hat{p}L}\ket{\tilde{\psi}}\\
         & = \exp\left[-\frac{i}{\hbar}mgLt_{0}\right]\bra{\tilde{\psi}}\hat{U}^{T\dagger}_{(t_{0},t_{0}+\tau)}~\hat{U}^{T}_{(t_{0},t_{0}+\tau)}|_{\Delta\hat{x}=-L}\ket{\tilde{\psi}}
    \end{split}
\label{eq:CoherencePathDiff}
\end{equation}
where $\ket{\tilde{\psi}} = \hat{U}^{F}\ket{\psi_{t = 0}}$. $\hbar q$ is the momentum kick for both the first two pulses. Note that one of the operators $\hat{U}^{T}$ in Eq.~\ref{eq:CoherencePathDiff} has been shifted by $L$ with respect to the other, reflecting the expected location of the atoms in the other path. We note that a typical (non-trapped) atom interferometer has $\hat{U}^{T}= \textbf{I}$ and the interferometer phase is just the prefactor in Equation~\ref{eq:CoherencePathDiff}.

Experimentally, the coherence $C$ is measured by obtaining the population of atoms of a particular state at the end of the interferometer sequence~\cite{AtomInterExpHolger}. From Eq.~\ref{eq:CoherencePathDiff}, it can be seen that   
$C$ can be observed as long as $gL \neq 0$; if there were no path difference, then $C = 1$. It must be noted that $L$ is not a fundamental experimental quantity; that is, it depends on experimentally controllable quantities $q$ and $t_{0}$. If $L$ is to be varied, we can either manipulate the pulse for a different value of $q$, or vary $t_{0}$ by changing the launch velocity of the atom cloud~\cite{AtomInterExpHolger}.
It is also important to recognize that $q$ is related to the wavelength of the laser light, which in turn is determined by the choice of atom~\cite{AtomInterLightPulses}.

In order to fully evaluate Eq.~\ref{eq:CoherencePathDiff} in the presence of trapping, the form of the atom trap unitary $\hat{U}^{T}$ has to be established. A diagram of the type of potential that we focus on is given in Fig.~\ref{fig:TweezerPotential}, but our formalism works for any trapping potential in principle. For each trap, a laser beam propagates perpendicular to the direction of gravity and the atom trap is formed at the beam waist. Thus the potential is tighter along the direction of gravity. The direction of the focal plane of the trap (i.e. the direction of the beam waist cross-section) is perpendicular to the direction of gravity. Henceforth, we refer to the direction along that of gravity as \textit{co-gravity} and that along the axis of laser propagation as \textit{axial}. 

The most general form of $\hat{U}^{T}$ is
\begin{equation}
   \hat{U}^{T}= \hat{T}\exp \left[ \frac{-i}{\hbar}\int_{t_{0}}^{t_{0}+\tau}dt~ (\hat{H}_{0}+\hat{V}(t)) \right].
\label{eq:PotentialUnitary}
\end{equation}
Here, $\hat{V}(t)$ is the complete potential, i.e., it includes the potential due to both traps, and $\hat{T}$ is the time-ordering operator. The shift in $L$ means that if we expand the potential in each of $\hat{U}^{T\dagger}$ and $\hat{U}^{T}|_{\Delta \hat{x}=-L}$ about $x = 0$, the net effect is that of two Taylor expansions about the bottom and top well (initially at $x = -L$, but now shifted to $x = 0$), respectively. For a trap that is formed by laser(s) with a Gaussian intensity profile, approximating the single-well potential in the plane of the beam cross-section is a well-established approach~\cite{OpticalTweezersTheoryAndPractice,OpticalTweezersAndTheirApps}. Explicitly, we can take the potentials to be~\cite{OpticalTweezersTheoryAndPractice}
\begin{multline}
    \ V(x) = -V_{t,d} \exp\left[\frac{-2 (x+L)^{2}}{w_{0,t}^{2}}\right]-V_{b,d} \exp\left[\frac{-2 x^{2}}{w_{0,b}^{2}}\right]\\ -mgx
\label{eq:PotentialShift1}
\end{multline}
and
\begin{multline}
    \ V(x-L) = -V_{t,d} \exp\left[\frac{-2 x^{2}}{w_{0,t}^{2}}\right]-V_{b,d} \exp\left[\frac{-2 (x-L)^{2}}{w_{0,b}^{2}}\right]\\ -mg(x-L), 
\label{eq:PotentialShift2}
\end{multline}
where $V_{(b/t),d}$ is the trap depth of the bottom and top traps and $w_{0,(b/t)}$ is the waist. In the limit where $L \gg w_{0,(b/t)}$, the potentials Taylor expanded about $x=0$ would be 
\begin{align}
    \ V(x) = \frac{1}{2}k_{b}x^{2}-mgx-V_{b,d} \\ V(x-L) = \frac{1}{2}k_{t}x^{2}-mgx+mgL-V_{t,d}
\label{eq:PotentialShiftsExpand}
\end{align}
with $k_{(b/t)}$ as the corresponding spring constant. We can see how a phase arises due to the terms $mgL$, $V_{b,d}$, and $V_{t,d}$. This will be explored in detail in the next section. To reiterate, the gravitational potential is given by $-mgx$ due to our choice of axis direction (see Fig.~\ref{fig:InterSequence}). Further, due to the approximations made in this section, the succeeding sections are for the case where the condition $L \gg w_{0,(b/t)}$ is experimentally met.

It must be emphasized that in Eq.~\ref{eq:CoherencePathDiff}, $\hat{U}^{T}$ is unspecified, and, as such, the expression is also applicable to moving traps, since any information about movement will be included in $\hat{U}^{T}$. 

\section{Concept for accelerating traps}
\label{MovingTrapConcept}

We now show how moving traps allows us to cover a larger space-time area, enabling measurements that can observe more spatial and temporal information and/or achieve better imprecision or sensitivity. Specifically accelerations, gradients of acceleration, and rotations will add to the interferometer phase, depending on the trap motion. While we focus on the setup with two traps, this approach is extensible to an array of traps, with varieties of conceivable paths. Here, we restrict ourselves to the simplest iteration of such a setup as shown in Fig.~\ref{fig:InterSequence_b}, which is a modified version of Fig.~\ref{fig:InterSequence_a}: the traps undergo a constant artificial co-gravity acceleration, are subsequently held for some time at their new positions, and afterwards returned to their initial positions. In this section, we investigate in detail the basic strengths and limitations of the concept, including the signals that can be obtained, limits of trap acceleration, possible experimental setups, and comparison to existing technology.


\subsection{The coherence of moving traps}
\label{MovingTrapCoherence}

It is instructive to understand the signal one can expect from an interferometer sequence with the inclusion of moving traps. Here, we work with time evolution operators to determine the final state of an atom in the trap after trap movement. The result is expressed as the product of a phase, a displacement operator, and the initial state of the atom in the trap. There is a difference in phase between the two arms, both due to trap separation and because the path taken by the two traps may not be identical. When the trap trajectories enable a good overlap at the interferometer output, it is this difference which constitutes a signal that can be measured. 

Starting after launch, the process of an atom entering and remaining in the trap is as follows. As the wave packet free falls for time $t_{0}$, it spatially disperses due to the distribution of momentum states. After it reaches the trap, the trap potential is ramped up - its spring constant is increased such that the potential becomes spatially narrower. The atom is then held in a moving trap for some time, after which it is released once more. These processes can be mathematically expressed by decomposing the unitary in Eq.~\ref{eq:PotentialUnitary} as 
\begin{equation}
        \ \hat{U}^{T} = \hat{U}^{T}_{\textrm{rd}}\hat{U}^{T}_{\textrm{move}}\hat{U}^{T}_{\textrm{ru}},
\label{eq:PotentialUnitaryDecompose}
\end{equation}
where $\hat{U}^{T}$ has been separated into ``ramp up" (ru), moving trap (move), and ``ramp down" (rd) components.
For now, we consider the approximate scenario where ramp up results in the atom reaching the state $\ket{0}$ of a conventional simple harmonic oscillator, such that $\hat{U}^{T}_{\textrm{ru}}\ket{\tilde{\psi}}=\ket{0}$ for both arms. We later adapt the calculations for a thermal state. If $\ket{\psi_{b,f}}=\hat{U}^{T}_{\textrm{move}}\ket{0}$ denotes the spatial state of the atom in the bottom arm after the movement of the trap, in a harmonic oscillator approximation of the trap potential, $\ket{\psi_{b,f}}$ after time $t_{m}$ is explicitly given by
\begin{multline}
    \ \ket{\psi_{b,f}} = \hat{T}\exp[-\frac{i}{\hbar} \int_{0}^{t_{m}} dt~( \frac{\hat{p}^{2}}{2m}+\frac{1}{2}m\omega_{0}^{2}(\hat{x}-s_{b}(t))^{2}\\
    -mg\hat{x}-V_{b,d})] \ket{0},
\label{eq:MovingTrapState}
\end{multline}
where $s_{b}(t)$ is the displacement of the trap in the $x$ direction and $\omega_{0}$ is the constant trap frequency. This is simply the time evolution operator applied to the initial state for the case of a Hamiltonian composed of the potential due to gravity and that of a harmonic oscillator with a moving trap center. The calculations are done in the laser frame, in which the proper time is the coordinate time. This can be solved exactly in this approximation as detailed in Appendix \ref{AtomInMovingTrap}. We obtain the final state expressed in terms of the displacement operator, $D(\beta) = \exp[\beta \hat{a}^{\dagger}-\beta^{*}\hat{a}]$~\cite{DisplacementOperator}, to be: 
\begin{multline}
    \ \ket{\psi_{b,f}} = \Phi_{sc,b} ~\exp\left[i(\phi_{b,1}+\phi_{b,2}-\frac{\omega_{0} t_{m}}{2}+\frac{V_{b,d}t_{m}}{\hbar})\right] \\  D(\alpha_{\textrm{bot}}) \ket{0}
 \label{eq:MovingTrapStateEval}
\end{multline}
where
\begin{multline}
    \ \alpha_{\textrm{bot}} = \sqrt{\frac{m\omega_{0}}{2\hbar}}\tilde{s}_{b}(t_{m})-\sqrt{\frac{m\omega_{0}}{2\hbar}}\int_{0}^{t_{m}}dt~~\Dot{\tilde{s}}_{b}(t)e^{i\omega_{0}(t-t_{m})}, 
\label{eq:DAlphaBot}
\end{multline}
\begin{equation}
    \ \phi_{b,1} = \frac{m\omega_{0}}{2\hbar}\tilde{s}_{b}(t_{m})\int_{0}^{t_{m}} dt~~ \Dot{\tilde{s}}_{b}(t)\sin[\omega_{0}(t-t_{m})],
\label{eq:phi1}
\end{equation}
and
\begin{equation}
    \ \phi_{b,2} = \frac{m\omega_{0}}{2\hbar}\int_{0}^{t_{m}}\int_{0}^{t_{2}} dt_{1} dt_{2}~~ \Dot{\tilde{s}}_{b}(t_{1})\Dot{\tilde{s}}_{b}(t_{2})\sin[\omega_{0}(t_{1}-t_{2})].
\label{eq:phi2}
\end{equation}
Here $\tilde{s}_{b}(t) = s_{b}(t)+\frac{g}{\omega_{0}^{2}}$ (see Appendix \ref{AtomInMovingTrap} for the purpose and physical meaning of this shift).
The movement of the atom is captured by the displacement term $D(\alpha_{\textrm{bot}})$, which is comprised of the final displacement of the trap $s_{b}(t_{m})$ and an integral that includes the velocity of trap movement $\dot{s}_{b}(t)$ multiplied by an oscillating term. Additionally, we see the appearance of the non-trivial phase terms $\phi_{b,1}$ and $\phi_{b,2}$, due to the phase space area that the coherent state covers over the trapping time. While there is an adiabatic limit where $\Phi_{sc,b}$ dominates the signal, we focus on large accelerations in the succeeding sections.

$D(\alpha_{\textrm{bot}})$ is the term most important for the semiclassical atom dynamics in a moving trap. We can look at a simple example where $\Dot{s}_{b}(t) = (s_{0}/2)\omega_{m} \sin(\omega_{m}t)$, where $s_{0}$ is the maximum displacement of the trap and $\omega_{m}$ characterizes how fast $\Dot{s}_{b}(t)$ changes over time. In this case, the integral $I_{\alpha_{\textrm{bot}}}=\int_{0}^{t_{m}}dt~~\Dot{\tilde{s}}_{b}(t)e^{i\omega_{0}(t-t_{m})}$ in Eq.~\ref{eq:DAlphaBot} evaluates to 
\begin{equation}
    \ I_{\alpha_{\textrm{bot}}} = \frac{s_{0}\omega_{m}^{2}(1-e^{-\frac{2i\pi \omega_{0}}{\omega_{m}}})}{2(\omega_{0}^{2}-\omega_{m}^{2})}
\label{eq:DAlphaBotEx}
\end{equation}
when $t_{m}=2\pi/\omega_{m}$. In the limit where $\omega_{0} \gg \omega_{m}$, that is, in the limit where $\omega_{0} t \gg 1$ \textit{and} $\Dot{s}_{b}(t)$ is slowly varying, the integrand evaluates to zero. The remaining term within the displacement operator signifies that the atom has been displaced with the trap. Experimentally, it is recommended to be in the limit $\omega_{0} \gg \omega_{m}$. Physically, this limit means that the time period for which we move the trap must be much greater than the time period of atom oscillations in the trap. 

We now evaluate $\ket{\psi_{t,f}}=\hat{U}^{T}_{\textrm{move}}|_{\Delta\hat{x}=-L}\ket{0}$ in order to understand the resulting phase due to interference of the two arms. The calculation is essentially identical to the previous one, with the exception of an additional phase term, because, as we saw in Eq.~\ref{eq:PotentialShiftsExpand}, after Taylor expansion, a shifted potential is of the same form as an unshifted one, up to a constant. In this section, for simplicity, we will consider the top and bottom traps to be identical; then, the trap frequency $\omega_{0}$ is the same for both traps. We thus have 
{\small
\begin{equation}
    \begin{split}
    \ \ket{\psi_{t,f}} = & \exp\left[i(\phi_{t,1}+\phi_{t,2}-\frac{\omega_{0} t_{m}} {2}+\frac{(-mgL+V_{t,d})t_{m}}{\hbar
    })\right] \\
    & \times \Phi_{sc,t} ~D(\alpha_{\textrm{top}}) \ket{0},
    \end{split}
 \label{eq:MovingTrapStateEvalTop}
\end{equation}}
where $D(\alpha_{\textrm{top}})$, $\Phi_{sc,t}$, $\phi_{t,1}$, and $\phi_{t,2}$ have the exact same form as $D(\alpha_{\textrm{bot}})$, $\phi_{b,1}$, and $\phi_{b,2}$, except with the subscript $b$ now being replaced by $t$. Note that we also take $\tilde{s}_t(0) = 0$ and $s_{t}(0)=-g/\omega_{0}^{2}$. The additional phase term $-mgLt_{m}/\hbar$ arises from the initial trap separation. Since we consider identical traps, $V_{t,d}=V_{b,d}$.

With $\ket{\psi_{b,f}}$ and $\ket{\psi_{t,f}}$ in hand, under the assumption that the ramp up and down of the traps is identical for both traps (i.e. that the time-dependent spring constant is identical), the coherence can be expressed as
\begin{equation}
        C = e^{i \phi_{\textrm{TOT}}} |\bra{\alpha_{\textrm{bot}}}\alpha_{\textrm{top}}\rangle|,
\label{eq:CoherencePathDiff12}
\end{equation}
with 
\begin{equation}
    \ \ln[|\bra{\alpha_{\textrm{bot}}}\alpha_{\textrm{top}}\rangle|] =-|\alpha_{\textrm{top}}-\alpha_{\textrm{bot}}|^{2}/2
\label{eq:DAlphaBotSqr}
\end{equation}
and
\begin{equation}
    \begin{split}
        \ \phi_{\textrm{TOT}} = &  \phi_{t,1}-\phi_{b,1}+\phi_{t,2}-\phi_{b,2} -\frac{mgL(\tau+t_{0})}{\hbar} \\
        & +\frac{1}{\hbar}\int_{0}^{t_{m}}dt~ mg[s_{t}(t)-s_{b}(t)] 
     + \Im[\alpha_{\textrm{bot}}^{*}\alpha_{\textrm{top}}]
    \end{split}
\label{eq:PhiTotExact}
\end{equation}
where $\tau = t_{\rm ru} + t_{m} + t_{\rm rd}$, the phases from ramp up and down arising in an analogous manner to $-mgLt_{m}/\hbar$. If the top and bottom traps were non-identical, we would have additional phase terms; in particular, we would have the phase $(V_{t,d}-V_{b,d})t_{m}/\hbar$, which shows that relative intensity (trap depth) fluctuations will lead to a reduction of coherence. We address non-identical traps in Section~\ref{TrapCenterNoise}.

Now it can be seen clearly that the total phase measured at the end is composed of: a) the difference between phases of the two arms, such as $\phi_{t,1}-\phi_{b,1}$; b) a phase due to the initial trap separation as in $-mgL(\tau+t_{0})/\hbar$; c) the difference between the paths taken by the traps as in the integrand $s_{t}(t)-s_{b}(t)$; and d) the phase from the overlap of the coherent states. The term with $s_{t}(t)-s_{b}(t)$ constitutes the space-time area enclosed by the two arms when the trap is moving. 

In the limit of $\omega_{0} t \gg 1$, we expect the $\phi$ terms to average to zero, such that we have
\begin{multline}
        \ \phi_{\textrm{TOT}} = -\frac{mgL(\tau+t_{0})}{\hbar}
         +\frac{1}{\hbar}\int_{0}^{t_{m}}dt~ mg[s_{t}(t)-s_{b}(t)].
\label{eq:PhiTot}
\end{multline}
Furthermore, we obtain explicit expressions for the exponents of the second exponential,
\begin{equation}
    \ \ln[|\bra{\alpha_{\textrm{bot}}}\alpha_{\textrm{top}}\rangle|] = -\frac{m\omega_{0}}{4\hbar} \left[s_{b}(t_{m})-s_{t}(t_{m})\right]^{2}.
\label{eq:DAlphaBotSqrSimpl}
\end{equation}
As such, in this limit, the coherence will only depend on the initial trap separation $L$ and the moving trap paths (or final positions) $s_{b}(t)$($s_{b}(t_{m})$) and $s_{t}(t)$ ($s_{t}(t_{m})$). If the moving traps are returned to their starting points at $t_{m}$, then the decay term of Eq.~\ref{eq:DAlphaBotSqrSimpl} is zero.

In practice, the initial state in the trap will not be $\ket{0}$, so it is instructive to understand the effects of starting in a thermal state. For the case where the density matrix of the starting state is given by~\cite{ThermalStateRep}
\begin{equation}
        \ \rho_{\textrm{therm}} = \int~ d^{2}\beta~ \frac{e^{-|\beta|^{2}/\Bar{n}}}{\pi \Bar{n}} \ket{\beta}\bra{\beta}
\label{eq:rhoTherm}
\end{equation}
we find that the coherence is of the form (see Appendix~\ref{AtomInMovingTrap})
\begin{equation}
        C_{\textrm{therm}} = e^{i \phi_{\textrm{TOT}}} |\bra{\alpha_{\textrm{bot}}}\alpha_{\textrm{top}}\rangle|^{(\frac{\Bar{n}}{2}+1)},
\label{eq:CoherencePathDiffTherm}
\end{equation}
where $\phi_{\textrm{TOT}}$, $\alpha_{\textrm{bot}}$, and $\alpha_{\textrm{top}}$ are as given previously, and $\Bar{n}$ is the thermal occupancy. In the limit $\omega_{0} t \gg 1$, Eq. \ref{eq:DAlphaBotSqrSimpl} and \ref{eq:CoherencePathDiffTherm} suggest that noise arises from an initial thermal state only in the case where there is insufficient control over the tweezers such that at the end of trap movement $s_{b}(t_{m}) \neq s_{t}(t_{m})$. In such a situation, there is additional decay of the coherence, which increases with the temperature.

Note that the equations in this section, with the exception of Equations \ref{eq:PhiTot} and \ref{eq:DAlphaBotSqrSimpl}, are exact and without approximation. All effects of any kind of motion, whether adiabatic or non-adiabatic, are contained in Equations \ref{eq:phi1}, \ref{eq:phi2}, and \ref{eq:PhiTotExact}. In the succeeding sections we are in a regime where $\omega_{0}\sim\omega_{m}$, and as such, not in the adiabatic limit discussed here for which Equations \ref{eq:PhiTot} and \ref{eq:DAlphaBotSqrSimpl} apply (we find this non-adiabatic regime to be acceptable in the context of atom loss prevention, though we work with the adiabatically simplified results in our sensitivity analysis). Therefore, we expect to see additional phases and decays as given by the integral terms of Equations \ref{eq:DAlphaBot}, \ref{eq:phi1} and \ref{eq:phi2}.

\subsection{Physical limits of trap motion: keeping the atoms trapped}
\label{PhysicalLimits}

\begin{table}[htb]
\caption{\label{tab:PhysLimitsParameters}
Unless otherwise mentioned, the values of parameters used in Section~\ref{PhysicalLimits} are as in this table.
} 
\begin{ruledtabular}
\begin{tabular}{ccc}
\textrm{Parameter}&
\textrm{Symbol}&
\textrm{Value}\\
\colrule
acceleration of free fall & $g$ & 9.81 m/s$^{2}$ \\
mass of cesium & $m$ & $2.2 \times 10^{-25}$ kg~\cite{SteckCesium}\\
trap wavelength & $\lambda$ & 935.6 nm~\cite{935.6}\\
trap beam waist & $w_{0}$ & 38 $\mu$m \\
\end{tabular}
\end{ruledtabular}
\end{table}

A key question is how fast the trap can be moved while maintaining an interferometer signal. This question has two parts: the physical constraints due to the shape of the potential and the limitations of experimental apparatus. In this section, we address the first, specifically, by identifying a maximum acceleration $a_{\textrm{max}}$ before atom loss.

\begin{figure*}
    \centering
    \subfloat[]{     
     \includegraphics[width=8.6 cm]{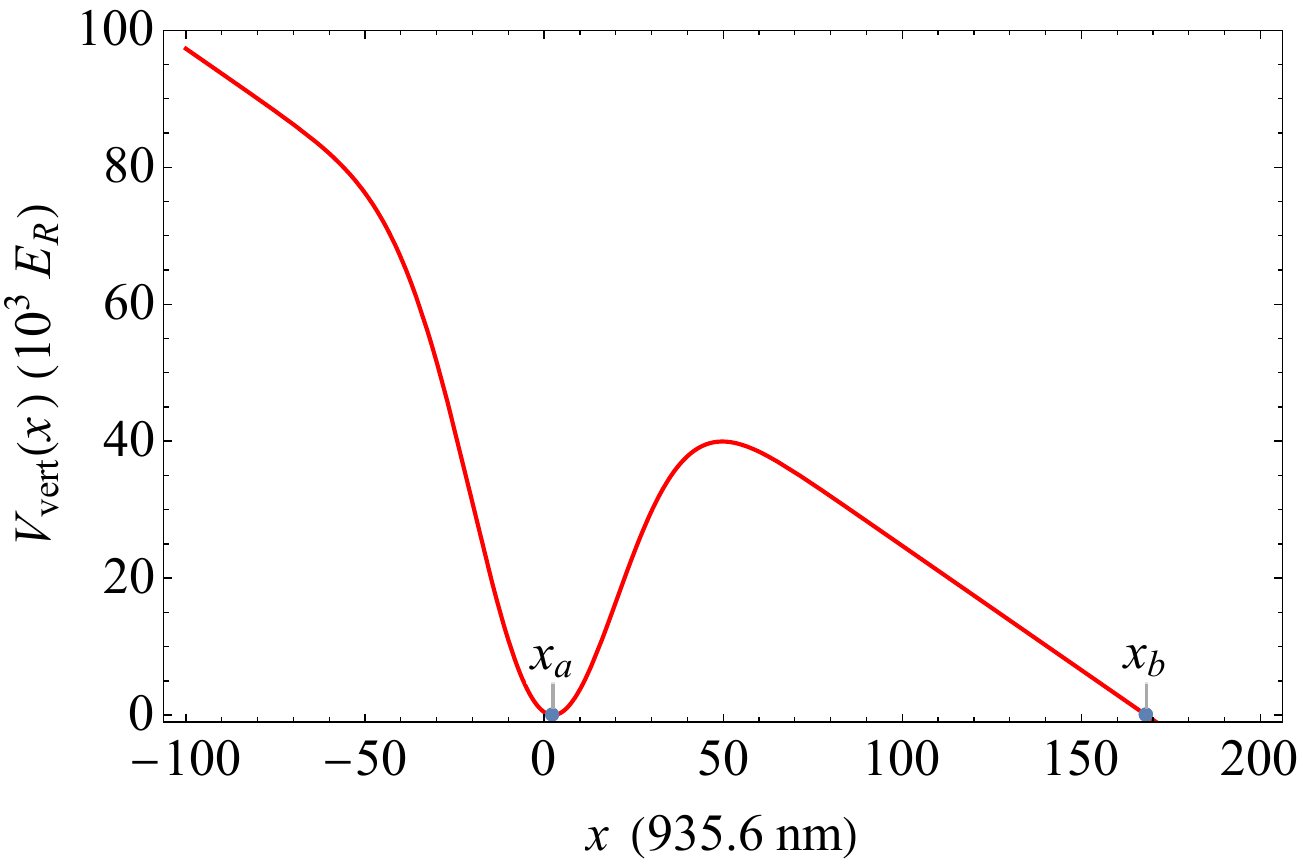}
     \label{fig:Vvertical}
     }
     \hfill
      \subfloat[]{       
      \includegraphics[width=8.6 cm]{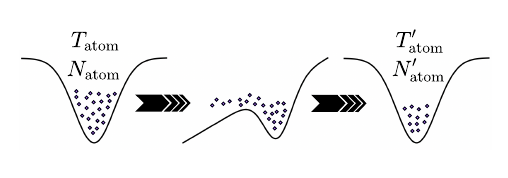}
      \label{fig:Effusivity}
    }
     \caption{\textbf{(a)} Plot of $V_{\textrm{vert}}(x)$ (Eq.~\ref{eq:TunnelingPotentialVert}), in units of recoil energy $E_{R}$, against $x$, in units of $\lambda$. $a = 2000$ m/s$^{2}$, $V_{d}/k_{\textrm{B}}= 5$ mK, and $E_{R}/k_{\textrm{B}}=0.083$ $\mu$K. $x_{a}$ is the location of the minimum and $x_{b}$ is the endpoint of the finite barrier.  \textbf{(b)} The trap initially has $N_{\rm atom}$ atoms at temperature $T_{\textrm{atom}}$. If the traps are accelerated for time $t_{m} < \tau_{E}(x_{0})$, where $x_{0}$ is a specific location, then atoms at $x_{0}$ will not have time to escape the trap. If certain atoms (at other values of x, with faster speeds), do escape, once acceleration stops, there will now be $N'_{\rm atom}$ atoms in the well with new temperature $T'_{\rm atom}$. Due to the non-zero effusivity time, even if the trap is accelerated such that the well completely disappears, it is possible to stop acceleration before all the atoms escape.}
     \label{fig:TrapTiltEffects}
\end{figure*}

As shown in Fig.~\ref{fig:Vvertical}, for a potential with a Gaussian profile, a constant acceleration appears as a constant force on the atom, thereby tilting the potential well. The higher the acceleration, the greater the tilt and the lower the resultant height of the barrier between the trap center and the continuum. In this scenario, loss of atoms can occur due to two main reasons; there is atom loss that can be described by classical physics, due to atoms escaping from the well as shown in Fig.~\ref{fig:Effusivity} (more elaboration on Fig.~\ref{fig:Effusivity} is at the end of this section); then, there is the quantum effect of tunneling which can occur due to the reduction of the barrier height and width when the potential is tilted. The aim is to understand how pervasive each of these effects are. We find that only the first is appreciable for typical parameters as we now show.

Unless otherwise mentioned, the values of parameters used in this section are as in Table~\ref{tab:PhysLimitsParameters}.


\begin{figure*}
    \centering
    \subfloat[]{
         \includegraphics[width=8.6 cm]{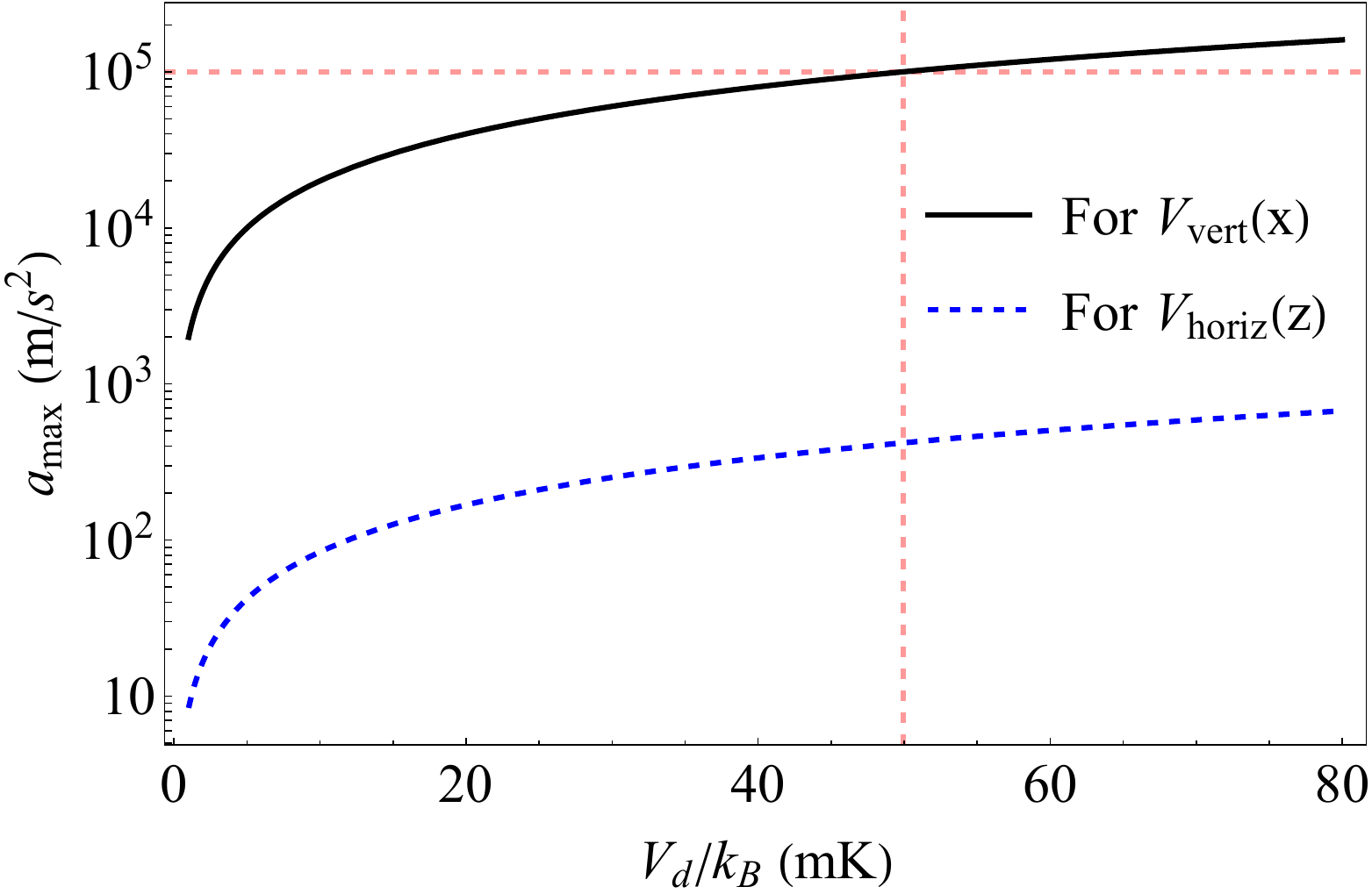}
         \label{fig:AvsULogPlot}
        }
     \hfill
     \subfloat[]{
         \includegraphics[width=8.6 cm]{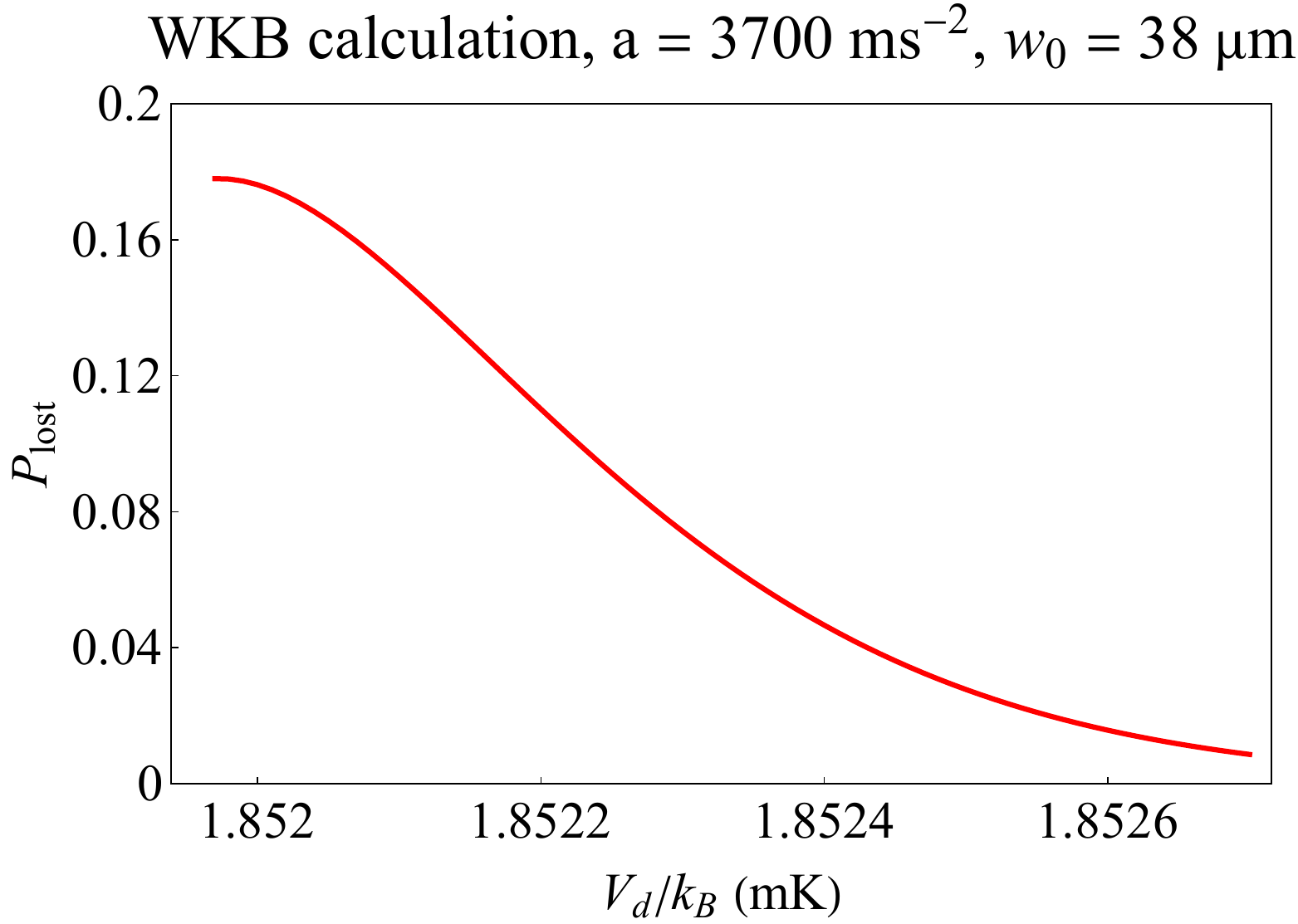}
         \label{fig:WKBplot}
      }
        \caption{\textbf{(a)} Plot of $a~(\text{m/}\text{s}^{2})$ against $V_{d}/k_{\textrm{B}}$ for the classical limit of the well (dis)appearing. Black is for $V_{\textrm{vert}}(x)$ (Eq.~\ref{eq:TunnelingPotentialVert}) and blue (dashed) is for $V_{\textrm{horiz}}(z)$ (Eq.~\ref{eq:TunnelingPotentialHoriz}), with $a_{\rm max}$ as given by Equations~\ref{eq:amaxvert} and \ref{eq:amaxhoriz}. It can be seen that co-gravity accelerations of up to order $10^{5}$  can be achieved at $V_{d}/k_{\textrm{B}} \approx 50$ mK. However, the achievable axial acceleration is much smaller for the same $V_{d}$. \textbf{(b)} Plot of $P_{\textrm{lost}}$ against $V_{d}/k_{\textrm{B}}$ for $a=3700$ m/s$^{2}$ around the limit of the well's appearance for $V_{\rm vert}(x)$. Here, $P_{\textrm{lost}}$ is the probability of losing an atom over a time duration of 0.6 ms. The data points have been obtained by varying $V_{d}/k_{\textrm{B}}$ in steps of order 0.01 $\mu$K and $C_{0}$ has been set to 1. The WKB approximation is expected to break down going towards the lower trap depths. Note the small range of the axes; there is a significant change in $P_{\textrm{lost}}$ over just 0.6 $\mu$K. This highlights the fact that tunneling effects are not expected to be of concern. For comparison, using Eq.~\ref{eq:amaxvert}, the classical limit at which the well (dis)appears is $V_{d}/k_{\textrm{B}} = 1.85179$ mK.}
        \label{fig:AccelTrapDynamics}
\end{figure*}

A simple approximation for $a_{\textrm{max}}$ is 
\begin{equation}
    \ a_{\textrm{max}} \approx \frac{V_{d}}{w_{0}m}
\label{eq:MaxAccel}
\end{equation}
where $V_{d}$ is the trap depth, $w_{0}$ the trap waist, and $m$ the mass of the atom. $a_{\textrm{max}}$ represents the work $ma_{\textrm{max}}w_{0}$ necessary to match the trap depth $V_{d}$. Practically, this is the order of magnitude when the barrier becomes close to zero. For $(V_{d}/k_{\textrm{B}}) = 1$ mK~\cite{Brandt}, we obtain $a_{\textrm{max}}$ to be of order $10^{3}$ m/s$^2$. Comparing this value to those obtained with the upcoming more rigorous calculations, we find that this simple estimate provides an accurate order of magnitude.

For a single trap, the co-gravity potential can be considered to be~\cite{OpticalTweezersTheoryAndPractice}
\begin{equation}
    \ V_{\textrm{vert}}(x) = -V_{d} \exp\left[\frac{-2 x^{2}}{w_{0}^{2}}\right]-m(a+g)x - V_{\textrm{vert,min}},
\label{eq:TunnelingPotentialVert}
\end{equation}
where for convenience we subtract off the local potential minimum, $V_{\textrm{vert,min}}$. It is depicted in Fig.~\ref{fig:Vvertical} for $ a = 2000 $ m/s$^2$ and $V_{d}/k_{\textrm{B}}= 5$ mK. Perpendicular to gravity, the axial potential of a single trap due to only the intensity gradient and an artificial acceleration is~\cite{OpticalTweezersTheoryAndPractice,TrapHorizPot,TrapVolume}
\begin{equation}
    \ V_{\textrm{horiz}}(z) = -V_{d} \frac{1}{\left[1+(\frac{\lambda z}{\pi w_{0}^{2}})^{2}\right]}+maz,
\label{eq:TunnelingPotentialHoriz}
\end{equation}
where $\lambda$ is the trap wavelength. This potential is relevant for the case where we achieve axial acceleration by changing the focal plane of the trap (which is not the only possible method).

We can calculate the critical accelerations $a_{\textrm{max}}$ for which the stable minimum disappears for both potentials $V_{\textrm{vert}}(x)$ and $V_{\textrm{horiz}}(z)$, that is, find the classical limits of the well's (dis)appearance. We obtain that 
\begin{equation}
    \ a_{\textrm{\rm max,vert}} = \frac{2}{\sqrt{e}} \frac{V_d}{m w_0} - g
\label{eq:amaxvert}    
\end{equation}
and 
\begin{equation}
    \ |a_{\textrm{max,horiz}}| = \frac{3 \sqrt{3}}{8\pi} \frac{V_d \lambda}{m w_0^2}.
\label{eq:amaxhoriz}
\end{equation}
Fig.~\ref{fig:AvsULogPlot} shows a plot of these equations against $V_{d}/k_{\rm B}$. It can be seen that co-gravity accelerations of up to order $10^{5}$ m/s$^2$ can be achieved at $V_{d}/k_{\rm B} \approx 50$ mK. However, the achievable axial acceleration is much smaller for the same $V_{d}$. This is most simply explained by the fact that the trap potential is less tight in the axial direction, thereby necessitating alternative methods to be employed, as discussed in the next section. There could also be some effect owing to the exclusion of the radiation pressure term in $V_{\textrm{horiz}}$~\cite{OpticalTweezersTheoryAndPractice}. It must be noted that the limits are for atoms located at the waist of the Gaussian beam; the potential is less tight moving away from the waist, and, hence, the atoms will have lower limits of acceleration.


To understand the prevalence of tunneling, for a potential $V(x)$ as shown in Fig.~\ref{fig:Vvertical}, we can work with the WKB approximation around the area of the potential well to determine the limits of acceleration. The probability per unit time $\Lambda$ that a trapped particle of low energy passes through the barrier is given by~\cite{WKBtunneling} 
\begin{equation}
    \ \Lambda = A_{0} \exp\left[-\frac{B_{0}}{\hbar}\right],~A_{0}=C_{0}\omega_{p}\sqrt{\frac{B_{0}}{2\pi \hbar}}
\label{eq:TunnelingProbability}
\end{equation}
with 
\begin{equation}
    \ B_{0} = 2 \int_{x_{a}}^{x_{b}} dx~ \sqrt{2mV(x)},
\label{eq:WKBintegral}
\end{equation}
and 
\begin{equation}
    \ \omega_{p}=\sqrt{\frac{1}{m} \left(\frac{d^{2}V(x)}{dx^{2}}\right)_{x=x_{a}}}
\label{eq:OmegaP}
\end{equation}
Here, $C_{0}$ is an order unity geometric prefactor that we do not explicitly calculate. $x_{a}$ is the location of the minimum and $x_{b}$ is the point at which the barrier ends. The total probability of losing atoms over time $t$ is then given by
\begin{equation}
    \ P_{\textrm{lost}}=1-e^{-\Lambda t}.
\label{eq:Plost}
\end{equation}

\begin{figure}[htb]
        \includegraphics[width = 8.6 cm]{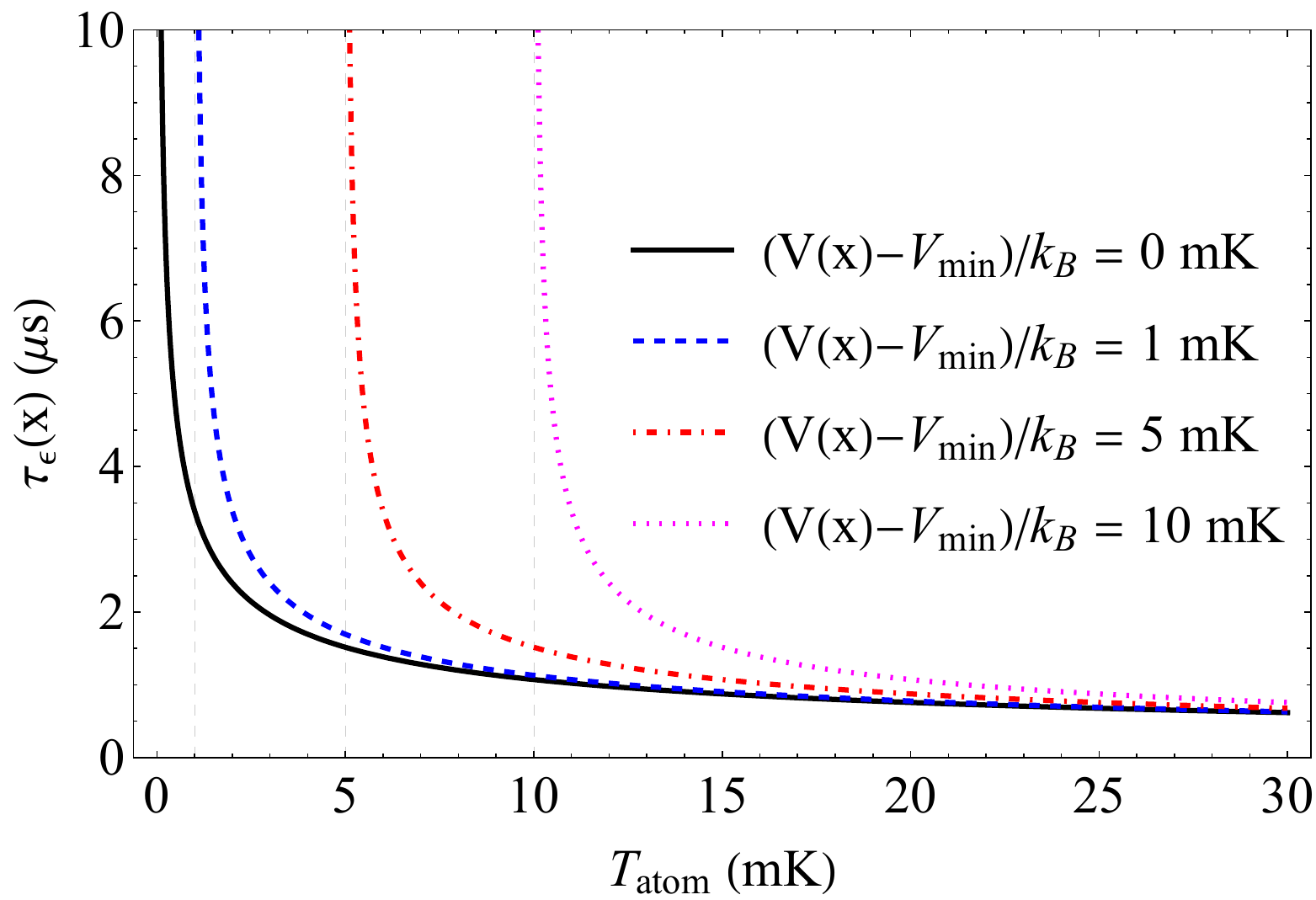}
        \caption{A plot of the effusivity time $\tau_{\epsilon}(x)$ against the temperature of the atom $T_{\textrm{atom}}$ for different values of $(V(x)-V_{\textrm{min}})/k_{\textrm{B}}$. It can be seen that $\tau_{\epsilon}(x)$ is on the order of $\mu$s. A graph such as this can be used to either estimate the effusivity time over the course of a single well, or compare the effusivity times between wells with different trap depths. Asymptotes occur when $T_{\textrm{atom}}$ approaches $(V(x)-V_{\textrm{min}})/k_{\textrm{B}}$, simply encapsulating the fact that the phenomenon does not apply to cases where $T_{\textrm{atom}}<(V(x_{\rm barrier})-V_{\textrm{min}})$, the parametric region for which tunneling effects have been investigated.}
        \label{fig:EffusivityTime}
\end{figure}

Tunneling effects can be considered for co-gravity and axial accelerations. $V_{\textrm{vert}}(x)$ can be substituted into Eq.~\ref{eq:WKBintegral} and integrated between $x_{a}$ and $x_{b}$ to obtain tunneling probabilities, which can then be used to understand the pervasiveness of tunneling. Fig.~\ref{fig:WKBplot} shows the variation of $P_{\textrm{lost}}$ with $V_{d}/k_{\rm B}$ around the limit of the well's appearance for $a=3700$ m/s$^2$. Note the small range of the axes; there is a significant drop in $P_{\textrm{lost}}$ over just 0.6 $\mu$K. This highlights the fact that tunneling effects are essentially negligible, because they can be drastically reduced within only a 0.6 $\mu$K variation in $V_{d}/k_{\rm B}$. The WKB approximation is expected to start becoming inapplicable towards the lowest trap depths of the plot, because the simple harmonicity that Eq.~\ref{eq:OmegaP} depends on starts to disappear. With the small probability values found, we will not consider tunneling effects further.

As verified with these numerical results, trap depths for which the tunneling is negligible can be conveniently found for the case of co-gravity acceleration. For axial acceleration, however, it appears as though alternative methods have to be employed, as will be discussed in the next section.

Atom loss is also dependent on the temperature of the atoms. In particular, we can define the effusivity time $\tau_{E}$ to be
\begin{equation}
    \ \tau_{E} (x) \equiv  \frac{w_{0}}{\left[\frac{2}{m} (k_{\textrm{B}} T_{\textrm{atom}}-(V(x) - V_{\textrm{min}}))\right]^{1/2}}.
\label{eq:EffusivityTime}
\end{equation}
$\tau_{E}$ is an estimate of the time it takes for atoms of temperature $T_{\textrm{atom}}$ to escape past the well, given the potential energy cost $V(x)- V_{\textrm{min}}$ associated with being away from the bottom of the potential well, where $V_{\textrm{min}}$ is the value of the potential well minimum. The difference between the atom temperature and the potential energy cost gives the characteristic excess kinetic energy and, hence, the characteristic velocity for atoms at different locations in the trap. This velocity sets the timescale on which they escape. Atoms are expected to be at their highest speed at $x_{\textrm{min}}$, where the potential has a local minimum, and lowest at $x_{\rm barrier}$, where there is a local potential maximum. As such, $\tau_E(x_{\textrm{min}})$ and $\tau_E(x_{\rm barrier})$ give us a lower and upper bound for the time taken by atoms of temperature $T_{\textrm{atom}}$ to leave the trap. These times also bound the duration for which the trap can be accelerated; if the trap is accelerated (i.e., tilted) for times less than these bounds, the atoms will not have time to escape from the trap, and thermal atom loss can be prevented. Fig.~\ref{fig:Effusivity} illustrates these concepts, while Fig.~\ref{fig:EffusivityTime} is a plot of $\tau_{\epsilon}(x)$ for different values of $(V(x) - V_{\textrm{min}})/k_{\textrm{B}}$. It can be seen that $\tau_{\epsilon}(x)$ is on the order of $\mu$s. Of course, $\tau_{\epsilon}(x)$ is only well-defined with the assumption that $T_{\textrm{atom}} > (V(x_{\rm barrier})-V_{\textrm{min}})/k_{\textrm{B}}$; for $T_{\textrm{atom}} < (V(x_{\rm barrier})-V_{\textrm{min}})/k_{\textrm{B}}$, tunneling effects were investigated previously. 

Note that the evaluations in this section, including that of $a_{\rm{max}}$, only analyze classical limits. We may also attempt to understand a quantum limit of $a_{\rm{max}}$, specifically in relation to the quantum speed limit as in the paper by Lam et al~\cite{LamEtAl}. For our work, the quantum speed limit characterizes the fidelity of the final state after trap movement, that is, whether at the end of trap acceleration, the state is that which we intended to achieve. A simple adaptation of the results in Fig. 3 of Lam et al implies that we can expect the fidelity of our final trap state and the interferometry contrast to be maximized for a single-direction time $t_{m}/2 \gtrsim 1.5 (2 \pi/\omega_{0})$, which is satisfied for most trap depths (unlike in Lam et al, our maximum achievable fidelity may be less than 1 since we do not optimize trap paths as they do).

\subsection{Experimental considerations of trap acceleration}

As just discussed, typical trap parameters suggest accelerations of order $10^{5}$ m/s$^2$ are likely to still lead to an appreciable interferometer signal. We now investigate the feasibility of reaching such high accelerations for extended times by leveraging a setup that employs an acousto-optic deflector (AOD)~\cite{Duocastella2021}. We discover that such an apparatus can provide co-gravity accelerations of up to $10^{3}$ m/s$^2$-$10^{5}$ m/s$^2$. 
However, the need for a tight focus limits the total distance over which accelerations can be applied. This distance is set by the Field of View (FOV) of the objective lens used to focus the trap beam wavefront and the quality of the trap beam wavefront itself. For an unaberrated incoming trap beam wavefront, this distance is simply the FOV of the objective.

Light from an infinite source cannot be focused down to a point by an ideal lens due to the effects of diffraction. This is antithetical to the prediction from geometrical optics. This diffraction-limited focused spot size is inversely proportional to the NA of the ideal lens. However, in the case of a real lens,
diffraction-limited performance is achievable only within the lens's FOV. Beyond the FOV of real lenses, lens aberrations
dominate the effects due to diffraction. This results in increased focal spot size (weaker traps for the same optical trap
power and detuning) and the optical system is said to be aberration-limited. While lenses that simultaneously possess large
FOV (greater space-time area) and large NA (tighter dipole traps and higher $a_{\rm max}$) are desirable, these lens parameters are, unfortunately,
inversely related to each other~\cite{FOV1,FOV2}.

\begin{figure}
    \centering
      \subfloat[]{       
      \includegraphics[width=8.6 cm]{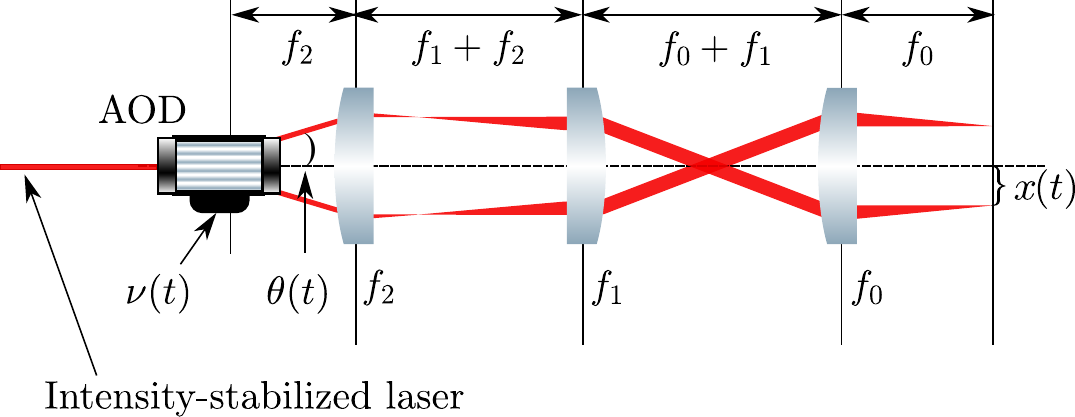}
      \label{fig:AODlensSetup_a}
     }
     \hfill
     \subfloat[]{     
     \includegraphics[width=8.6 cm]{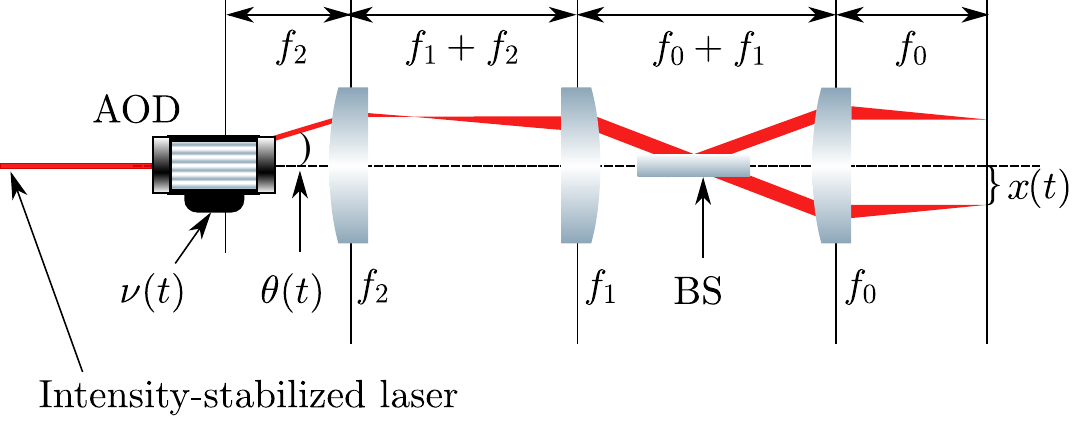}
      \label{fig:AODlensSetup_b}
    }
    \caption{AOD and lens assembly arranged in a $6f$ configuration in order to move two optical tweezers with minimal common-mode noise. \textbf{(a)} Electronically generating the tweezers using an amplitude modulated waveform $\nu(t)$ \textbf{(b)} Optically generating the tweezers using a phase grating, or custom 50:50 beam splitter, or a polarizing prism like Rochon/Wollaston/S\'enarmont prism, or a metasurface. }  
    \label{fig:AODlensSetup}
\end{figure}

A typical setup for accelerating an optical tweezer in the focal plane of an objective lens ($f_0$) in shown in Fig.~\ref{fig:AODlensSetup}. In order to minimize the relative intensity and pointing noise between the bottom arm tweezer and top arm tweezer (Section~\ref{NoisSection}), both tweezers must be generated simultaneously using one element from an intensity-stabilized laser beam source~\cite{Vahlbruch2018,Kwee2010,Buchler1998} and also share the same optical beam path. This element can either be electronic or optical. The tweezer beams can be generated electronically by driving the AOD with an amplitude modulated waveform (Fig.~\ref{fig:AODlensSetup_a}). In order to generate the tweezer beams using an optical element, one may use phase gratings, custom 50:50 beam splitters, polarizing prisms of the Rochon/Wollaston/S\'enarmont type, or even metasurfaces (Fig.~\ref{fig:AODlensSetup_b}). 

Arbitrary motion of the tweezer in the focal plane is facilitated by the electronically engineered radio-frequency (rf) waveform $\nu(t)$ driving the AOD. When the AOD is driven by the rf waveform, it deflects the beam by an angle $\theta(t)$ (Fig.~\ref{fig:AODlensSetup}). $\theta(t)$ is then mapped to a tweezer position $x(t)$ by the 6f optical system constructed out of a 3 lens assembly~\cite{goodman2005introduction,doi:https://doi.org/10.1002/9783527699223.ch12}. The lenses must be positioned appropriately, as this facilitates the angle-to-position mapping, which can be quantitatively expressed via the ABCD matrix analysis~\cite{ABCDmatrix,ABCDmatrix2} as follows:

\begin{equation}
x(t)=-\frac{f_0}{M}~\theta(t)
\label{eq:RayMatrix}
\end{equation}
where $M=f_1/f_2$ is the magnification. $\theta(t)$ and $x(t)$ are measured with respect to the optic axis of the $6f$ system. This angle-to-position mapping can be readily extended to the other axes by using a 2-axis AOD allowing for arbitrary 2D tweezer motion in the focal plane. We only treat one axis here for simplicity.   

The relation between $\theta(t)$ and the radio-frequency (rf) drive $\nu(t)$ to an AOD operating in the Bragg regime is $
    \theta(t)=2 \sin ^{-1} \frac{\lambda}{2 \lambda_{\mathrm{rf}}(t-\tau)}\simeq\frac{\lambda\nu(t-\tau)}{v_{\mathrm{rf}}},
$ where $v_{\mathrm{rf}}$ is the velocity of sound in the AOD crystal, $\lambda$ is the wavelength of light in the AOD crystal, $\nu(t)$ is the rf frequency waveform, and $\tau$ is the access time of the AOD i.e. the finite time (set by the speed of sound in the AOD crystal) for a change in the rf frequency $\nu(t)$ to traverse the entirety of the input laser beam spot size~\cite{SteckClassicalOptics}. $\tau$ is typically $\sim1$ $\mu s$ for an input beam with a diameter of 1 mm. Substituting this expression for $\theta(t)$ in Eq.~\ref{eq:RayMatrix}, we get the following important relation between $x(t)$ and $\nu(t)$:
\begin{equation}
    x(t)=-\frac{\lambda f_0}{ Mv_{\mathrm{rf}}}\nu(t-\tau).
    \label{main0}
\end{equation}
 This governing equation (Eq.~\ref{main0}) is what maps the motion of the tweezer in the focal plane of the objective lens to an electronically engineered rf waveform $\nu(t)$. For instance, if the functional form of $\nu(t)$ is parabolic in time ($\nu(t)\propto t^2$), then the tweezer moves at a constant acceleration. Of course, the motion of the tweezer is limited to within the FOV of the objective lens ($|x(t)|\le$FOV) in order to guarantee to diffraction-limited performance from the objective.

We can program either an Arbitrary Waveform Generator (AWG)~\cite{Note} or a Direct Digital Synthesizer (DDS)~\cite{Murphy2004,Devices1999} to digitally generate the desired frequency chirp $\nu(t)$. AWGs have faster update rates than DDS sources and have more flexibility over DDS sources in engineering waveforms. AWGs are therefore typically more expensive than DDS sources.  If one decides to use a DDS source to engineer the $\nu(t)$ waveform, the waveform will be discretized in units of time of duration $\eta$ ($\eta$ must be longer than the frequency switching time of the DDS). During the time step of duration $\eta$, the DDS outputs a single frequency tone. It is therefore important to set a bound on the maximum tolerable $\eta$ for a particular tweezer motion application.   

Let us consider the case of a tweezer accelerating at a rate $a$. For each time step (of duration $\eta$) in the chirp waveform, $\nu(t)$ is a constant frequency tone. A constant frequency tone maps to a specific tweezer position via Eq.~\ref{main0}. Therefore, a change in $\nu(t)$ to a different constant frequency tone leads to a change in the tweezer position. For the atoms to follow these jumps in tweezer positions and avoid trap loss, it is imperative that the overlap integral between the ground state wavefunctions of the atoms in the tweezer for two consecutive tweezer positions is close to 1. 

The ground state wavefunction of an atom with mass $m$ in a harmonic potential well with trap frequency $\omega$ at $y=0$
is given by $u_0(y)=\left(\frac{m \omega}{\pi \hbar}\right)^{\frac{1}{4}} e^{-m \omega (y)^2 / 2 \hbar}$.  When the harmonic potential well is suddenly displaced by $\Delta y$, the probability that the atom will follow this sudden displacement should be close to 1 i.e.
$\int^\infty_{-\infty}{u_0(y)u_0(y-\Delta y)} dy=e^{-\frac{\Delta y^2 m \omega}{4 \hbar}}\sim 1.$ For $y(n\eta)=a(n\eta)^2/2$, this implies 
$n\ll \frac{2}{a\tau^2}\sqrt{\frac{\hbar}{m\omega}}$ for $n\gg1$. Using $n=\frac{1}{\tau}\sqrt{\frac{2\textrm{FOV}}{a}}$ as the number of time steps needed to accelerate the tweezer over the FOV of the objective lens, we get a bound on $\eta$: $\eta\ll\sqrt{\frac{2\hbar}{am\omega \textrm{FOV}}}.$ For $m=133 \textrm{ amu}, \omega=2\pi\times10 \textrm{ kHz}, \textrm{FOV}=200 \mu \textrm{m}, a= 1000 \textrm{ ms}^{-2}$, we get $\eta\ll276$ ns. This criterion is satisfied for most high-end DDS sources, as their frequency switching times are in the tens of nanoseconds range~\cite{Caruso2003, Zhang2022, New, Ridenour}. We'd also like to point to the favourable $1/\sqrt{a}$ scaling for $\eta$. With complete waveform control and faster update rates facilitated by AWGs, accelerations in the 10$^{5}$ m/s$^2$ range seem feasible. Furthermore, the bound on $\eta$ is relatively strict as it was derived for the largest jump in tweezer position, which only occurs at the end of the chirp.

Lastly, we'll need to ensure that the maximum frequency excursion ($\nu_{\textrm{chirp}}$) during the chirp is well within the bandwidth (BW) of the AOD ($\nu_{\textrm{chirp}}=\frac{M\textrm{FOV}v_{\mathrm{rf}}}{\lambda f_0}\le\textrm{BW})$. For an FOV$=200\mu$m, $M=4$, $\lambda=\lambda_0/n_0=409$ nm, $v_{\mathrm{rf}}=$ 650 m/s, $f_0=$30 mm, we get $\nu_{\textrm{chirp}}=42.4$ MHz, which is within the typical AOD bandwidths ($>50$ MHz)~\cite{Duocastella2021}. Here $\lambda_0=900$ nm is the wavelength of the light in vacuum and $n_0=2.2$ is the refractive index of the TeO$_2$ AOD crystal.  

In spite of the content of the discussions so far, trap movement is not necessarily restricted to being along the focal plane. Indeed, being able to realize fast movement perpendicular to the focal plane will allow for a greater flexibility in the spatial region that can be sampled. Varifocal lenses are well suited for that particular application~\cite{Kang2020,Duocastella2021}. Varifocal lenses are lenses with tunable focal lengths and are typically placed $5f$ away from the objective back focal plane in a $6f$ optical system, just like in the optical arrangement illustrated in Fig.~\ref{fig:AODlensSetup}. In fact, the varifocal lens takes the place of the AOD in the $6f$ optical system and this placement helps with preserving magnification when the focus is tuned~\cite{Kang2020}. One can therefore pair a 2-axis AOD with a varifocal lens for arbitrary three-dimensional control over the tweezer position. This can be  done by integrating the 2-axis AOD and varifocal lens into a $10f$ arrangement, which is basically a $4f$ arrangement plus a $6f$ arrangement. The Fourier plane of the AOD is first mapped downstream to the varifocal lens plane by a $4f$ optical arrangement ~\cite{goodman2005introduction, doi:https://doi.org/10.1002/9783527699223.ch12}. Then the varifocal lens plane (plus the mapped AOD Fourier plane) is placed $5f$ upstream from the back focal plane of the objective in a $6f$ optical arrangement (Fig.~\ref{fig:AODlensSetup}). It may then be possible to achieve large accelerations perpendicular to the focal plane of the objective using a state-of-the-art Tunable Acoutic Gradient (TAG) lens, a type of varifocal lens, that can change its focal length in a few tens of microseconds~\cite{Mermillod-Blondin:08,Kang2020}. This is in addition to the large accelerations already possible in the 2D focal plane of the objective facilitated by the 2-axis AOD. However, one will need to consider the effects of aberrations and aperture size when using a varifocal lens for controlling the tweezer axial focus shift. Large axial focus shifts with respect to the default front focal plane of the objective will compromise its diffraction-limited performance and lead to weaker traps~\cite{Kang2020}.  

Another method could involve the use of an optical lattice, rather than a dipole trap, to move atoms perpendicular to the gravitational field, such as in the paper by Schmid et al~\cite{OpticalLatticeTransport}. Their lattice is created using counter-propagating Gaussian and Bessel beams. The Bessel beam provides the vertical force countering gravity. The detuning between the two beams is varied using an acousto-optic modulator (AOM) in order to move the atoms perpendicular to gravity. The velocity of the lattice is given by $v = (\lambda/2)\Delta\nu$, where $\lambda$ is the lattice wavelength and $\Delta\nu$ the detuning. Since $\Delta\nu$ has to be maintained such that it is within the bandwidth of the AOM, the maximum velocity that can be achieved is around $6$ m/s, in accordance with an AOM bandwidth of $15$ MHz. With a lattice depth of about 20 $E_{R}$, the authors are able to achieve a ``critical" acceleration of $\approx$ 2600 m/s$^2$, where the ``critical" acceleration is defined to be that up to which at most 50$\%$ of atoms are lost after traversing a closed path of total distance 5 mm. They experimentally verify that this critical acceleration is sufficiently described with the classical limit $a = qV_d/m$, where $q$ is the wave vector. Additionally, they were able to transport atoms nearly 20 cm in one direction (within 200 ms), before atom loss, due to the decreasing vertical force of the Bessel beam (which is used to counter gravity), became significant.

Now that we have identified possible methods by which moving trap interferometers can be physically realized, it is informative to understand where our proposal stands with existing technology. Here, we make a comparison with 100$\hbar k$ atom interferometers. Chiow et al.~\cite{102hk} use 17 6$\hbar k$ pulses to relay a total momentum of 102$\hbar k$ to an atom, giving it a resultant change in velocity of approximately 1 m/s. The velocity transferred per pulse can be approximated to be 1/17 m/s $\approx$ 0.06 m/s (the atom's velocity loss due to gravity after a single pulse and, hence, that which the succeeding pulse must additionally account for, is of order 1 mm/s, so it will be ignored in this approximation). Taking a single pulse to last approximately 0.1 ms, based on the data given in Fig. 1 of Ref.~\cite{102hk}, it can be said that the technology presented in the paper gives an atom a velocity of 0.06 m/s within 0.1 ms. During the same period of time, the accelerations that can be achieved by moving tweezer traps using AODs can result in velocities ranging from 0.1 m/s to 10 m/s. Compared to this specific 100$\hbar k$ interferometer, AOD technology can result in higher velocities. However, there is certainly the limitation that the distance covered during the actual movement of the trap is restricted to the FOV of the setup. This can possibly be overcome by releasing the atom(s) at the end of their acceleration, such that their free fall would result in additional distance being covered. This would essentially be the moving trap version of having atoms in free fall after a pulse momentum transfer. With the velocities calculated here, an atom subjected to a single 6$\hbar k$ pulse would travel 180 $\mu$m in free fall, while an atom released from an accelerated trap would travel between 500 $\mu$m to 5 m, depending on the acceleration. Nevertheless, this release method requires the catching of atoms in a trap after free fall~\cite{ThrowCatch}.

As investigated in this section, existing technology can achieve accelerations between $10^{3}$-$10^{5}$ m/s$^2$. Furthermore, while we only focus on two traps in this paper, it may also be possible to have an array of traps, such that several pathways can be simultaneously covered.

\subsection{Sensitivity with accelerating traps}
\label{Sensitivity}

\begin{figure*}[htb]
        \includegraphics[width = 16 cm]{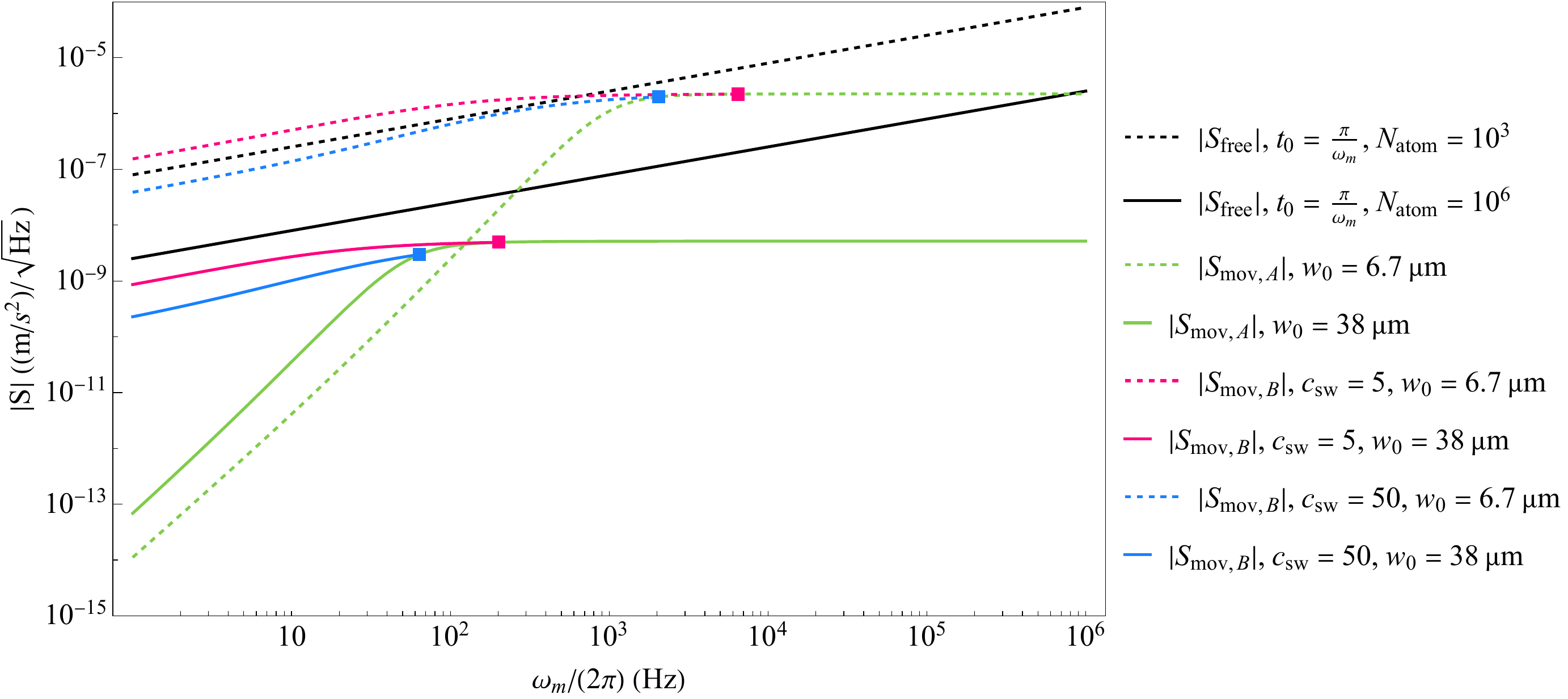}
        \caption{Plot of sensitivity $|S|$ against $\omega_{m}/(2\pi)$ for Equations~\ref{eq:sensitivityStat}, \ref{eq:InSensRegime1}, and \ref{eq:InSensRegime2} (color online). \textbf{Black} lines have been used for $|S_{\rm free}|$, \textbf{green} for $|S_{\rm mov,A}|$, \textbf{pink} for $|S_{\rm mov,B}|$ with $c_{sw} = 5$, and \textbf{blue} for $|S_{\rm mov,B}|$ with $c_{sw} = 50$. \textbf{Dashed} lines correspond to the effective use of $10^{3}$ atoms, while the \textbf{non-dashed} lines to $10^{6}$ atoms. We have set $\lambda = 935.6$ nm~\cite{935.6}, $\rho_{\rm MOT}= 1.1 \times 10^{11}$ cm$^{-3}$~\cite{MOTdensity}, $q = 2\pi/\lambda$, and $r(t_{0}) \rightarrow r_{\rm MOT}$. For cesium~\cite{SteckCesium}, $m=2.2 \times 10^{-25}$ kg, $\nu_{\rm atom} = 2106$ T(rad s$^{-1}$), $\gamma = 28.7$ M(rad s$^{-1}$), and $I_{\rm sat} = 24.98$ Wm$^{-2}$. For $|S_{\rm mov,(A/B)}|$, $L=10w_{0}$ and $P_{m}=1$ W. For $|S_{\rm free}|$, $t_{\rm shot}=2t_{0}$ and $L=380$ µm. Trap depths of $6.7$ µm and $38$ µm correspond to approximately $10^{3}$ and $10^{6}$ atoms in the trap. It can be seen that sensitivity improvements compared to non-trapped interferometers occur at the lower end of the range of $\omega_{m}/(2\pi)$. Unrestricted $s_{0}$, as shown by $|S_{\rm mov,A}|$, can provide several orders of magnitude of improvement. Here, for the lower end of $\omega_{m}/(2\pi)$, the lower the $w_{0}$, the better the sensitivity, since for low $\omega_{m}$, the accelerating trap term of Eq.~\ref{eq:InSensRegime1} takes over. Conversely, for $|S_{\rm mov,B}|$, sensitivity is better for higher $w_{0}$. Further, the square markers indicate that each line is truncated at $\omega_{\rm m,max}$.}
    \label{fig:SensPlots}
\end{figure*}
We can understand the advantage of the high accelerations discussed in the previous sections by considering the sensitivity. In the case of non-trapped interferometers, the sensitivity to gravitational acceleration is given by~\cite{Sensitivity}, 
\begin{equation}
    \ S_{\rm free} = \frac{\hbar}{\sqrt{\frac{N_{\rm atom}}{t_{\rm shot}}}mLt_{0}},
\label{eq:sensitivityStat}
\end{equation}
For high-bandwidth applications, $t_{0}$ is required to be shorter, thereby worsening the value of $S_{\rm free}$. In contrast, the general sensitivity for our scheme is 
\begin{equation}
    \ S_{\rm mov} = \frac{\hbar}{\sqrt{\frac{N_{\rm atom}}{t_{\rm shot}}}[-mL(\tau + t_{0})+\int_{0}^{t_{m}} dt~m(s_{t}(t)-s_{b}(t))]}
\label{eq:SensitivityMov}
\end{equation}
which, for constant acceleration and the case where $t_{m}=2\pi/\omega_{m}$ and $t_{\rm shot}=2\pi/\omega_{m} + 2 t_{0}$, reduces to

\begin{equation}
    \ |S_{\rm mov}| = \frac{\hbar}{\sqrt{\frac{N_{\rm atom}}{\frac{2\pi}{\omega_{m}}+2t_{0}}}[mL(\tau + t_{0})+\frac{2\pi m s_{0}}{\omega_{m}}]}
\label{eq:SensitivityMovConstAccel}
\end{equation}


The number of atoms which are actively used in the interferometer vary at the different steps of the sequence. For instance, when atoms are released from the MOT, there is thermal expansion of the atom cloud~\cite{CloudExpansion1} and not all of these atoms are trapped when the optical dipole trap is turned on in the middle of the sequence. Therefore, it is important to express $N_{\rm atom}$ in Eq.~\ref{eq:SensitivityMovConstAccel} in terms of experimental parameters that are initially known, namely, the density $\rho_{\rm MOT}$ of atoms in the MOT, and either the radius $r_{\rm MOT}$ of the atom cloud in the MOT right before release or the initial number of atoms $N_{\rm atom,MOT}$ in the MOT. It is also useful to express the time and distance terms in Eq.~\ref{eq:SensitivityMovConstAccel} in terms of the parameters of accelerating traps. Rewriting Eq.~\ref{eq:SensitivityMovConstAccel} as such means that it will be expressed more fundamentally in terms of MOT and dipole trap parameters, such that we can more conveniently grasp which values they must take in order to have the best sensitivities. In our analysis, we consider two main regimes: trapping $10^{3}$ and $10^{6}$ atoms.

For the purposes of simplifying the calculation, we approximate that the atom cloud is a uniform sphere with a uniform density distribution in the MOT right before release with radius $r_{\rm MOT}$, expands as a uniform sphere during free fall, and that this expansion is purely thermal due to a cloud temperature of $T_{\rm free}$. We can then express the radius $r(t_{0})$ of the atom cloud right before the optical trap is switched on as follows~\cite{CloudExpansion1,CloudExpansion2}:
\begin{equation}
    \ r(t_{0})^{2} = r_{\rm MOT}^{2}+\frac{k_{\rm B}T_{\rm free}}{m}t_{0}^{2}.
\label{eq:Cloudradius}
\end{equation}
Evaluating as in Appendix \ref{SensitivityElaboration}, for the case where the optical dipole trap is turned on abruptly (an approximation made for mathematical ease), the number of atoms $N_{\rm atom,trap}$ that gets trapped is given by
\begin{equation}
    \ N_{\rm atom,trap} \approx  0.417 \frac{\pi^{2}}{\lambda} \rho_{\rm MOT} \left(\frac{r_{\rm MOT}}{r(t_{0})}\right)^{3}  w_{0}^{4}. 
\end{equation}

We must next relate the spring constant of the trap to the laser's peak intensity $I_{m}$ and the atom's saturation intensity $I_{\rm sat}$ as follows~\cite{OpticalTweezersTheoryAndPractice,SteckCesium,SteckQuantumOptics}
\begin{equation}
   \ k = \chi \frac{I_{m}}{I_{\rm sat}}
\label{eq:SpringConstant}
\end{equation}
where 
\begin{equation}
   \ \chi = \frac{\hbar\Delta \gamma^{2}}{2w_{0}^{2}\left(\frac{\gamma^{2}}{4}+\Delta^{2}\right)}.
\label{eq:SpringConstantCoefficient}
\end{equation}
Here, $\Delta = \nu_{\rm atom}-\nu$ is the detuning, with $\nu_{\rm atom}$ being the resonant frequency of the atom and $\nu$ the frequency of the trap's laser, and $\gamma$ is the spontaneous emission rate. 


With all these parameters in hand, we can now understand the sensitivity for different cases in the limit $a_{\rm max} \gg g$.

The first is where we can achieve arbitrarily large values of $s_{0}$ for a fixed power $P_{m}=I_{m} \times \pi w_{0}^{2}$; this may require a setup other than the one proposed with an AOD. Here, we have the conditions that
\begin{align}
   \ c_{L} \equiv \frac{L}{w_{0}}=\text{constant},~c_{L} \gg 1 
   \label{eq:SensRegime1_a}\\
   t_{0} = \frac{m L}{\hbar q},~\tau \sim \frac{2\pi}{\omega_{m}},~\text{and}~ s_{0}=\frac{\chi_{s}}{\omega_{m}^{2}w_{0}^{3}}
\label{eq:SensRegime1_b}\\
   \chi_{s} = \frac{\chi_{c}P_{m}}{\pi\sqrt{e}mI_{\rm sat}},~\chi_{c}  = w_{0}^{2}\chi
\label{eq:SensRegime1_c}
\end{align}
$c_{L} \gg 1$ is achieved by either realizing an atom cloud launch velocity or initial laser pulse kick $q$ such that $L \gg w_{0}$. The times of ramp up and ramp down are chosen to be small enough such that $\tau \sim 2\pi\omega_{m}^{-1}$. The expression for $s_{0}$ is determined by $a_{\rm max,vert}$ of Eq.~\ref{eq:amaxvert} and the maximum gradient of $\Dot{s}(t) = -(s_{0}/2)\omega_{m} \sin(\omega_{m}t)$; as such $s_{0}$ of Eq.~\ref{eq:SensRegime1_b} is constrained by the classical limit of maximum acceleration.

With these, for the case where we have a fixed power $P_{m}$ and  can achieve arbitrarily large values of $s_{0}$, i.e, $s_{0}$ is not constrained by the optics, the inverse sensitivity is given by
{\small
\begin{multline}
   \ |S_{\rm mov,A}|^{-1} = \frac{1}{\hbar}\sqrt{0.417 \frac{\pi^{2}}{\lambda} \rho_{\rm MOT} \left(\frac{r_{\rm MOT}}{r(t_{0})}\right)^{3}w_{0}^{4}}\\
   ~~~~~~~~~ \times \sqrt{\frac{1}{\left(\frac{2\pi}{\omega_{m}}+\frac{2mL}{\hbar q}\right)}} 
~\left[mL\left(\frac{2\pi}{\omega_{m}}+\frac{mL}{\hbar q}\right)+\frac{2\pi m \chi_{s}}{\omega_{m}^{3}w_{0}^{3}} \right]
\label{eq:InSensRegime1}
\end{multline}}

The second case we can look at is where the power $P_{m}$ is variable in the setup, but $s_{0}$ is constrained by the optics. In particular, we have that
\begin{align}
   \ c_{sw} \equiv \frac{s_{0}}{w_{0}}=\text{constant}\\
L \gg w_{0},~\text{and}~\tau \sim \frac{2\pi}{\omega_{m}}
\label{eq:SensRegime2}
\end{align} 
That the ratio $s_{0}/w_{0}$ is fixed reflects a setup where the acceleration is implemented using AOD technology, because both $w_{0}$ and $s_{0}$ are set by the same optics. Since $s_{0}=\chi_{s}/(\omega_{m}^{2}w_{0}^{3})$, the fixed ratio requires that $\chi_{s}$ (hence, $P_{m}$, and, consequently, $V_{d}$) be changed with $\omega_{m}$ such that $\chi_{s}/\omega_{m}^{2} = c_{sw} w_{0}^{-4}$ for a particular $w_{0}$. However, this ratio can only be maintained up to a finite bandwidth $\omega_{m}$; as $\chi_{s}$ is proportional to $P_{m}$ and the achievable power has a maximum value $P_{\rm m,max}$, the maximum bandwidth $\omega_{\rm m,max}$ that can be sampled for a particular optics setup is
\begin{equation}
   \ \omega_{\rm m,max}=\frac{1}{w_{0}^{2}}\sqrt{\frac{\chi_{s}(P_{\rm m,max})}{c_{sw}}},
\label{eq:SensRegime2}
\end{equation}
where $w_{0}$ and $c_{sw}$ characterize the optical system.

Therefore, in the case where the power $P_{m}$ is variable in the setup, but $s_{0}$ is constrained by the optics, we have an inverse sensitivity of 
{\small
\begin{multline}
   \ |S_{\rm mov,B}|^{-1} = \frac{1}{\hbar}\sqrt{0.417 \frac{\pi^{2}}{\lambda} \rho_{\rm MOT} \left(\frac{r_{\rm MOT}}{r(t_{0})}\right)^{3}}\\
   \sqrt{\frac{w_{0}^{4}}{\left(\frac{2\pi}{\omega_{m}}+\frac{2mL}{\hbar q}\right)}}
~\left[ mL\left(\frac{2\pi}{\omega_{m}} +\frac{mL}{\hbar q}\right)+\frac{2\pi m c_{sw}w_{0}}{\omega_{m}}\right]
\label{eq:InSensRegime2}
\end{multline}}

Plots of Equations~\ref{eq:sensitivityStat}, \ref{eq:InSensRegime1}, and \ref{eq:InSensRegime2} are shown in Fig. \ref{fig:SensPlots}. It can be seen that sensitivity improvements compared to non-trapped interferometers occur at the lower end of the range of $\omega_{m}/(2\pi)$. Unrestricted $s_{0}$, as shown by $|S_{\rm mov,A}|$, can provide several orders of magnitude of improvement. Here, for the lower end of $\omega_{m}/(2\pi)$, the lower the $w_{0}$, the better the sensitivity, since for low $\omega_{m}$, the accelerating trap term of Eq.~\ref{eq:InSensRegime1} takes over. Conversely, for $|S_{\rm mov,B}|$, sensitivity is better for higher $w_{0}$. Further, the square markers indicate that each line is truncated at $\omega_{\rm m,max}$.

Note that the calculations in this section are correct up to factors of order unity corrections to $N_{\rm atom}$ used, since we have not included the process of creating a superposition of the initial atom cloud and having two traps in the calculation.

\section{The noisy atom trap}
\label{NoisSection}

\subsection{Understanding spring constant noise}
\label{SpringConstantNoiseMain}

Fluctuations in the laser intensity can lead to a time-dependent spring constant. In this paper, for the spring constant as given by Equations~\ref{eq:SpringConstant} and \ref{eq:SpringConstantCoefficient}, we expect the noise to be largely due to fluctuations in $I_{m}$.

Using the forms $k(t) = k_{0}+\delta k(t)$ and $I_{m}(t) = I_{m, 0}+\delta I_{m}(t)$ for the noise, where $\delta k(t)$ and $\delta I_{m}(t)$ are fluctuations from the ideal value, we can relate the distribution of the spring constant to that of the intensity as follows
\begin{multline}
   \ \langle \langle \delta k(t) \delta k(t+\Tilde{\tau}) \rangle \rangle = \left(\frac{\chi}{I_{\rm sat}}\right)^{2}
   \langle \langle \delta I_{m}(t) \delta I_{m}(t+\Tilde{\tau}) \rangle \rangle
\label{eq:deltakdeltaI}
\end{multline}
with $\Tilde{\tau} = t'-t$. That is, the auto-correlation function of the spring constant is proportional to that of the intensity.

We further have the relationship between the spectral density $S_{I_{m}}(f)$ of the intensity noise and the intensity $I_{m}$ of the laser~\cite{RelIntNoise},
\begin{equation}
   \ S_{I_{m}}(f) = \frac{2}{\langle\langle I_{m} \rangle\rangle ^{2}}\int_{-\infty}^{\infty}\langle \langle \delta I_{m}(t) \delta I_{m}(t+\Tilde{\tau}) \rangle \rangle e^{i2\pi f \Tilde{\tau}} d \Tilde{\tau},
\label{eq:IntensitySD}
\end{equation}
this being the normalized Fourier transform of the auto-correlation function of the intensity.

We can see that the relationship between $S_{k}(f)$ and $S_{I_{m}}(f)$ is 
\begin{equation}
   \ S_{k}(f) = S_{I_{m}}(f);
\label{eq:SISKrelation}
\end{equation}
the spectral density of the spring constant is identical to that of the peak intensity.  

\subsection{Intrinsic laser noise}
\label{IntrinsicNoiseMain}

Here, we obtain explicit expressions for the spring constant's noise distribution in the event that laser intensity fluctuations are caused by fluctuations in the photon number. 

From the fact that the coherent state's number statistics follow a Poisson distribution, we take the distribution of the number of photons in a cavity, given by $N$, to be
\begin{equation}
   \ \langle\langle N(t')N(t) \rangle\rangle - \langle\langle N(t) \rangle\rangle ^{2} \approx \Bar{N} e^{-\kappa|t'-t|},
\label{eq:NofPhotonsDist}
\end{equation}
where $\Bar{N}$ is the average number of photons in the cavity and $\kappa^{-1}$ is the coherence time of the light in that cavity.

We can use Equations~\ref{eq:ElectricFieldOperator} (relation between electric-field operator and number-of-photons operator) and \ref{eq:IntensityAndEField} (relation between intensity and electric field), with refractive index $n_{r} = 1$, to express the number of photons $N$ associated with the peak intensity $I_{m}$ to be $N = C_{I} I_{m}$, where
\begin{equation}
     \ C_{I} = \frac{\pi w_{0}^{2}l}{\hbar c 
     \nu},
\label{eq:CI}
\end{equation}
with $l$ being the length of the cavity and $c$ the speed of light. The average is then $\Bar{N} = C_{I} \Bar{I}_{m}$.
The distribution of $I_{m}(t)$ can also be derived as in Appendix~\ref{IntrinsicNoise}. These expressions can then be used, along with Eq.~\ref{eq:SpringConstant}, to show that  
\begin{multline}
   \ \langle\langle k(t')k(t) \rangle\rangle - \langle\langle k(t) \rangle\rangle ^{2} = \left(\frac{\chi}{I_{\rm sat}}\right)^{2} \frac{1}{\kappa}\frac{1}{C_{I}}\Bar{I}_{m}\kappa e^{-\kappa|t'-t|}.
\label{eq:kDist}
\end{multline}

Since $k(t) = k_{0}+\delta k(t)$, as mentioned previously, we can find the auto-correlation function of $\delta k(t)$ for the case where $\langle\langle \delta k(t) \rangle\rangle = 0$:
\begin{align}
   \ \langle\langle \delta k_{(b/t)}(t') \delta k_{(b/t)}(t) \rangle\rangle = \Gamma_{(b/t)} ~ \frac{\kappa}{2} e^{-\kappa|t'-t|}.
\label{eq:deltakDist}
\end{align}
where
\begin{align}
   \ \Gamma_{(b/t)} = 2\left(\frac{\chi}{I_{\rm sat}}\right)^{2} \frac{1}{\kappa}\frac{1}{C_{I}}\Bar{I}_{m}.
\label{eq:Gamma}
\end{align}
Here, $(b/t)$ indicates that the correlation is for either one of the bottom or top traps.

Between the two traps we have
\begin{multline}
   \ \langle\langle k_{b}(t')k_{t}(t) \rangle\rangle - \langle\langle k_{b}(t) \rangle\rangle \langle\langle k_{t}(t) \rangle\rangle = \langle\langle \delta k_b(t') \delta k_t(t) \rangle\rangle
\label{eq:kDistTwoTraps}
\end{multline}
and
\begin{equation}
   \ \langle\langle N_{b}(t')N_{t}(t) \rangle\rangle - \langle\langle N_{b}(t) \rangle\rangle \langle\langle N_{t}(t) \rangle\rangle \approx \Bar{N}_{bt} e^{-\kappa_{bt}|t'-t|}.
\label{eq:NofPhotonsDistTwoTraps}
\end{equation}
We can then express the auto-correlation function between the two traps to be 
\begin{equation}
    \begin{split}
     \ \langle\langle \delta k_{b}(t') \delta k_{t}(t) \rangle\rangle & = \Gamma_{bt} \frac{\kappa_{bt}}{2} e^{-\kappa_{bt}|t'-t|}.
    \end{split}
\label{eq:deltakbdeltaktDist}
\end{equation}
with
\begin{align}
   \ \Gamma_{bt} = \frac{2}{\kappa_{bt}}\frac{\chi_{b}\chi_{t}}{(I_{\rm sat})^{2}}\frac{1}{\sqrt{C_{I,b}C_{I,t}}}\sqrt{\Bar{I}_{m,b}\Bar{I}_{m,t}}~.
\label{eq:Gammabt}
\end{align}
$\sqrt{\Bar{I}_{m,b}\Bar{I}_{m,t}}$ comes from the  fact that we take $\Bar{N}_{bt}$ to be the geometric mean of the respective trap parameters, that is, $ \Bar{N}_{bt} = \sqrt{\Bar{N}_{b}\Bar{N}_{t}}$.

Looking at equations~\ref{eq:deltakDist} and \ref{eq:deltakbdeltaktDist}, it can be seen that the right hand side is proportional to $(\kappa_{(b/t)}/2) e^{-\kappa_{(b/t)}|t'-t|} \to \delta(t-t')$ in the limit $\kappa \rightarrow \infty$. That is, the spring constant noise distribution goes to that of white noise~\cite{GaussianWhiteNoiseBook,GaussianProcesses,GaussianLect2,GaussianDist}:
\begin{align}
    \langle\langle \delta k_{(b/t)}(t) \rangle\rangle &= 0,
    \label{eq:whitenoiseconditions1}\\
    \langle\langle \delta k_{b}(t)\delta k_{b}(t') \rangle\rangle &= \Gamma_{b} \delta(t-t'),
    \label{eq:whitenoiseconditions2}\\
    \langle\langle \delta k_{t}(t)\delta k_{t}(t') \rangle\rangle &= \Gamma_t \delta(t-t'),
    \label{eq:whitenoiseconditions3}\\
    \langle\langle \delta k_{b}(t)\delta k_{t}(t') \rangle\rangle &= \Gamma_{bt} \delta(t-t')
    \label{eq:whitenoiseconditions4}
\end{align}
These conditions, and, hence, the coefficients in Equations~\ref{eq:Gamma} and \ref{eq:Gammabt} will be used in the evaluation of noise-included coherence terms in the succeeding sections. \\

Using the quantities calculated in sections~\ref{SpringConstantNoiseMain} and \ref{IntrinsicNoiseMain}, we can obtain explicit expressions for the resulting coherences and noise variances for the cases of center of trap fluctuations, time-dependent spring constants, and trap depth fluctuations.

We further explore harmonic oscillator transitions which arise in the trap due to the time-dependent trap frequency, both during noiseless ramp up of the trap and the subsequent steady, but noisy, potential.  

\subsection{Center of trap and trap depth noise}
\label{TrapCenterNoise}
 
In this section, we investigate center of trap and trap depth noise. To incorporate these, we can adapt the moving trap analysis of Section~\ref{MovingTrapCoherence}. With noise, we have $\ket{\psi_{b,f}}_{\textrm{noise}}=\hat{U}^{T}_{\textrm{move}}\ket{0}$ being
\begin{multline}
    \ \ket{\psi_{b,f}}_{\textrm{noise}} = \hat{T}\exp[-\frac{i}{\hbar} \int_{0}^{t_{m}} dt~( \frac{\hat{p}^{2}}{2m}...\\
    +\frac{1}{2}m\omega_{0,b}^{2}\left(\hat{x}-s_{b}(t)+\eta_{b}(t)-\eta_{b}(0) \right)^{2} - mg\hat{x}-V_{b,d}(t))]\ket{0}.
\label{eq:MovingTrapStateNoise0}
\end{multline}
Note that we now have a time-dependent trap depth potential shift of $V_{b,d}(t)$. This time-dependency arises from the trap's spring constant fluctuations. However, in  order to directly adopt the calculation in Section~\ref{MovingTrapCoherence}, we keep the frequency $\omega_{0,b}$ of the simple harmonic potential term in the Hamiltonian time-independent. The fluctuations from the intended location of the trap center are given by $\eta_{b}(t)-\eta_{b}(0)$, which also states that at $t=0$ the location of the trap center is unperturbed~\cite{Heating2}. We will also consider the parameters for the top and bottom trap to be non-identical. Therefore, it is possible to reuse the calculations in Section~\ref{MovingTrapCoherence} by simply replacing the $s_{(b/t)}(t)$ and $\tilde{s}_{(b/t)}(t)$ with the new location shifts and keeping track of any new phases that arise due to trap parameter differences. 

Processing as in Appendix \ref{CenterOfTrapNoiseApp} (and under the same assumptions of ramp down as in the Section~\ref{MovingTrapCoherence}), we obtain the expressions for the coherence, which are of the same form as Equations~\ref{eq:CoherencePathDiff12} and \ref{eq:DAlphaBotSqr}, with notation $\tilde{\phi}_{\textrm{TOT}}$ replacing $\phi_{\textrm{TOT}}$, and $\tilde{\alpha}_{\textrm{top}}$ and $\tilde{\alpha}_{\textrm{bot}}$ replacing $\alpha_{\textrm{top}}$ and $\alpha_{\textrm{bot}}$. The new parameters $\tilde{\alpha}_{\textrm{bot}}(\tilde{\eta}_{b}(t),\Dot{\tilde{\eta_{b}}}(t))$, $\tilde{\phi}_{b,1}(\tilde{\eta_{b}},\Dot{\tilde{\eta_{b}}})$, and $\tilde{\phi}_{b,2}(\Dot{\tilde{\eta_{b}}})$ are functions of $\tilde{\eta_{b}}(t) = s_{b}(t) - \eta_{b}(t)+\eta_{b}(0)+g/\omega_{0,b}^{2}$.
We have,
\begin{equation}
    \begin{split}
        \ \tilde{\phi}_{\textrm{TOT}} = & \tilde{\phi}_{t,1}-\tilde{\phi}_{b,1}+\tilde{\phi}_{t,2}-\tilde{\phi}_{b,2} \\ 
       & -\frac{(\omega_{0,t}-\omega_{0,b}) t_{m}} {2}-\frac{mgL(\tau+t_{0})}{\hbar} \\
       & -\frac{1}{\hbar}\int_{0}^{t_{m}}dt~ mg[s_{\eta,t}(t)-s_{\eta,b}(t)] \\
       & +\frac{1}{\hbar}\int_{0}^{t_{m}}dt~\frac{1}{2}mg^{2}\left(\frac{1}{\omega_{0,t}^{2}}-\frac{1}{\omega_{0,b}^{2}}\right) \\
       & + \int_{0}^{t_{m}} dt~ \frac{(V_{t,d}(t)-V_{b,d}(t))}{\hbar}\\
       & + \Im[\tilde{\alpha}_{\textrm{bot}}^{*}\tilde{\alpha}_{\textrm{top}}]
    \end{split}
\label{eq:PhiTotNoise}
\end{equation}
where $s_{\eta,b}(t) = -s_{b}(t)+\eta_{b}(t)-\eta_{b}(0)$ and similarly for $s_{\eta,t}(t)$. The explicit form of $\tilde{\alpha}_{\textrm{bot}}^{*}\tilde{\alpha}_{\textrm{top}}$ for $\omega_{0,t}=\omega_{0,b}$ is given in Appendix~\ref{CenterOfTrapNoiseApp}.

Again note that in order to proceed with the calculation, we set $s_{b}(0)=-g/\omega_{0,b}^{2}$, so that $\tilde{s}_{b}(0) = 0$; similarly, we have $s_{t}(0)=-g/\omega_{0,t}^{2}$, so that $\tilde{s}_{t}(0) = 0$. However, the nuance arises that $s_{t}(0) \neq s_{b}(0)$ due to the different trap frequencies; as such, we are actually describing a diagram where the top trap starts at $x=-L-g/\omega_{0,t}^{2}$ and the bottom at $x=-g/\omega_{0,b}^{2}$. The calculation is exact for this case. However, for the purposes of visual representation, this means that the initial trap separation is not exactly $L$. Nevertheless, in the limit $L \gg |g/\omega_{0,t}^{2}-g/\omega_{0,b}^{2}|$, the initial trap separation can be approximated to be $L$. Therefore, we can again approximately think of the calculation in this section as simply describing Fig.~\ref{fig:InterSequence_b} translated by $-g/\omega_{0,b}^{2}$.

It is instructive to understand how the magnitude of the different noise terms in Eq.~\ref{eq:PhiTotNoise} compare.
First looking at the pointing noise, relabeled as
\begin{equation}
    \ \tilde{\phi}_{{\textrm{TOT,A}}}(t_{m}) = -\frac{1}{\hbar}\int_{0}^{t_{m}}dt~ mg[s_{\eta,t}(t)-s_{\eta,b}(t)],
\label{eq:PointingNoise1}
\end{equation}
we evaluate the variance of $\tilde{\phi}_{{\textrm{TOT,A}}}(t_{m})$ in the case of noise described by Brownian motion. We do this under certain conditions: we take the auto-correlation function of the noise in terms of its power spectral density to be~\cite{RelIntNoise}
\begin{equation}
    \  \langle\langle \eta_{(t/b)}(t)\eta_{(t/b)}(t+\tau)\rangle\rangle = \int_{-\infty}^{\infty} df~ S_{\eta,(t/b)}e^{-i2\pi f\tau};
\label{eq:TrapCenterFourier}
\end{equation}
we consider the noise of the top and bottom traps to be uncorrelated, such that $\ \langle\langle \eta_{t}(t)\eta_{b}(t+\tau)\rangle\rangle = 0$; we define the mean of the noise to be zero for both traps, such that $\ \langle\langle \eta_{(b/t)}(t)\rangle\rangle = 0$; we label $\tau = t'-t$. 

We proceed with the calculation in Appendix \ref{CenterOfTrapNoiseApp} to obtain a general expression for the noise as given by Equation~\ref{eq:TrapCenterPSD}. As a concrete example, we evaluate Equation~\ref{eq:TrapCenterPSD} for the case where trap center noise is modeled by Brownian motion, such that the spectral density has the relation $S_{\eta,(t/b)}(f) = D_{(t/b)}/(\pi^{2}f^{2})$~\cite{BrownianPSD,BrownianPSD2,BrownianMotion} -- here, the different prefactor compared to that in Ref~\cite{BrownianPSD} is due to the difference in the Fourier transform convention between Eq.~\ref{eq:TrapCenterFourier} and Reference~\cite{BrownianPSD}. With this, the variance of $\tilde{\phi}_{{\textrm{TOT,A}}}$ is 
\begin{equation}
    \ \sigma_{\tilde{\phi}_{{\textrm{TOT,A}}}}^{2} = \frac{4}{3}\left(\frac{mg}{\hbar}\right)^{2}(D_{t}+D_{b})~ t_{m}^{3},
\label{eq:BrownianSquareMean}
\end{equation}
where $D_{(t/b)}$ is the diffusion coefficient, corresponding to the mathematical relations as given in Ref.~\cite{BrownianPSD}, for the top and bottom traps. It can be seen that $\sigma_{\tilde{\phi}_{{\textrm{TOT,A}}}}^{2}$ varies linearly with $D_{t}$ and $D_{b}$, and cubically with $t_{m}$.

Next, we look at trap depth noise, specifically the term
\begin{equation}
    \ \tilde{\phi}_{{\textrm{TOT}},B}(t_{m}) = \int_{0}^{t_{m}} dt~ \frac{(V_{t,d}(t)-V_{b,d}(t))}{\hbar}
\label{eq:TrapDepthNoise1}
\end{equation}
In order for the trap depth noise to be zero, we want $V_{t,d}(t)=V_{b,d}(t)$, which occurs when both the laser correlations and the optics used are identical. This can be understood explicitly by considering the more fundamental parameters that make up the trap depths $V_{t,d}(t)$ and $V_{b,d}(t)$ or by looking at the variance of the noise.

\begin{figure}
    \centering
      \subfloat[]{       
      \includegraphics[width=8.6 cm]{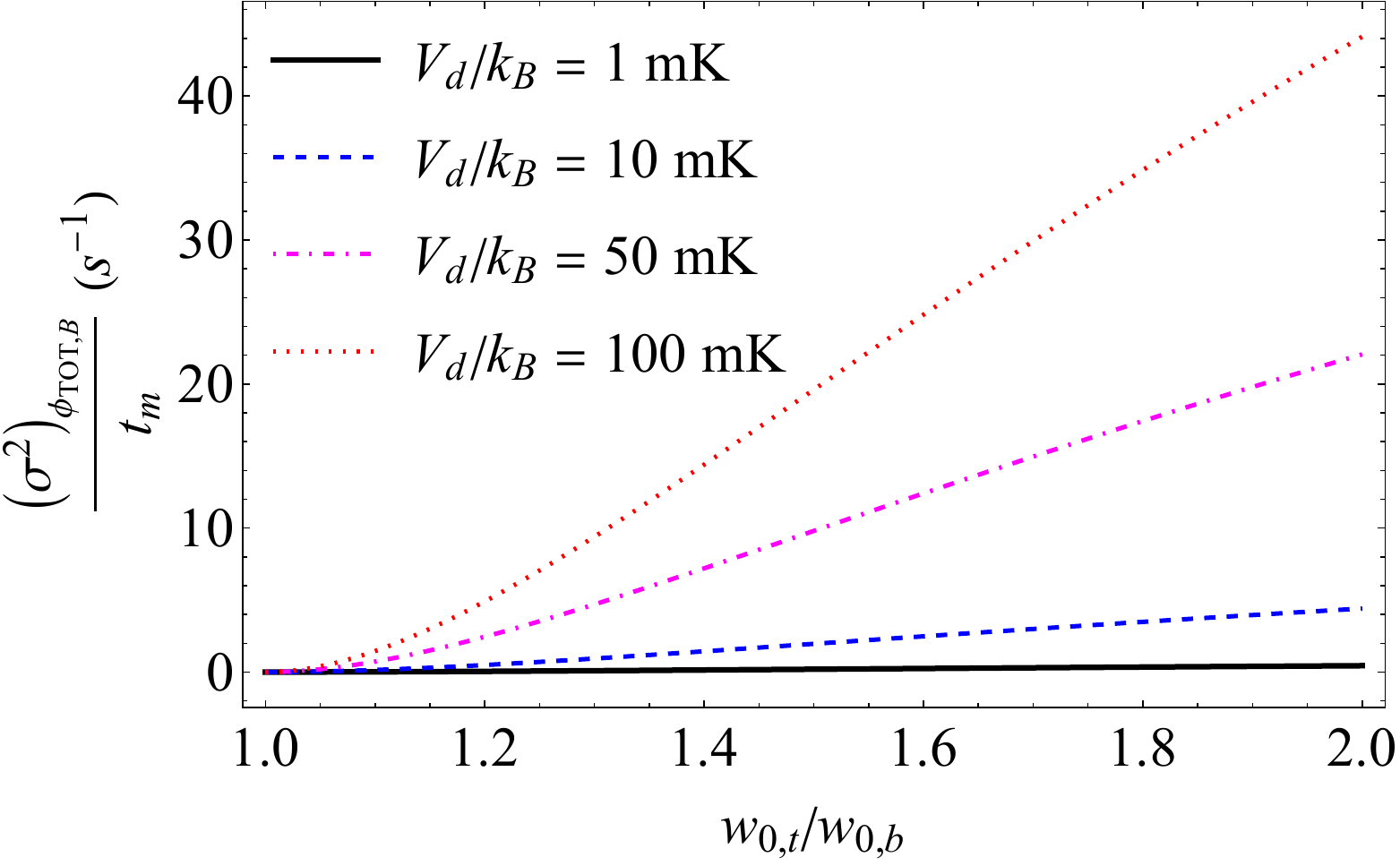}
      \label{fig:TrapDepthNoiseWaistRatios}
     }
     \hfill
     \subfloat[]{     
     \includegraphics[width=8.6 cm]{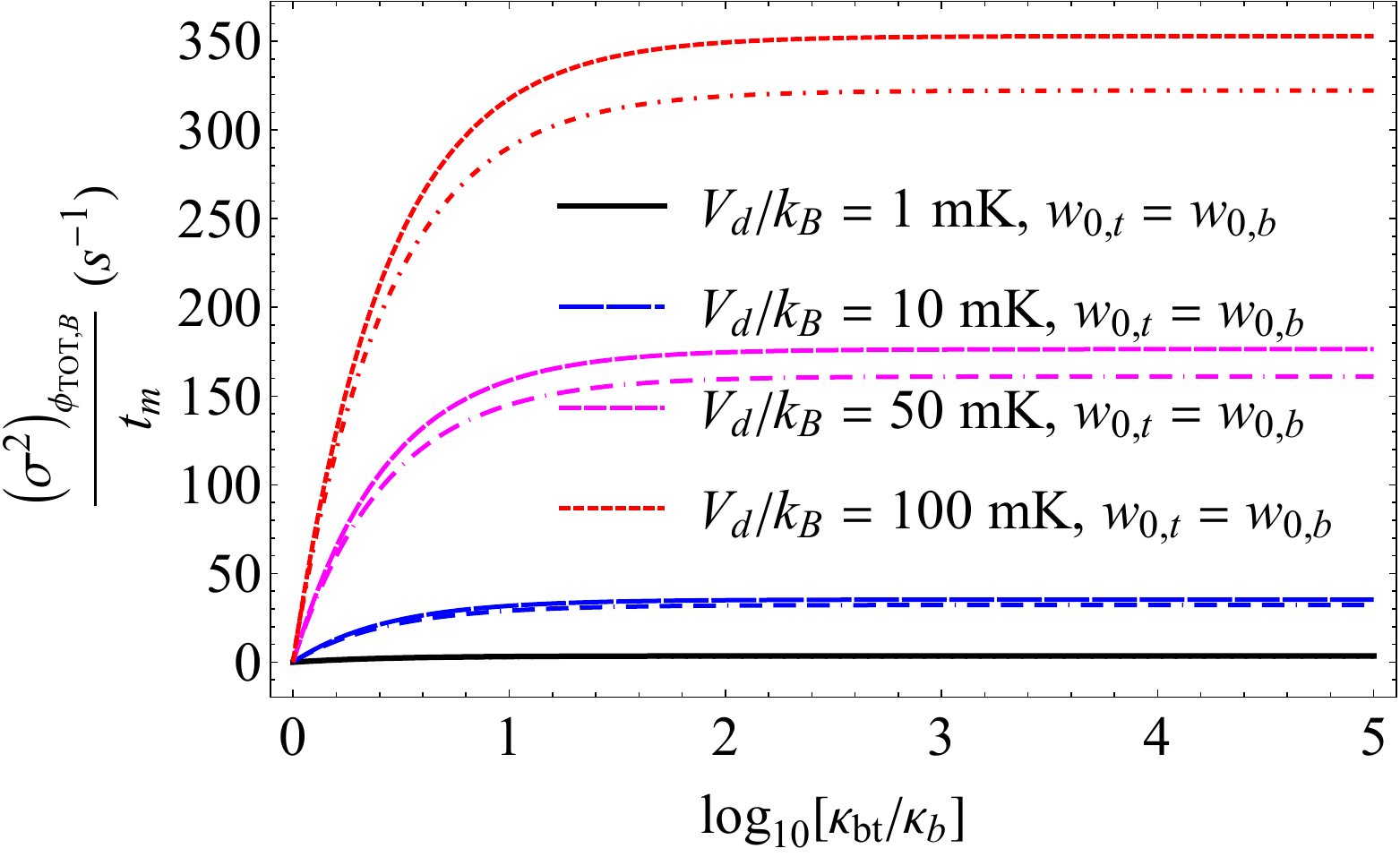}
      \label{fig:TrapDepthNoiseFullRange}
    }
    \caption{The parameters used are $\nu_{b} = \nu_{t}=2015$ T(rad s$^{-1}$) from $\lambda=935.6$ nm~\cite{935.6}, $w_{0,b} = 38$ $\mu$m, $\kappa_{(b/t)} = (\pi c)/(l_{(b/t)}F)$ as derived in Appendix~\ref{IntrinsicNoise}, $l_{b} = 200 $ $\mu$m, $l_{t}=l_{b}$, $F = 10^{4}$, and, for cesium~\cite{SteckCesium}, $\nu_{\rm atom} = 2106$ T(rad s$^{-1}$), $\gamma = 28.7$ M(rad s$^{-1}$), and $I_{\rm sat} = 24.98$ Wm$^{-2}$. \textbf{(a)} Plot of $\sigma_{\tilde{\phi}_{{\textrm{TOT}},B}}^{2}/t_{m}$ against $w_{0,t}/w_{0,b}$ for $\kappa_{b}=\kappa_{t}=\kappa_{bt}$. \textbf{(b)} Plot of $\sigma_{\tilde{\phi}_{{\textrm{TOT}},B}}^{2}/t_{m}$ against $\log_{10}(\kappa_{bt}/\kappa_{b})$. We have set $\kappa_{b} = \kappa_{t}$. The lines with dots are for $w_{0,t} = 1.1~w_{0,b}$ (dotted black line and straight black line cannot be distinguished in the plot). Same color indicates same trap depth (in black and white version, this corresponds to each pair of lines composed of a dashed line and a line-with-dots which starts deviating from this dashed line).}  
    \label{fig:TrapDepthNoise}
\end{figure}



In the case where the noise arises from laser intensity fluctuations due to fluctuations in the photon number, we can use Equations~\ref{eq:whitenoiseconditions1} -- \ref{eq:whitenoiseconditions4} and the relation $V_{d}(t)=k(t)w_{0}^{2}/4$~\cite{OpticalTweezersTheoryAndPractice} to arrive at the variance 
\begin{multline}
    \ \frac{\sigma_{\tilde{\phi}_{{\textrm{TOT}},B}}^{2}}{t_{m}}
 = \frac{1}{16 \hbar^{2}}(\Gamma_{t}w_{0,t}^{4}+\Gamma_{b}w_{0,b}^{4}-2\Gamma_{bt}w_{0,t}^{2}w_{0,b}^{2})
\label{eq:TrapDepthNoise2}
\end{multline}
with $\Gamma_{t},~\Gamma_{b},$ and $\Gamma_{bt}$ as defined previously. The variance goes to zero when two conditions are satisfied: a) $\Gamma_{t}=\Gamma_{b}=\Gamma_{bt}$, which occurs when $\kappa_{t}=\kappa_{b}=\kappa_{bt}$, as can be achieved by using the same beam to create the two traps, \textit{and} b) $w_{0,t}=w_{0,b}$, which occurs when the optical elements used to create the trap are identical. This is explicitly shown in Fig.~\ref{fig:TrapDepthNoise}, which shows plots of $\sigma_{\tilde{\phi}_{{\textrm{TOT}},B}}^{2}/t_{m}$ against $w_{0,t}/w_{0,b}$ for $\kappa_{t}=\kappa_{b}=\kappa_{bt}$, and against $\log_{10}[\kappa_{bt}/\kappa_{b}]$ for $\kappa_{t}=\kappa_{b}$. It can be seen that as $w_{0,t}$ approaches $w_{0,b}$ and as $\kappa_{bt}$ approaches $\kappa_{b}$, the variance of the noise goes to zero, as expected.

\subsection{Trap hamiltonian with only spring constant noise}

Moving on to the time-dependent spring constant, in order to make certain effects of the phenomenon clear, we can write the Hamiltonian~\cite{Heating1,Heating2} 
\begin{equation}
    \ \hat{H} = \frac{\hat{p}^{2}}{2m}+\frac{1}{2}k(t)\hat{x}^{2},
\label{eq:TimeDepH}
\end{equation}
in an alternate form. Note that, for mathematical simplicity, we conduct the upcoming noise calculations for the case of stationary traps and the $-mg\hat{x}$ term has been excluded. We apply a unitary transformation to Eq.~\ref{eq:TimeDepH} to obtain an expression which consists of the simple harmonic potential with a modified mass and spring constant, in addition to a term which drives harmonic oscillator transitions. 

Applying the unitary $\hat{U} = \exp(-i\alpha(t) \{\hat{x},\hat{p}\}/2$), we get the effective Hamiltonian, $\hat{H}_{\rm eff}$, for the state $\hat{U}^\dagger \ket{\beta}$ to be
\begin{equation}
    \ \hat{H}_{\rm eff} = e^{-2 \hbar \alpha}\frac{\hat{p}^{2}}{2m}+\frac{k(t)}{2}e^{2\hbar\alpha}\hat{x}^{2}-\frac{\hbar\Dot{\alpha}}{2}\{\hat{x},\hat{p}\},
\label{eq:SHOHamiltonianSqueezed}
\end{equation}
We can see that there is now an effective mass $m_{\rm eff} = e^{2\hbar \alpha} m$ and an effective spring constant $k_{\rm eff}=k(t)e^{2\hbar\alpha}$. An additional time-dependent term involving $\{\hat{x},\hat{p}\}$ also appears and will lead to oscillator transitions.

$\alpha$ can be expressed in terms of both the unperturbed and time-dependent spring constant, $k_{0}$ and $k(t)$. Factoring out $e^{-2\hbar \alpha}$, we can define $e^{4\hbar \alpha} k(t) \equiv k_{0}$, with which we can rewrite $\hat{H}_{\rm eff}$ to be 
\begin{equation}
    \ \hat{H}_{\rm eff} = e^{-2\hbar \alpha}\left(\frac{\hat{p}^{2}}{2m}+\frac{k_{0}}{2}\hat{x}^{2}\right)-\frac{\hbar\Dot{\alpha}}{2}\{\hat{x},\hat{p}\}.
\label{eq:SHOHamiltonianSqueezed}
\end{equation}
Since the first two terms are conveniently in the form of the conventional simple harmonic oscillator potential, we can straightforwardly express $\hat{H}_{\rm eff}$ in terms of the creation and annihilation operators~\cite{Sakurai},
\begin{equation}
   \ \hat{H}_{\rm eff}= \hbar\sqrt{\frac{k(t)}{k_{0}}}\omega_{0} \left(\Hat{a}^\dagger \Hat{a} + \frac{1}{2}\right)-\frac{i\hbar^{2}\Dot{\alpha}}{2}(\hat{a}^{\dagger 2}-\hat{a}^{2}).
\label{eq:AdiabaticEffectiveHamiltonian}
\end{equation}

After this unitary transformation, $\Hat{a}^\dagger$ and $\Hat{a}$ become time-independent; for instance, in
\begin{equation}
   \ \Hat{a} = \sqrt{\frac{m_{\rm eff}\omega_{\rm eff}}{2\hbar}} \left( \hat{x} + \frac{i\hat{p}}{m_{\rm eff}\omega_{\rm eff}}\right),
\label{eq:anniop}
\end{equation}
where $\omega_{\rm eff} = e^{-2\hbar \alpha} \omega_{0}$~\cite{Sakurai}, we can see that $m_{\rm eff}\omega_{\rm eff} = m\omega_{0}$, and, therefore, time-independent. This is in contrast to the $m\omega(t)$ factors we would have had with the original Hamiltonian in Equation~\ref{eq:TimeDepH}.

Note that $k(t)$ can be any arbitrary function, since we can always solve for the corresponding $\alpha$. For our case, we consider two specific forms, those during ramp up and hold. When $k(t)$ details the ramp being turned on, Eq.~\ref{eq:AdiabaticEffectiveHamiltonian} describes the conventional harmonic oscillator ground state wavepacket being trapped in an increasingly narrowing harmonic potential. When an atom is trapped in the ``steady state" potential, $k(t) = k_{0} + \delta k(t)$, where $\delta k(t)$ is the noise due to spring constant fluctuations, which will be explored in detail later. Technically, the ramp up also includes $\delta k(t)$, but we do not investigate this in the paper. 

If we consider $k(t)$ to be adiabatic, such that the $\Dot{\alpha}$ term is relatively negligible, we can neglect the $\{\hat{x},\hat{p}\}$ term, resulting in
\begin{equation}
   \ \hat{H}_{\rm eff} \approx \hbar\sqrt{\frac{k(t)}{k_{0}}}\omega_{0} \left(\hat{a}^\dagger \hat{a} + \frac{1}{2}\right).
\label{eq:AdiabaticEffectiveHamiltonianApprox}
\end{equation}
Eq.~\ref{eq:AdiabaticEffectiveHamiltonianApprox} can be used to calculate the suppression of the coherence purely due to a time-dependent spring constant, as will be done in a later section.

The term $\{\hat{x},\hat{p}\}$ can lead to transitions between the ground and second excited state of the conventional harmonic oscillator, and will be separately considered in the next section.

\subsection{Harmonic oscillator transitions}
\label{SHOtransitions}

The effect of the $\{\hat{x},\hat{p}\}$ term can be independently analyzed. The terms square in $\hat{a}^\dagger$ will drive transitions from the harmonic oscillator ground state to the second excited state. It is instructive to investigate the prevalence of such transitions and their effect on observed results. 

The transition probabilities can be obtained using time-dependent perturbation theory. For 
$k(t)$, the probability of transition to the lowest order is~\cite{Sakurai} 
\begin{equation}
   \ p^{(1)}(t) = \left|\frac{\sqrt{2}}{8} \int_{t_{a}}^{t_{b}} \frac{\Dot{k}}{k} \exp\left[2i\int_{0}^{t'}
   \sqrt{\frac{k}{k_{0}}}\omega_{0} ~dt\right] dt' \right|^{2}
\label{eq:TransitionProbability}
\end{equation}
which can be applied to the cases of ramp up and hold separately. 

If we consider ramp up to be exponential and noiseless, $k(t) = k_{0}e^{\zeta (t-t_{\rm ru})}$, such that $\zeta$ characterizes the rate of the process and at $t=t_{\rm ru}$ the maximum potential with spring constant $k_{0}$ is reached. The transition probability $p_{\rm ru}^{(1)}(t)$ will then be 
\begin{multline}
   \ p^{(1)}_{\rm ru}(t) = |\frac{\sqrt{2}}{8}\zeta \exp\left[\frac{-4i\omega_{0}}{\zeta}e^{-\frac{\zeta }{2}t_{\rm ru}}\right]\\  \times \int_{t_{0}}^{t_{0}+t_{\rm ru}}\exp\left[\frac{4i\omega_{0}}{\zeta}e^{\frac{\zeta}{2}(t'-t_{\rm ru})}\right] dt' |^{2}
\label{eq:TransitionProbabilityRamp}
\end{multline}
This appears to be a rather complex expression, involving oscillating functions with exponential arguments. It can be seen that the three characteristic parameters $\omega_{0}$, $\zeta$, and $t_{\rm ru}$ all appear in this argument. It is found that the integral evaluates to a sum of exponential integral functions. If we impose certain limits, for instance, that the ramp up is adiabatic such that $\zeta t_{\rm ru} \ll 1$, then we can express $ p^{(1)}_{\rm ru}(t)$ as a series in $\zeta t_{\rm ru}$. Up to fourth order in $t_{\rm ru}$, the expression straightforwardly evaluates to
\begin{equation}
   \ p_{\rm ru}^{(1)}(t) \approx \frac{\zeta^2}{32}\left(t_{\rm ru}^2+\omega_{0}^2\left(-2+e^{\frac{\zeta t_{0}}{2}}\right)^{2}t_{\rm ru}^{4}\right).
\label{eq:TransitionProbabilityEval}
\end{equation}
We can see that at this order, $t_{\rm ru}$ is raised to the even powers of two and four. $p_{\rm ru}^{(1)}(t)$ also, rather interestingly, depends on $t_{0}$, the time for which the atoms free fall after launch; the longer the free fall, the higher the probability of transition during the ramp.

During hold, $k(t) = k_{0}+\delta k(t)$, and Eq.~\ref{eq:TransitionProbability} becomes
\begin{multline}
   \ p^{(1)}_{h}(t) = |\frac{\sqrt{2}}{8} \int_{t_{0}+t_{\rm ru}}^{t_{0}+t_{\rm ru}+t_{t}} \frac{\frac{d}{dt'}(\delta k(t'))}{k_{0}+\delta k(t')}\\  
   \times \exp\left[2i\omega_{0}\int_{0}^{t'}\sqrt{1+\frac{\delta k(t)}{k_{0}}} dt \right] dt'|^{2}
\label{eq:TransitionProbabilityHold}
\end{multline}
Here, $t_{t}$ is the duration of the ``steady" trap potential. Again, we have a rather complicated integrand involving the time derivative of the noise $\delta k(t)$ and an oscillating term with an integral parameter. The details of evaluating this can be found in Appendix~\ref{AppB}, but a summary is given here. First, we consider the magnitude of the noise to be extremely small, such that $\delta k(t)/k_{0} \ll 1$. This allows us to work without the square root in the integral. Afterwards, integration by parts will remove the derivative of $\delta k(t)$. Next, we make an approximation where we consider the contribution by the $\exp\left[\frac{i\omega_{0}}{k_{0}}\int_{0}^{t'} \delta k(t)~dt \right]$ to be negligible. After Taylor expanding the logarithmic term and neglecting terms of order $(\delta k(t')/k_{0})^{2}$, we have for the probability of transitions during hold,
\begin{equation}
   \ p^{(1)}_{h}(t) = \frac{1}{8 m k_{0}} \iint_{t_{0}+t_{\rm ru}}^{t_{0}+t_{\rm ru}+t_{t}} dt' dt~ \delta k(t') \delta k(t) e^{2i\omega_{0}(t'-t)} 
\label{eq:TransitionProbabilityHoldEval}
\end{equation}

It can be see that there is a product of the noise term $\delta k(t)$ within the integral. As such, a meaningful quantity to calculate would be the mean of $p^{(1)}_{h}(t)$ taken over several experimental runs, 
\begin{multline}
   \ \langle \langle p^{(1)}_{h}(t) \rangle \rangle = \frac{1}{8 m k_{0}} \iint_{t_{0}+t_{\rm ru}}^{t_{0}+t_{\rm ru}+t_{t}} dt' dt~[\langle \langle \delta k(t') \delta k(t) \rangle \rangle \\
   e^{2i\omega_{0}(t'-t)}].
\label{eq:AvgTransitionProbabilityHold}
\end{multline}
For white noise, where $\langle \langle \delta k(t') \delta k(t) \rangle \rangle = \Gamma \delta (t - t')$, for some $\Gamma$, we get
\begin{equation}
   \ \langle \langle p^{(1)}_{h}(t) \rangle \rangle = \frac{\Gamma t_{t}}{8 m k_{0}}
\label{eq:TransitionProbabilityHoldWhiteNoise}
\end{equation}
As can be seen, for white noise, we obtain a very simple form where $\ \langle \langle p^{(1)}_{h}(t) \rangle \rangle$ is linear in $t_{t}$. The specific form of the quantity $\Gamma$ depends on the exact source of the intensity noise.

We can now numerically evaluate the probabilities using the parameters $t_{0}=0.5$ ms~\cite{AtomInterExpHolger}, $t_{t}=0.6$ ms, $t_{\rm ru}=100$ µs, $\zeta=0.1$ µs, $V_{d}/k_{\rm B} = 1$ mK, $\nu=2015$ T(rad s$^{-1}$) from $\lambda=935.6$ nm~\cite{935.6}, $w_{0} = 38$ $\mu$m, $\kappa = (\pi c)/(lF)$ as derived in Appendix~\ref{IntrinsicNoise}, $l = 200 $ $\mu$m, $F = 10^{4}$, and, for cesium~\cite{SteckCesium}, $m =2.2 \times 10^{-25}$ kg, $\nu_{\rm atom} = 2106$ T(rad s$^{-1}$), $\gamma = 28.7$ M(rad s$^{-1}$), and $I_{\rm sat} = 24.98$ Wm$^{-2}$. We find that $p_{\rm ru}^{(1)}(t) \approx 8.6 \times 10^{-24}$, indicating that harmonic oscillator transitions during ramp up are negligible.  
$\Gamma$ for intrinsic noise due to photon number fluctuations was calculated in Section~\ref{IntrinsicNoiseMain}, and for this case, we find that $\ \langle \langle p^{(1)}_{h}(t) \rangle \rangle \approx 1.3 \times 10^{-12}$, indicating that harmonic oscillator transitions during hold are negligible.

Going back to the harmonic oscillator transitions when the atoms are trapped, we can rewrite the mean of the hold transition probability, Eq.~\ref{eq:AvgTransitionProbabilityHold}, using the spectral density of the spring constant, given by Eq.~\ref{eq:SISKrelation}:
\begin{multline}
   \ \langle \langle p^{(1)}_{h}(t) \rangle \rangle = \frac{1}{32 \pi m k_{0}} \langle \langle k \rangle \rangle^{2}\\
   \times
   \int_{-\infty}^{\infty} d\omega~S_{k}(\omega/2\pi)\left[\frac{2-2\cos[(\omega-2\omega_{0})\,t_{t}]}{(\omega-2\omega_{0})^{2}}\right]
\label{eq:AvgTransitionProbabilityHold2}
\end{multline}
where $\omega = 2\pi f$. It can be seen that the integrand is peaked about $\omega = 2\omega_{0}$, at which point it takes its highest value of $S_{k}(\omega/2\pi) t_{t}^{2}$. 


\subsection{Parametric dephasing}

Here, we investigate noise in the case where the effect only comes from the time-dependent spring constant. To reiterate, the calculations are for the case of stationary traps with the exclusion of the $-mg\hat{x}$ term. Considering Eq.~\ref{eq:AdiabaticEffectiveHamiltonianApprox} when the atom is trapped after ramp up, that is, when $k(t) = k_{0}+\delta k (t)$, under the assumptions that ramp up and ramp down are identical, the coherence operator of Eq.~\ref{eq:CoherencePathDiff} is given by
\begin{equation}
   \ [\hat{U}^{T\dagger}_{{\textrm{hold}}}~\hat{U}^{T}_{{\textrm{hold}}}|_{\Delta \hat{x}=-L}]^{*} = \exp\left[i \hat{a}^{\dagger} \hat{a} \Delta \phi_{{\textrm{eff}}} + \frac{i}{2}\Delta\phi_{{\textrm{eff}}}\right].
\label{eq:AdiabaticCoherenceTerm}
\end{equation}
Note that here we have also excluded the phase due to uncorrelated trap depths, whose noise effects we already evaluated and was given by Equation~\ref{eq:TrapDepthNoise2}. The $*$ indicates that this is the coherence operator subject to the assumptions and exclusions stated here.
We have
\begin{equation}
\begin{split}
    \ \Delta\phi_{{\textrm{eff}}} & = \int_{t_{0}+t_{r}}^{t_{0}+t_{r}+t_{t}} \omega_{0,b}\sqrt{\frac{k_{0,b}+\delta k_{b}(t)}{k_{0,b}}} dt\\
    & - \int_{t_{0}+t_{r}}^{t_{0}+t_{r}+t_{t}} \omega_{0,t}\sqrt{\frac{k_{0,t}+\delta k_{t}(t)}{k_{0,t}}} dt.
\end{split}
\label{eq:WEffAnalytical}
\end{equation}

The mean of Eq.~\ref{eq:AdiabaticCoherenceTerm} is taken between the arbitrary state $\sum_{n=0}^{\infty} c_{n} \ket{n}$, where $\ket{n}$ is the $n^{th}$ simple harmonic oscillator eigenstate. With the number operator $N = a^{\dagger} a$~\cite{Sakurai} as per usual, in the case where $\Delta\phi_{{\textrm{eff}}}$ has a Gaussian distribution, the mean of the noisy part of the coherence term takes the form
\begin{multline}
   \ \langle\langle~\langle   [\hat{U}^{T\dagger}_{{\textrm{hold}}}~\hat{U}^{T}_{{\textrm{hold}}}|_{\Delta \hat{x}=-L}]^{*} \rangle~\rangle\rangle \\ = \sum_{n=0}^{\infty} |c_{n}|^{2}
   \exp\left[-\frac{(n+\frac{1}{2})^{2}}{2}(\langle\langle (\Delta\phi_{{\textrm{eff}}})^{2}\rangle\rangle-\langle\langle \Delta\phi_{{\textrm{eff}}}\rangle\rangle^{2})\right]
\label{eq:AdiabaticCoherenceTermMean}
\end{multline}
As previously mentioned, we can consider the case where the ramp up process takes the wave packet to the ground state of the harmonic oscillator, that is, $\hat{U}^{T}_{\textrm{ru}}\ket{\tilde{\psi}}=\ket{0}$, where $\ket{\tilde{\psi}} = \hat{U}^{F}\ket{\psi_{t = 0}}$ from Equation~\ref{eq:CoherencePathDiff}. In this instance, only the $n = 0$ term of Eq.~\ref{eq:AdiabaticCoherenceTermMean} will remain, giving
\begin{multline}
   \ \langle\langle~\langle [\hat{U}^{T\dagger}_{{\textrm{hold}}}~\hat{U}^{T}_{{\textrm{hold}}}|_{\Delta \hat{x}=-L}]^{*}\rangle~\rangle\rangle \\ = \exp\left[\frac{-1}{8}(\langle\langle (\Delta\phi_{{\textrm{eff}}})^{2}\rangle\rangle-\langle\langle \Delta\phi_{{\textrm{eff}}}\rangle\rangle^{2})\right],
\label{eq:AdiabaticCoherenceTermMean2}
\end{multline}
where $|c_{0}|^2$ has been taken to be 1 to be consistent with the fact that we consider only the state $\ket{0}$ to be produced. 

Taking ($\delta k(t)/k_{0}) \ll 1$ and expanding the integrands of Eq.~\ref{eq:WEffAnalytical} up to first order, we get  
\begin{equation}
   \ \langle\langle\Delta\phi_{{\textrm{eff}}}\rangle\rangle = (\omega_{0,b}-\omega_{0,t})t_{t}
\label{eq:WEffMean}
\end{equation}
\begin{equation}
\begin{split}
    \ \langle\langle(\Delta\phi_{{\textrm{eff}}})^{2}\rangle\rangle & = (\omega_{0,b}-\omega_{0,t})^{2}t_{t}^{2}+\frac{\omega_{0,b}^{2}\Gamma_{b}t_{t}}{4k_{0,b}^{2}}+\frac{\omega_{0,t}^{2}\Gamma_{t}t_{t}}{4k_{0,t}^{2}}\\
    & -\frac{\omega_{0,b}\omega_{0,t}\Gamma_{bt}t_{t}}{2 k_{0,b} k_{0,t} },
\end{split} 
\label{eq:WEffSquare}
\end{equation}
which are unitless quantities. They have been derived under the white noise conditions of Equations~\ref{eq:whitenoiseconditions1} -- \ref{eq:whitenoiseconditions4}. 


Substituting Equations~\ref{eq:WEffMean} and \ref{eq:WEffSquare} into Eq.~\ref{eq:AdiabaticCoherenceTermMean2}, under the condition that $\omega_{0,b}\approx\omega_{0,t}$, we get
\begin{equation}
\begin{split}
   \ \langle\langle~\langle [\hat{U}^{T\dagger}_{{\textrm{hold}}}~\hat{U}^{T}_{{\textrm{hold}}}|_{\Delta \hat{x}=-L}]^{*}\rangle~\rangle\rangle  & = \exp\left[-\phi_{coh}\right]\\
   & = \exp\left[-\frac{(\Gamma_{t}+\Gamma_{b}-2\Gamma_{bt})t_{t}}{32 m k_{0,b}}\right].
\end{split}
\label{eq:AdiabaticCoherenceTermMean3}
\end{equation}
This is then the most general explicit expression for the coherence. Its specific form for a particular setup will depend on the forms that $\Gamma_{t}, \Gamma_{b}$, and $\Gamma_{bt}$ take.

\begin{figure}[htb]
        \includegraphics[width = 8.6 cm]{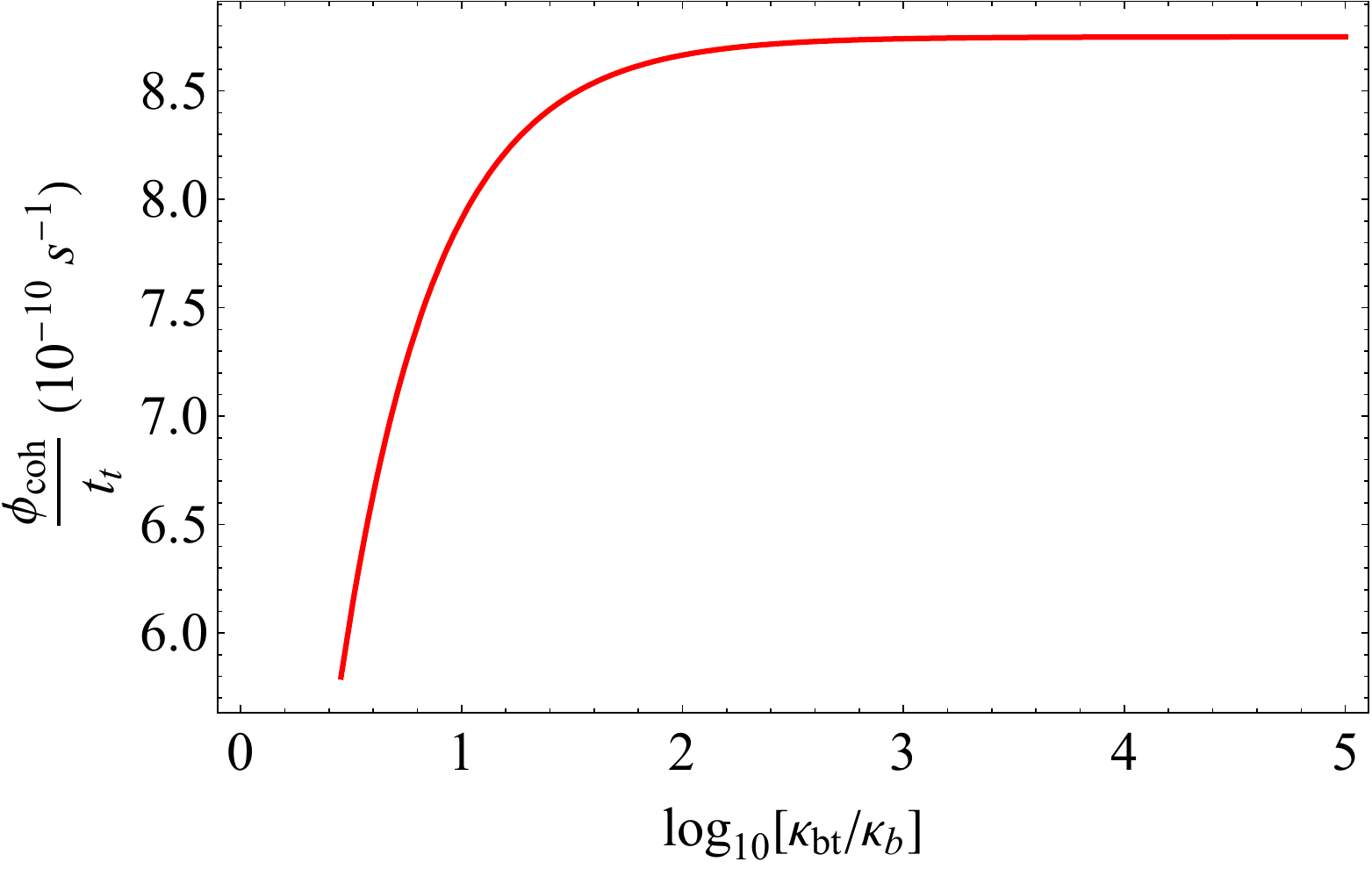}
        \caption{Plot of $\phi_{coh}/t_{t}$ from Eq. \ref{eq:AdiabaticCoherenceTermMean3} against $\log_{10}[\kappa_{bt}/\kappa_{b}]$. $m = 2.2 \times 10^{-25}$ kg, $\nu_{\rm atom} = 2106$ T(rad s$^{-1}$), $\gamma = 28.7$ M(rad s$^{-1}$), and $I_{\rm sat} = 24.98$ Wm$^{-2}$ for cesium~\cite{SteckCesium}. $\nu_{b} = \nu_{t} = 2015$ T(rad s$^{-1}$) from $\lambda=935.6$ nm~\cite{935.6}, $w_{0,b} = 38$ $\mu$m, $w_{0,t} = 1.1 w_{0,b}$, $\kappa_{(b/t)} = (\pi c)/(l_{(b/t)}F)$ as derived in Appendix~\ref{IntrinsicNoise}, $l_{b} = 200 $ $\mu$m, $l_{t}=l_{b}$, $F = 10^{4}$, and $\kappa_{b}=\kappa_{t}$.}
    \label{fig:CoherenceDecayRateVsFreq}
\end{figure}

We evaluate for the case where the intensity noise is caused by variations in photon number. A plot of the decay rate $\phi_{coh}/t_{t}$ is shown in  Figure~\ref{fig:CoherenceDecayRateVsFreq}. Comparing to the plots of trap depth noise in Fig.~\ref{fig:TrapDepthNoise}, the suppression of the coherence during hold due to \textit{only} the effects of Eq. \ref{eq:AdiabaticCoherenceTerm} is relatively negligible.

\section{Outlook}

In this work we have considered the potential benefits and challenges of using optical tweezers to implement a trapped atom interferometer where the traps can be moved throughout the interference process. We find that previously unreached accelerations (and acceleration sensitivities) may be feasible for such interferometers, opening up the potential for high bandwidth observations. However, two major technical hurdles are in the way of achieving these outcomes. First, the dephasing due to intensity fluctuations of the trapping light is a substantial limit to this design, and likely requires using a single beam design to make trap-to-trap variations minimal. Second, the most impressive improvements in sensitivity require moving well beyond the typical image plane of an acousto-optic deflector. Thus investigation of alternative beam steering technologies is in order. Regardless of these technical hurdles, the ability to control the trap motion means that there may be opportunities for the use of these devices to trace interesting spacetime areas, including for rotation sensing and higher-order gravity gradients, as well as for time dependent signals; future work will be necessary to fully identify the most important regimes of interest.

We remark that a similar work was suggested around the time of our first version's submission to arXiv~\cite{SimilarAtomInter}. 

\section{Acknowledgements}
We thank C.~Panda, H.~Müller, W.~D.~Phillips, D.~Carney, W.~McGehee, M. Anderson, and A. Duspayev for helpful conversations. Partial support for this was provided by the Heising-Simons Foundation.

\appendix

\section{Derivation of interferometer sequence final state}
\label{InterFinState}

To obtain Equations~\ref{eq:Coherence} through \ref{eq:UBot}, a mathematical description for the setup is first written in terms of unitary operators. Since our experimental setup makes use of two internal states of the atom, we will follow the effect of the interferometer sequence for a two-level atom, where the relevant states are labeled as $\ket{g}$ and $\ket{e}$. The unitaries for the $\pi/2$ and $\pi$ pulses, including both the spin and spatial components, with the spatial part consisting of exponential momentum kicks provided by the pulses, are given by~\cite{AtomInterLightPulses,JonInterPublished}
\begin{equation}
    \begin{split}
        \ \hat{U}^{\pi/2}(t) & = \frac{1}{\sqrt{2}} \textbf{I}-\frac{i}{\sqrt{2}}(\ket{e}\bra{g}\otimes e^{-iq\hat{x}}\\
       & + \ket{g}\bra{e}\otimes e^{iq\hat{x}})
    \end{split}
\label{eq:Pi2Pulse}
\end{equation}
\begin{equation}
    \begin{split}
        \ \hat{U}^{\pi}(t) & = -i(\ket{e}\bra{g}\otimes e^{-iq\hat{x}}
    \\ & +\ket{g}\bra{e}\otimes e^{iq\hat{x}})
    \end{split}
\label{eq:PiPulse}
\end{equation}
where $\textbf I$ is the identity on both the internal and spatial degrees of freedom, and $\hbar q$ is the magnitude of the momentum kick. From these unitaries, it can be seen that the $\pi/2$ pulse results in the superposition of the states $\ket{g}$ and $\ket{e}$ (along with lower and higher momenta) when acting on either one of $\ket{g}$ or $\ket{e}$, while the $\pi$ pulse takes each of these states to the other.

We can now proceed following a similar method to that in Reference~\cite{JonInterPublished}. Using the pulse unitaries, we can obtain the final state $\ket{\chi_{f}}$ for the experimental sequence described previously. If we start with $\ket{\chi_{t=0}} = \ket{g}\otimes\ket{\psi_{t=0}}$, where $\ket{\psi_{t=0}}$ is the initial spatial component, we find that
\begin{equation}
    \begin{split}
    \ \ket{\chi_{f}} & = \hat{U}^{\pi/2}\hat{U}^{F}\hat{U}^{\pi}_{\textrm{bot}}\hat{U}^{T}\hat{U}^{\pi}_{\textrm{top}}\hat{U}^{F}\hat{U}^{\pi/2}\ket{\chi_{t=0}}\\
    & = \ket{g} \otimes \frac{-1}{2}[\hat{U}_{\textrm{top}}+\hat{U}_{\textrm{bot}}]\ket{\psi_{t=0}}\\
     & + \ket{e}\otimes \frac{i}{2}e^{-iq\hat{x}}[\hat{U}_{\textrm{top}}-\hat{U}_{\textrm{bot}}]\ket{\psi_{t=0}}
    \end{split}
\label{eq:FinalState}
\end{equation}
Here, $\hat{U}^{\pi}_{\rm top/bot}$ is a unitary such that the resulting effect is that of a $\pi$ pulse acting only on the top or bottom arm. For example, after the first $\hat{U}^{F}$, we have the state 
\begin{equation}
    \begin{split}
    \ \ket{\chi_{U^{F}}} & = \frac{1}{\sqrt{2}}\ket{g}\otimes \hat{U}^{F}\ket{\psi_{t=0}}-\frac{i}{\sqrt{2}}\ket{e}\otimes \hat{U}^{F}e^{-iq\hat{x}}\ket{\psi_{t=0}}
    \end{split}
\label{eq:FirstUF}
\end{equation}
Applying $\hat{U}^{\pi}_{\textrm{top}}$ to this gives 
\begin{equation}
    \begin{split}
    \ \ket{\chi_{U^{\pi}_{\textrm{top}}}} & = \frac{1}{\sqrt{2}}\ket{g}\otimes \hat{U}^{F}\ket{\psi_{t=0}}
    \\ & -\hat{U}^{\pi}\frac{i}{\sqrt{2}}\ket{e}\otimes \hat{U}^{F}e^{-iq\hat{x}}\ket{\psi_{t=0}} \\
    & = \frac{1}{\sqrt{2}}\ket{g}\otimes \hat{U}^{F}\ket{\psi_{t=0}}\\
    & -\frac{1}{\sqrt{2}}\ket{g}\otimes e^{iq\hat{x}} \hat{U}^{F}e^{-iq\hat{x}}\ket{\psi_{t=0}}
    \end{split}
\label{eq:UPiTop}
\end{equation}
Both the first two pulses have a momentum kick of magnitude $\hbar q$. This is necessary if both arms are to be at zero momentum at time $t_{0}$. While the momentum kicks for the last pair of pulses can be, in principle, arbitrary, choosing them to be $\hbar q$ means that they would be separated in time by $t_{0}$ as well. 

Equations~\ref{eq:UTop} and \ref{eq:UBot} were obtained by identifying the spatial unitary sequences corresponding to state $\ket{g}$ of the top and bottom arms, as signified in Equation~\ref{eq:FinalState}.
From Eq.~\ref{eq:FinalState}, we obtain the probability of measuring state $\ket{g}$ of the atoms at the end of the interferometer sequence, given by $P_{g}$, and the expression for the coherence, Equation~\ref{eq:Coherence}.

Next, we elaborate on the derivation of  Equation~\ref{eq:CoherencePathDiff}. Taking $\hat{U}^{F}$ to be 
\begin{equation}
    \ \hat{U}^{F} = \exp\left[-\frac{i}{\hbar}\left(\frac{\hat{p}^{2}}{2m}-mg\hat{x}\right)t_{0}\right],
\label{eq:UFgravity}
\end{equation}
the last three terms in Eq.~\ref{eq:UTop} give
\begin{multline}
     \ \hat{U}_{\textrm{top}} = \hat{U}^{F}_{(t_{0}+\tau,2t_{0}+\tau)} \hat{U}^{T}_{(t_{0},t_{0}+\tau)}\\ \times \exp\left[-\frac{i}{\hbar}\left(\frac{(\hat{p}-\hbar q)^{2}}{2m}- mg\hat{x}\right)t_{0}\right],
\label{eq:UTop2}
\end{multline}  
and the first three terms in Eq.~\ref{eq:UBot},
\begin{multline}
    \ \hat{U}_{\textrm{bot}} =\exp\left[-\frac{i}{\hbar}\left(\frac{(\hat{p}-\hbar q)^{2}}{2m}- mg\hat{x}\right)t_{0}\right]\\ \times \hat{U}^{T}_{(t_{0},t_{0}+\tau)} \hat{U}^{F}_{(0,t_{0})}
\label{eq:UBot2}
\end{multline}
Using these, the coherence can then be evaluated to get Equation~\ref{eq:CoherencePathDiff}. 

\section{Calculation of final state of atom in moving trap}
\label{AtomInMovingTrap}

This section elaborates on the calculation in Section~\ref{MovingTrapConcept}.

Starting with Eq. \ref{eq:MovingTrapState}, we first conduct a manipulation as follows:
{\small
\begin{equation}
    \ \frac{1}{2}m\omega_{0}^{2}(\hat{x}-s_{b}(t))^{2}-mg\hat{x} = \frac{1}{2}m\omega_{0}^{2}(\hat{x}-\tilde{s}_{b}(t))^{2} - mgs_{b}(t)-\frac{mg^{2}}{2\omega_{0}^{2}},
\label{eq:MovingTrapSub}
\end{equation}}
where $\tilde{s}_{b}(t) = s_{b}(t)+\frac{g}{\omega_{0}^{2}}$. The final state can now be written as
\begin{multline}
    \ \ket{\psi_{b,f}} = \exp\left[\frac{i}{\hbar}V_{b,d}t_{m}\right]\Phi_{sc,b} \\
    \times \hat{T}\exp\left[-\frac{i}{\hbar} \int_{0}^{t_{m}} dt~\left(\frac{\hat{p}^{2}}{2m}+\frac{1}{2}m\omega_{0}^{2}(\hat{x}-\tilde{s}_{b}(t))^{2}\right)\right]\ket{0},
\label{eq:MovingTrapState2}
\end{multline}
which is in terms of the explicit integral phase due to $s_{b}(t)$ and a time ordered operator. Here,
\begin{equation}
    \ \Phi_{sc,b} = \exp\left[\frac{i}{\hbar}\int_{0}^{t_{m}}dt~ \left(mgs_{b}(t)+\frac{mg^{2}}{2\omega_{0}^{2}}\right)\right]
\label{eq:SCphaseBot}
\end{equation}

For future ease of calculation, we set $s_{b}(0)=-g/\omega_{0}^{2}$, so that $\tilde{s}_{b}(0) = 0$. This simply means that we are considering a diagram which is that of Fig.~\ref{fig:InterSequence_b} subjected to a translation along the x-axis by $-g/\omega_{0}^{2}$ (zero gravitational potential is still at $x=0$). With $\tilde{s}_{b}(0) = 0$ and 
\begin{equation}
    \ H_{0} =  \frac{\hat{p}^{2}}{2m}+\frac{1}{2}m\omega_{0}^{2}\hat{x}^{2},
\label{eq:H0}
\end{equation}
Eq.~\ref{eq:MovingTrapState2} can now be evaluated. The time-ordered element of Eq.~\ref{eq:MovingTrapState2}, 
\begin{equation}
    \ket{\tilde{\psi}_{f}} = \hat{T}\exp\left[-\frac{i}{\hbar} \int_{0}^{t_{m}} dt\left(\frac{\hat{p}^{2}}{2m}+\frac{1}{2}m\omega_{0}^{2}(\hat{x}-\tilde{s}_{b}(t))^{2}\right)\right]\ket{0},
\end{equation}
can be split into a time-ordered product of exponential terms consisting of the Hamiltonian $\hat{H}_{0}$ shifted in space as follows:
\begin{multline}
    \ \ket{\tilde{\psi}_{f}} = e^{-\frac{i}{\hbar}\hat{p}\tilde{s}_{b}(t_{m}-\Delta t)}e^{-\frac{i}{\hbar}\hat{H}_{0}\Delta t}e^{\frac{i}{\hbar}\hat{p}\tilde{s}_{b}(t_{m}-\Delta t)}\\
    \times e^{-\frac{i}{\hbar}\hat{p}\tilde{s}_{b}(t_{m}-2\Delta t)}e^{-\frac{i}{\hbar}\hat{H}_{0}\Delta t}e^{\frac{i}{\hbar}\hat{p}\tilde{s}_{b}(t_{m}-2\Delta t)}...\\
    \times ... e^{-\frac{i}{\hbar}\hat{p}\tilde{s}_{b}(0)}e^{-\frac{i}{\hbar}\hat{H}_{0}\Delta t}e^{\frac{i}{\hbar}\hat{p}\tilde{s}_{b}(0)} \ket{0},
\end{multline}
where $\Delta t = t_{m}/N$ in the limit $N \rightarrow \infty$, $\tilde{s}_{b}(t_{m}-n\Delta t)$ is $\tilde{s}_{b}$ evaluated at $t_{m}-n\Delta t$ with $n$ an integer, and $\hat{H}_{0}$ is as in Eq.~\ref{eq:H0}, 
\begin{equation}
    \ \hat{H}_{0} =  \frac{\hat{p}^{2}}{2m}+\frac{1}{2}m\omega_{0}^{2}\hat{x}^{2}.
\end{equation}

We can combine the exponents of the consecutive terms $e^{\frac{i}{\hbar}\hat{p}\tilde{s}_{b}(t_{m}-\Delta t)} e^{-\frac{i}{\hbar}\hat{p}\tilde{s}_{b}(t_{m}-2\Delta t)}$ to be $e^{\frac{i}{\hbar}\hat{p}\Dot{\tilde{s}}_{b}(t_{m}-2\Delta t)\Delta t}$, such that
\begin{multline}
    \ \ket{\tilde{\psi}_{f}} = e^{-\frac{i}{\hbar}\hat{p}\tilde{s}_{b}(t_{m}-\Delta t)}e^{-\frac{i}{\hbar}\hat{H}_{0}\Delta t}e^{\frac{i}{\hbar}\hat{p}\Dot{\tilde{s}}_{b}(t_{m}-2\Delta t)\Delta t}\\
    \times e^{-\frac{i}{\hbar}\hat{H}_{0}\Delta t}e^{\frac{i}{\hbar}\hat{p}\Dot{\tilde{s}}_{b}(t_{m}-3\Delta t)\Delta t}e^{-\frac{i}{\hbar}\hat{H}_{0}\Delta t}e^{\frac{i}{\hbar}\hat{p}\Dot{\tilde{s}}_{b}(t_{m}-4\Delta t)\Delta t}...\\
    \times ... e^{-\frac{i}{\hbar}\hat{H}_{0}\Delta t}e^{\frac{i}{\hbar}\hat{p}\Dot{\tilde{s}}_{b}(0)\Delta t}e^{-\frac{i}{\hbar}\hat{H}_{0}\Delta t}e^{\frac{i}{\hbar}\hat{p}\tilde{s}_{b}(0)} \ket{0}.
\label{eq:MovingTrapState3}
\end{multline}

Unitaries of integer multiples of $\pm \hat{H}_0 \Delta t$ can now be inserted. This serves to ``push" the various $e^{-\frac{i}{\hbar}\hat{H}_{0}\Delta t}$ terms in Eq.~\ref{eq:MovingTrapState3} all the way to the right. Explicitly, this manipulation works as follows:
\begin{multline}
    \ \ket{\tilde{\psi}_{f}} = e^{-\frac{i}{\hbar}\hat{p}\tilde{s}_{b}(t_{m}-\Delta t)}e^{\frac{i}{\hbar}\hat{H}_{0}\Delta t} e^{-\frac{i}{\hbar}\hat{H}_{0}\Delta t}\\
    e^{-\frac{i}{\hbar}\hat{H}_{0}\Delta t}e^{\frac{i}{\hbar}\hat{p}\Dot{\tilde{s}}_{b}(t_{m}-2\Delta t)\Delta t}e^{\frac{i}{\hbar}\hat{H}_{0}2\Delta t}\\ e^{-\frac{i}{\hbar}\hat{H}_{0}2\Delta t}
  e^{-\frac{i}{\hbar}\hat{H}_{0}\Delta t}e^{\frac{i}{\hbar}\hat{p}\Dot{\tilde{s}}_{b}(t_{m}-3\Delta t)}e^{\frac{i}{\hbar}\hat{H}_{0}3\Delta t}\\e^{-\frac{i}{\hbar}\hat{H}_{0}3\Delta t}e^{-\frac{i}{\hbar}\hat{H}_{0}\Delta t}e^{\frac{i}{\hbar}\hat{p}\Dot{\tilde{s}}_{b}(t_{m}-4\Delta t)\Delta t}e^{\frac{i}{\hbar}\hat{H}_{0}4\Delta t}\\
    e^{-\frac{i}{\hbar}\hat{H}_{0}4\Delta t} ...~ e^{-\frac{i}{\hbar}\hat{H}_{0}(N-1)\Delta t}e^{-\frac{i}{\hbar}\hat{H}_{0}\Delta t}e^{\frac{i}{\hbar}\hat{p}\Dot{\tilde{s}}_{b}(0)\Delta t}e^{\frac{i}{\hbar}\hat{H}_{0}N\Delta t}\\e^{-\frac{i}{\hbar}\hat{H}_{0}N\Delta t} e^{-\frac{i}{\hbar}\hat{H}_{0}\Delta t}e^{\frac{i}{\hbar}\hat{p}\tilde{s}_{b}(0)} \ket{0}.
\label{eq:MovingTrapState4}
\end{multline}
It can be seen that the unitaries inserted are of the form $e^{\frac{i}{\hbar}\hat{H}_{0}n\Delta t}e^{-\frac{i}{\hbar}\hat{H}_{0}n\Delta t}$, with $n$ increasing to the right. Looking at the second to the fourth lines of Eq.~\ref{eq:MovingTrapState4}, it can be seen that the operator $\hat{p}$ is being evolved for time $-n\Delta t$ through the Hamiltonian $\hat{H}_{0}$. Denoting this as $\hat{p}_{H}(-n\Delta t)$, we get
\begin{multline}
    \ \ket{\tilde{\psi}_{f}} = e^{-\frac{i}{\hbar}\hat{p}\tilde{s}_{b}(t_{m}-\Delta t)} e^{\frac{i}{\hbar}\hat{H}_{0}\Delta t} e^{\frac{i}{\hbar}\hat{p_{H}}(-2\Delta t)\Dot{\tilde{s}}_{b}(t_{m}-2\Delta t)}\\
    e^{\frac{i}{\hbar}\hat{p_{H}}(-3\Delta t)\Dot{\tilde{s}}_{b}(t_{m}-3\Delta t)}...~ e^{\frac{i}{\hbar}\hat{p_{H}}(\Delta t - t_{m})\Dot{\tilde{s}}_{b}(\Delta t)\Delta t}\\
    e^{\frac{i}{\hbar}\hat{p_{H}}(-t_{m})\Dot{\tilde{s}}_{b}(0)\Delta t}e^{-\frac{i}{\hbar}\hat{H}_{0}N\Delta t}e^{-\frac{i}{\hbar}\hat{H}_{0}\Delta t}e^{\frac{i}{\hbar}\hat{p}\tilde{s}_{b}(0)} \ket{0}.
\label{eq:MovingTrapState5}
\end{multline}

This can now be expressed in integral form in the limit $\Delta t \rightarrow 0$ as
\begin{equation}
    \ \ket{\tilde{\psi}_{f}} = e^{-\frac{i}{\hbar}\hat{p}\tilde{s}_{b}(t_{m})}e^{\frac{i}{\hbar}\int_{0}^{t_{m}} dt~ \hat{p}_{H}(t-t_{m})\Dot{\tilde{s}}_{b}(t)}e^{-\frac{i}{\hbar}\hat{H}_{0}t_{m}}e^{\frac{i}{\hbar}\hat{p}\tilde{s}_{b}(0)} \ket{0}
\end{equation}

With
\begin{equation}
    \ \hat{\Tilde{x}} = \sqrt{\frac{\hbar}{2m\omega_{0}}}(\hat{a}^{\dagger}+\hat{a}),  
\end{equation}
leading to $\hat{p}_{H}(-\Delta t)$ simply taking the form of momentum evolved through simple harmonic motion~\cite{Sakurai,Clerk},
\begin{equation}
    \ \hat{p}_{H}(t) = \frac{1}{i}\sqrt{\frac{m\omega_{0}\hbar}{2}}(\hat{a}e^{-i\omega_{0} t}-\hat{a}^{\dagger}e^{i\omega_{0} t}).
\end{equation}
$\ket{\tilde{\psi}_{f}}$ can now be expressed in terms of the ladder operators~\cite{DisplacementOperator}, 
\begin{equation}
    \ \ket{\tilde{\psi}_{f}} = e^{-\frac{i}{\hbar}\hat{p}\tilde{s}_{b}(t_{m})}e^{\int_{0}^{t_{m}} dt~ [\alpha(t)\hat{a}^{\dagger}-\alpha(t)^{*}\hat{a}]}e^{-\frac{i}{\hbar}\hat{H}_{0}t_{m}}e^{\frac{i}{\hbar}\hat{p}\tilde{s}_{b}(0)} \ket{0}
\label{eq:MovingTrapStateLadderOperators}
\end{equation}
with 
\begin{equation}
    \ \alpha(t) = -\sqrt{\frac{m\omega_{0}}{2\hbar}}\Dot{\tilde{s}}_{b}(t)e^{i\omega_{0}(t-t_{m})}.
\end{equation}

From $D(\beta_{1})D(\beta_{2})=e^{i\Im(\beta_{1}\beta_{2}^{*})}D(\beta_{1}+\beta_{2})$~\cite{DisplacementOperator}, we obtain the relation
\begin{multline}
    \ e^{\int_{0}^{t_{m}} dt~ [\alpha(t)\hat{a}^{\dagger}-\alpha(t)^{*}\hat{a}]} = e^{i\int_{0}^{t_{m}} dt_{2} \int_{0}^{t_{2}} dt_{1}~ \Im[\alpha(t_{1})\alpha(t_{2})^{*}]}\\
    \times D\left(\int_{0}^{t_{m}} dt~\alpha(t)\right)
\end{multline}
Plugging this into Eq.~\ref{eq:MovingTrapStateLadderOperators} and simplifying, additionally using the conditions $\tilde{s}_{b}(0) = 0$ and $H_{0}\ket{0}=(\hbar \omega_{0}/2) \ket{0}$~\cite{Sakurai}, we are finally left with
\begin{multline}
    \ \ket{\tilde{\psi}_{f}} = \exp\left[i(\phi_{b,1}+\phi_{b,2}-\frac{\omega_{0} t_{m}}{2})\right]\\
    \times D\left(\sqrt{\frac{m\omega_{0}}{2\hbar}}\tilde{s}_{b}(t_{m})-\sqrt{\frac{m\omega_{0}}{2\hbar}}\int_{0}^{t_{m}}dt~~\Dot{\tilde{s}}_{b}(t)e^{i\omega_{0}(t-t_{m})}\right) \ket{0},
\end{multline}
where
\begin{equation}
    \ \phi_{b,1} = \frac{m\omega_{0}}{2h}\tilde{s}_{b}(t_{m})\int_{0}^{t_{m}} dt~~ \Dot{\tilde{s}}_{b}(t)\sin[\omega_{0}(t-t_{m})]
\end{equation}
and 
\begin{equation}
    \ \phi_{b,2} = \frac{m\omega_{0}}{2h}\int_{0}^{t_{m}}\int_{0}^{t_{2}} dt_{1} dt_{2}~~ \Dot{\tilde{s}}_{b}(t_{1})\Dot{\tilde{s}}_{b}(t_{2})\sin[\omega_{0}(t_{1}-t_{2})].
\end{equation}

Eq.~\ref{eq:MovingTrapState2} can finally be expressed in terms of the displacement operator as follows: 
\begin{multline}
    \ \ket{\psi_{b,f}} = \Phi_{sc,b} ~\exp\left[i(\phi_{b,1}+\phi_{b,2}-\frac{\omega_{0} t_{m}}{2}+\frac{V_{b,d}t_{m}}{\hbar})\right] \\  D(\alpha_{\textrm{bot}}) \ket{0}
 \label{eq:MovingTrapStateEval}
\end{multline}
where
\begin{multline}
    \ \alpha_{\textrm{bot}} = \sqrt{\frac{m\omega_{0}}{2\hbar}}\tilde{s}_{b}(t_{m})-\sqrt{\frac{m\omega_{0}}{2\hbar}}\int_{0}^{t_{m}}dt~~\Dot{\tilde{s}}_{b}(t)e^{i\omega_{0}(t-t_{m})}.
\label{eq:DAlphaBotApp}
\end{multline}

If the initial state in the trap is not $\ket{0}$, but is instead a coherent state $\ket{\beta}$, the calculations in this section follow up to and including Eq.~\ref{eq:MovingTrapStateLadderOperators} (with $\ket{0}$ replaced by $\ket{\beta}$). From that point onwards, the evaluation proceeds as follows.

Hamiltonian evolution acts on $\ket{\beta}$ as \cite{CoherentStatesSHO}
\begin{equation}
    \ e^{-\frac{i}{\hbar}\hat{H}_{0}t_{m}}\ket{\beta}=\ket{e^{-i\omega_{0}t_{m}}\beta}=\beta(t_{m})
\end{equation}
The final state for the bottom trap will now be given by
{\small
\begin{multline}
    \ \ket{\psi_{b,f}}_{\beta} = \Phi_{sc,b} ~\exp\left[i(\phi_{b,1}+\phi_{b,2}+\frac{V_{b,d}t_{m}}{\hbar}+\Im(\alpha_{\textrm{bot}}\beta^{*}(t_{m}))\right] \\  D(\alpha_{\textrm{bot}}+\beta(t_{m})) \ket{0}.
 \label{eq:MovingTrapStateEvalTherm}
\end{multline}}
Comparing with Eq. \ref{eq:MovingTrapStateEval}, it can be seen that $\ket{\beta}$ simply resulted in shifting the final coherent state and introducing a different phase. We can now write down the updated coherence for $\ket{\beta}$, 
\begin{multline}
    \ C_{\beta}=e^{i\phi_{\textrm{TOT}}}\exp[i{\Im[(\alpha_{top}-\alpha_{bot})\beta^{*}(t_{m})]}]\\
    |\bra{\alpha_{\textrm{bot}}+\beta(t_{m})}\alpha_{\textrm{top}}+\beta(t_{m})\rangle|
 \label{eq:MovingTrapStateEvalTherm2}
\end{multline}
This can be used to evaluate the coherence for the case where the initial trap state is a thermal state. That is, for the thermal state distribution~\cite{ThermalStateRep}
\begin{equation}
    \ P(\beta,\beta^{*}) = \frac{1}{\pi \Bar{n}}e^{-\frac{|\beta|^{2}}{\Bar{n}}},
\end{equation}
the coherence is
\begin{multline}
    \ C_{\textrm{therm}} = e^{i\phi_{\textrm{TOT}}}|\bra{\alpha_{\textrm{bot}}+\beta(t_{m})}\alpha_{\textrm{top}}+\beta(t_{m})\rangle|\\
   \times \int~ d^{2}\beta~ \frac{e^{-|\beta|^{2}/\Bar{n}}}{\pi \Bar{n}} \exp[i{\Im[(\alpha_{\rm{top}}-\alpha_{\rm{bot}})\beta^{*}(t_{m})]}].
\end{multline}
This can be evaluated to obtain
\begin{multline}
        C_{\textrm{therm}} = e^{i \phi_{\textrm{TOT}}}e^{-\frac{\Bar{n}}{4}|\alpha_{\textrm{top}}-\alpha_{\textrm{bot}}|^{2}} \\
        \times |\bra{\alpha_{\textrm{bot}}+\beta(t_{m})}\alpha_{\textrm{top}}+\beta(t_{m})\rangle|.
\label{eq:CoherencePathDiffThermApp}
\end{multline}
Combining this with the fact that $|\bra{\alpha_{\textrm{bot}}+\beta(t_{m})}\alpha_{\textrm{top}}+\beta(t_{m})\rangle|=|\bra{\alpha_{\textrm{bot}}}\alpha_{\textrm{top}}\rangle|$ and using the relation in Eq. \ref{eq:DAlphaBotSqr}, we get Equation \ref{eq:CoherencePathDiffTherm}.

\section{Sensitivity elaboration}
\label{SensitivityElaboration}

Following from Eq. \ref{eq:Cloudradius}, the density of atoms $\rho_{\rm atom}(t_{0})$ right before the trap is turned on is approximated by
\begin{equation}
    \ \rho_{\rm atom}(t_{0}) = \rho_{\rm MOT} \left(\frac{r_{\rm MOT}}{r(t_{0})}\right)^{3}
\end{equation}    
The volume of a far-off resonance trap can be taken to be~\cite{TrapVolume,TrapHorizPot}
\begin{equation}
    \ V_{\rm trap} = \frac{c_{V}}{\lambda} w_{0}^{4} 
\end{equation}
where
\begin{equation}
    \ c_{V} = -\pi^{2}\ln(1-\eta)\sqrt{\frac{\eta}{1-\eta}}
\end{equation}
where $\eta = k_{\rm B}T_{\rm trap}/V_{d}$ is a constant, whose value we take to be 0.4 as was also given in Ref.~\cite{TrapVolume}, such that
\begin{equation}
    \ V_{\rm trap} = 0.417 \frac{\pi^{2}}{\lambda} w_{0}^{4} 
\end{equation}
For mathematical convenience, making the approximation that the optical dipole trap is turned on abruptly, the number of atoms $N_{\rm atom,trap}$ that get trapped is given by $N_{\rm atom,trap} \approx \rho_{\rm atom}(t_{0}) \times V_{\rm trap}$, such that, 
\begin{equation}
    \ N_{\rm atom,trap} \approx  0.417 \frac{\pi^{2}}{\lambda} \rho_{\rm MOT} \left(\frac{r_{\rm MOT}}{r(t_{0})}\right)^{3}  w_{0}^{4}. 
\end{equation}



\section{Intrinsic laser noise elaboration}
\label{IntrinsicNoise}

In this section, we show the derivation of results of Section~\ref{IntrinsicNoiseMain}. If we consider the peak electric field to be $E_{p}$, we can approximate the electric field operator $\hat{E_{p}}$ in terms of the operator for the number of photons, $\hat{N}$, to be~\cite{Meystre}
\begin{equation}
   \ |\hat{E_{p}}| = \sqrt{\frac{2\hbar\nu(\hat{N}+\frac{1}{2})}{\epsilon_{0}V_{OL}}}
\label{eq:ElectricFieldOperator}
\end{equation}
where $\nu$ is the (noiseless) frequency of the light creating the optical dipole trap, $V_{OL}$ is the volume of the cavity in which the trap is located, and $\hat{N}\ket{n} = N\ket{n}$, with $\ket{n}$ being harmonic oscillator eigenstates and $N$ being the total number of photons. Note that the original formalism in Ref.~\cite{Meystre} is for a one-dimensional cavity, but we very approximately apply it to a Gaussian beam.

With the distribution of $N$ as given in Eq.~\ref{eq:NofPhotonsDist}, we can use the relationship between the intensity and the electric field~\cite{OpticalTweezersTheoryAndPractice},
\begin{equation}
   \ I_{m} = \frac{1}{2}n_{r}\epsilon_{0} c |E_{p}|^{2},
\label{eq:IntensityAndEField}
\end{equation}
to find the distribution of $I_{m}(t)$:
\begin{multline}
   \ \langle\langle I_{m}(t')I_{m}(t) \rangle\rangle - \langle\langle I_{m}(t) \rangle\rangle ^{2} = \left(\frac{ \hbar \nu}{\pi w_{0}^{2}} \right)^{2} \left(\frac{c}{l} \right)^{2} \left(\frac{n_{r}}{ \sqrt{\kappa}} \right)^{2}\Bar{N} \\
   \kappa e^{-\kappa|t'-t|},
\label{eq:IntensityDistAppend}
\end{multline}
where $l$ is the length of the cavity and $n_{r}$ is the cavity's refractive index. Here, $V_{OL} = \pi w_{0}^{2}l$.
Instead of explicitly deriving this expression, we can make a qualitative argument to obtain it. The coefficients consist of the square of the energy per unit area of a single photon divided by the time taken for the photon to cross the cavity (hence, the square intensity of a single photon), multiplied by the distribution of photons in the cavity as given by Equation~\ref{eq:NofPhotonsDist}.

We can use the relation $N=C_{I}I_{m}$, and Equations~\ref{eq:CI} (expression for $C_{I}$), and \ref{eq:SpringConstant} (spring constant in terms of $I_{m}/I_{\rm sat}$), to convert Eq.~\ref{eq:IntensityDistAppend} to Equation~\ref{eq:kDist} (the spring constant noise distribution).

To conduct numerical evaluations, in the case of a cavity, we can express $\kappa$ in terms of the cavity's finesse $F$. Using the following equations from Klein~\cite{Finesse} for a Fabry-Pérot cavity,
\begin{equation}
   \ Q = -2\pi\nu_{c,0}\frac{E}{dE/dt},~Q = \frac{\nu_{c,0}}{\Delta \nu_{c}},~\text{and} ~ F = \frac{c/(2nl)}{\Delta \nu_{c}},
\label{eq:CavityQandFinesse}
\end{equation}
where $Q$ is quality factor of the cavity, $E$ is the energy in the cavity, $\nu_{c,0}$ is the cavity's resonant frequency, $\Delta \nu_{c}$ is the linewidth of the cavity's resonance curve, $l$ its length, and $n$ is an integer, we find that 
\begin{equation}
   \ E =  E_{0}\exp\left[-\kappa t\right]
\label{eq:EnergyTimeDependence}
\end{equation}
where $\kappa = (\pi c)/(nlF)$ and is the constant of energy decay. We set $n=1$.

\section{Center of trap and trap depth noise elaboration}
\label{CenterOfTrapNoiseApp}

Eq.~\ref{eq:MovingTrapStateNoise0} can be rewritten as
{\small
\begin{multline}
    \ \ket{\psi_{b,f}}_{\textrm{noise}} = \tilde{\Phi}_{sc,b} \exp\left[\frac{i}{\hbar}\int_{0}^{t_{m}} dt~ V_{b,d}(t)\right]\\
    ~~~~~\times \hat{T}\exp\left[-\frac{i}{\hbar} \int_{0}^{t_{m}} dt~\left( \frac{\hat{p}^{2}}{2m}+\frac{1}{2}m\omega_{0,b}^{2}\left(\hat{x}-\tilde{\eta_{b}}(t) \right)^{2}\right)\right]\ket{0},
\label{eq:MovingTrapStateNoise1}
\end{multline}}
where $\tilde{\eta_{b}}(t) = s_{b}(t) - \eta_{b}(t)+\eta_{b}(0)+g/\omega_{0,b}^{2}$ with the subscript $b$ denoting that the quantities are for the ``bottom" trap. Here,
{\small
\begin{equation}
    \ \tilde{\Phi}_{sc,b} = \exp\left[\frac{i}{\hbar}\int_{0}^{t_{m}}dt~ \left(-mgs_{\eta,b}(t)+\frac{mg^{2}}{2\omega_{0,b}^{2}}\right)\right]
\label{eq:SCphaseBotNoise}
\end{equation}}
where $s_{\eta,b}(t) = -s_{b}(t)+\eta_{b}(t)-\eta_{b}(0)$.
The time-ordered exponential is of the same form as that evaluated in Section \ref{MovingTrapCoherence}, so its results can be used directly, with $-s_{b}(t)$ being replaced by $s_{\eta,b}(t)$ and $\tilde{s}_{b}(t)$ by $\tilde{\eta}_{b}(t)$. As such, we have the new parameters $\tilde{\alpha}_{\textrm{bot}}(\tilde{\eta}_{b}(t),\Dot{\tilde{\eta_{b}}}(t))$, $\tilde{\phi}_{b,1}(\tilde{\eta_{b}},\Dot{\tilde{\eta_{b}}})$, and $\tilde{\phi}_{b,2}(\Dot{\tilde{\eta_{b}}})$.
Similarly, we obtain the expression for $\ket{\psi_{t,f}}_{\textrm{noise}}=\hat{U}^{T}_{\textrm{move}}|_{\Delta\hat{x}=-L}\ket{0}$, where $\tilde{\alpha}_{\textrm{top}}$ has the same form as $\tilde{\alpha}_{\textrm{bot}}$, with the exception that the  subscripts are now $t$ instead of $b$.

For $\tilde{\phi}_{\textrm{TOT}}$ as given in Eq.~\ref{eq:PhiTotNoise}, setting $\omega_{0,t}=\omega_{0,b}$ to somewhat simplify the expression, $\tilde{\alpha}_{\textrm{bot}}^{*}\tilde{\alpha}_{\textrm{top}}$ is given by

{\small
\begin{multline}
        \ \tilde{\alpha}_{\textrm{bot}}^{*}\tilde{\alpha}_{\textrm{top}} = \frac{m\omega_{0}}{2\hbar}\tilde{\eta}_{b}(t_{m})\tilde{\eta}_{t}(t_{m}) \\
         -\frac{m\omega_{0}}{2\hbar}\int_{0}^{t_{m}} dt~[\tilde{\eta}_{b}(t_{m})\dot{\tilde{\eta}}_{t}(t)+\tilde{\eta}_{t}(t_{m})\dot{\tilde{\eta}}_{b}(t)]\cos[\omega_{0}(t-t_{m})] \\
         + \frac{m\omega_{0}}{2\hbar}\int_{0}^{t_{m}}\int_{0}^{t_{m}}dt~ dt'~ \dot{\tilde{\eta}}_{b}(t)\dot{\tilde{\eta}}_{t}(t')\cos[\omega_{0}(t-t_{m})]\cos[\omega_{0}(t'-t_{m})] \\
        + \frac{m\omega_{0}}{2\hbar}\int_{0}^{t_{m}}\int_{0}^{t_{m}}dt~ dt'~ \dot{\tilde{\eta}}_{b}(t)\dot{\tilde{\eta}}_{t}(t')\sin[\omega_{0}(t-t_{m})]\sin[\omega_{0}(t'-t_{m})] \\
        -i\frac{m\omega_{0}}{2\hbar}\int_{0}^{t_{m}} dt~ [\tilde{\eta}_{b}(t_{m})\dot{\tilde{\eta}}_{t}(t)-\tilde{\eta}_{t}(t_{m})\dot{\tilde{\eta}}_{b} (t)]\sin[\omega_{0}(t-t_{m})]\\
        +i\frac{m\omega_{0}}{2\hbar}\int_{0}^{t_{m}}\int_{0}^{t_{m}}dt~ dt'~\dot{\tilde{\eta}}_{b}(t)\dot{\tilde{\eta}}_{t}(t')\cos[\omega_{0}(t-t_{m})]\sin[\omega_{0}(t'-t_{m})] \\
        -i\frac{m\omega_{0}}{2\hbar}\int_{0}^{t_{m}}\int_{0}^{t_{m}}dt~ dt'~\dot{\tilde{\eta}}_{b}(t)\dot{\tilde{\eta}}_{t}(t')\sin[\omega_{0}(t-t_{m})]\cos[\omega_{0}(t'-t_{m})]
\label{eq:AlphaStarAlpha}
\end{multline}}

Continuing the center of trap noise calculation, with conditions as given in Section \ref{TrapCenterNoise}, we can write the variance of $\tilde{\phi}_{{\textrm{TOT,A}}}(t_{m})$ to be 
\begin{multline}
       \ \sigma_{\tilde{\phi}_{{\textrm{TOT,A}}}}^{2} = \langle\langle \tilde{\phi}_{{\textrm{TOT,A}}}(t_{m})^{2}\rangle\rangle - \ \langle\langle \tilde{\phi}_{{\textrm{TOT,A}}}(t_{m})\rangle\rangle^{2} = \\
        \left(\frac{mg}{\hbar}\right)^{2}\int_{0}^{t_{m}}\int_{0}^{t_{m}} dt~ dt' [\langle\langle \eta_{t}(0)^{2}\rangle\rangle 
        - \langle\langle \eta_{t}(0)\eta_{t}(t')\rangle\rangle \\  - \langle\langle \eta_{t}(t)\eta_{t}(0)\rangle\rangle + \langle\langle \eta_{t}(t)\eta_{t}(t')\rangle\rangle] \\
        +~\text{terms~for~}\eta_{b}.   
\label{eq:TrapCenterSquareMean}
\end{multline}
$\sigma_{\tilde{\phi}_{{\textrm{TOT,A}}}}^{2}$ characterizes the variance in the interferometer space-time area covered.

Now we have an expression for $\sigma_{\tilde{\phi}_{{\textrm{TOT,A}}}}^{2}$ in terms of the correlation functions of the noise terms $\eta_{t}$ and $\eta_{b}$. Plugging in appropriate versions of Eq.~\ref{eq:TrapCenterFourier} by setting none, one, or both of $t$ and $t'$ to 0, we find the general expression
\begin{multline}
       \ \sigma_{\tilde{\phi}_{{\textrm{TOT,A}}}}^{2} = \\
        \left(\frac{mg}{\hbar}\right)^{2}\int_{-\infty}^{\infty} df [S_{\eta,t}(f)+ S_{\eta,b}(f)] \left[\frac{1}{2 \pi^{2}f^{2}} -\frac{\cos(2\pi f t_{m})}{2\pi^{2}f^{2}} ... \right] \\
        \left[...+t_{m}^{2}-\frac{t_{m}\sin(2\pi f t_{m})}{\pi f}\right].
\label{eq:TrapCenterPSD}
\end{multline}

\section{Calculation of transition probability during hold}
\label{AppB}

Here, we detail out the calculation for the probability of transition from the harmonic oscillator ground to second excited state that occurs when the atom is on hold in the trap, as discussed in Section~\ref{SHOtransitions}. Here, $k(t) = k_{0}+\delta k(t)$, and Eq.~\ref{eq:TransitionProbability} becomes
\begin{multline}
   \ p^{(1)}_{h}(t) = |\frac{\sqrt{2}}{8} \int_{t_{0}+t_{r}}^{t_{0}+t_{r}+t_{t}} \frac{\frac{d}{dt'}(\delta k(t'))}{k_{0}+\delta k(t')}\\  
   \times \exp\left[2i\omega_{0}\int_{0}^{t'}\sqrt{1+\frac{\delta k(t)}{k_{0}}} dt \right] dt'|^{2}
\label{eq:TransitionProbabilityHold}
\end{multline}
We can work without the square root in the integral by taking $(\delta k(t)/k_{0}) \ll 1$, such that $p_{h}^{(1)}(t)$ is approximated to be
\begin{multline}
   \ p^{(1)}_{h}(t) = |\frac{\sqrt{2}}{8} \int_{t_{0}+t_{r}}^{t_{0}+t_{r}+t_{t}} \frac{\frac{d}{dt'}(\delta k(t'))}{k_{0}+\delta k(t')}\\  
   \times \exp[2i\omega_{0}t']\exp\left[\frac{i\omega_{0}}{k_{0}}\int_{0}^{t'} \delta k(t)~dt \right] dt'|^{2}
\label{eq:TransitionProbabilityHoldTaylor}
\end{multline}
To remove the derivative of $\delta k(t)$ we can do integration by parts to rewrite the equation as 
\begin{multline}
   \ p^{(1)}_{h}(t) = |\frac{\sqrt{2}}{8}[\ln\left(\frac{k_{0}+\delta k(t')}{k_{0}}\right)\exp[2i\omega_{0}t'] \\ 
  \times \exp\left[\frac{i\omega_{0}}{k_{0}}\int_{0}^{t'} \delta k(t)~dt \right]]^{t_{0}+t_{r}+t_{t}}_{t_{0}+t_{r}}\\   -\frac{\sqrt{2}}{8}\int_{t_{0}+t_{r}}^{t_{0}+t_{r}+t_{t}} \ln\left(\frac{k_{0}+\delta k(t')}{k_{0}}\right)\left(2i\omega_{0}+\frac{i\omega_{0}\delta k(t')}{k_{0}}\right)\\  
  \times \exp[2i\omega_{0}t']\exp\left[\frac{i\omega_{0}}{k_{0}}\int_{0}^{t'} \delta k(t)~dt \right] dt'|^{2}
\label{eq:TransitionProbabilityHoldTaylor}
\end{multline}
Here, we can make an approximation where we consider the contribution by $\exp\left[\frac{i\omega_{0}}{k_{0}}\int_{0}^{t'} \delta k(t)~dt \right]$ to be negligible, such that
\begin{multline}
   \ p^{(1)}_{h}(t) \approx |\frac{\sqrt{2}}{8}\left[\ln\left(\frac{k_{0}+\delta k(t')}{k_{0}}\right)\exp[2i\omega_{0}t']\right]^{t_{0}+t_{r}+t_{t}}_{t_{0}+t_{r}}\\  
   -\frac{\sqrt{2}}{8}\int_{t_{0}+t_{r}}^{t_{0}+t_{r}+t_{t}}dt'~ \ln\left(\frac{k_{0}+\delta k(t')}{k_{0}}\right)\left(2i\omega_{0}+\frac{i\omega_{0}\delta k(t')}{k_{0}}\right)\\  
  \times \exp[2i\omega_{0}t']|^{2}
\label{eq:TransitionProbabilityHoldExpApprox}
\end{multline}
After Taylor expanding the logarithmic term and neglecting terms of order $(\delta k(t')/k_{0})^{2}$, we finally get
\begin{equation}
   \ p^{(1)}_{h}(t) = \frac{1}{8 m k_{0}} \iint_{t_{0}+t_{r}}^{t_{0}+t_{r}+t_{t}} \delta k(t') \delta k(t) e^{2i\omega_{0}(t'-t)} dt' dt.
\label{eq:TransitionProbabilityHoldEval}
\end{equation}

\bibliography{apssamp}

\end{document}